\documentclass[tightenlines,aps,floatfix,preprint,
nofootinbib]{revtex4}
\usepackage{float,epsfig,axodraw,pstricks,graphicx}
\begin{document}

\title{
Numerical evaluation of loop integrals 
}

\author{
Charalampos Anastasiou\thanks{e-mail:babis@phys.ethz.ch}} 
\affiliation{
          Institute for Theoretical Physics,\\ 
          ETH, 8093 Z\"urich, Switzerland}
\author{Alejandro Daleo
        \thanks{e-mail: daleo@physik.unizh.ch}}
\affiliation{Institute for Theoretical Physics\\
          University of Z\"urich,\\ 8057 Z\"urich, Switzerland}  
\preprint{ZU-TH 22/05}

\begin{abstract}
We present a new method for the numerical  evaluation of arbitrary 
loop integrals in dimensional regularization. We first derive 
Mellin-Barnes integral representations and 
apply an algorithmic technique, based on the Cauchy theorem, 
to extract the divergent parts in the $\epsilon \to 0$ 
limit. We then perform an $\epsilon$-expansion and evaluate the integral 
coefficients of the expansion numerically.  The method yields 
stable results in physical kinematic regions avoiding intricate 
analytic continuations. It can also be applied  to evaluate 
both scalar and tensor integrals without employing reduction methods. 
We demonstrate our method with specific examples of infrared divergent 
integrals with many kinematic scales, such as  two-loop and three-loop box 
integrals and  tensor integrals of rank six for the one-loop hexagon 
topology.  
\end{abstract}

\maketitle

\section{Introduction}\label{sec:introduction}
Perturbative methods are indispensable in order to  establish consistent  
theories of particle interactions, and to predict quantitatively their 
experimental manifestations. 
The anticipation of new phenomena in modern experiments and theoretical 
extensions of the  Standard Model, 
requires cross-sections for complicated processes. 
This has  driven a remarkable progress in the development of new computational 
methods. At the one-loop level, the calculation  of  cross-sections with 
five external legs is gradually becoming  a routine activity 
(e.g.~\cite{nlo_example}).  
At two-loops, there have been 
recent successful computations of amplitudes with four external legs and up 
to three kinematic scales
(e.g.~\cite{nnlo_example}). 
At three-loops and beyond, amplitudes with up to one parametric variable have 
also been computed (e.g.~\cite{bnnlo_example}). 

Many new processes with higher 
final-state multiplicity, number of loops, and kinematic scales,
have been identified to be important at 
the TeV energy frontier.  The aim of our paper is to provide a new 
method which can be used to compute loop amplitudes for such, more 
complicated, processes.

Loop integrations are cumbersome due to the  presence of 
infrared singularities. Loop integrals  
with many kinematic scales have, in addition, 
a complicated analytic structure with respect to their kinematic parameters.
The tensor structure in gauge theories is also  an issue, since
it proliferates the number of terms.
A general method  for computing arbitrary loop integrals 
should extract their infrared and ultraviolet singularities,
and treat simply  kinematic discontinuities and threshold 
singularities.  For practical applications, it  should also be able to 
handle tensor integrals efficiently.   
There is no method which addresses satisfactorily all these issues; 
known techniques can compute a limited number of 
amplitudes where  some simplifications occur in special cases.

Following a traditional approach to calculate loop amplitudes, one
reduces the number of terms to a few master integrals and  
computes the latter analytically. One-loop integrals can be 
reduced using the classical method of Passarino and 
Veltman~\cite{pvreduction}. For generic  multi-loop computations one can 
derive reduction identities from integration by parts~\cite{ibp} 
and the invariance of scalar integrals under Lorentz  transformations~\cite{li}. 
There is a large variety of approaches for evaluating the master 
integrals analytically. For example, one can compute 
integrals with a simple singularity structure and a small number of kinematic 
scales by integrating directly their Feynman parameter representation. 
For more complicated cases one can use advanced techniques, such
as the method of differential equations~\cite{kotikov,lance,li}. 

This approach can fail, however, if we apply it  
to complicated processes. The reduction algebra
is hard, and the expressions of amplitudes in 
terms of master integrals may have spurious 
singularities which hamper their numerical evaluation.  The 
extraction of  $\epsilon$ poles  in the master integrals, the evaluation 
of the  coefficients of the $\epsilon$-expansion in terms of known 
analytic functions, and the analytic 
continuation of the latter to physically interesting kinematic regions are 
also involved.  It is, thus, very well motivated to improve or 
replace the ``traditional scheme'' and to develop new automated methods. 

One-loop amplitudes can be 
entirely determined in four dimensions in terms of basic functions 
such as logarithms and  polylogarithms that appear in the 
one-loop scalar box.  New methods introduce sophisticated algorithms 
for the reduction to the basic functions;  by either numerical or 
analytical techniques, they control the appearance of 
spurious singular terms and minimize the size of the intermediate 
expressions~\cite{glovergiele,denner,heinrich_oneloop,Binoth:2001vm,weinzierl,pittau}. 
Recently, cross-sections for $e^+e^- \to 4 \mbox{ fermion}$ processes 
at NLO were computed~\cite{denner_result} using such techniques. 
A method, which is inspired by techniques  for real 
radiation at NLO, renders one-loop graphs numerically 
integrable~\cite{nagy} with the introduction of universal subtraction
terms. A different approach uses unitarity, dualities, and 
analyticity properties for the determination of one-loop 
amplitudes~\cite{bdk}.   

Beyond one-loop, there has been a significant progress in the automation  
of the reduction methods to master integrals~\cite{laporta,air,grobner}. 
The infrared structure of multi-loop integrals is substantially more 
complex than at one-loop; the development of methods for their 
numerical evaluation is more difficult. Nevertheless, there is significant 
progress in this direction~\cite{passarino}. 
A powerful numerical method 
for multi-loop calculations  is the method of  sector 
decomposition~\cite{sector_loop,Binoth:2003ak,Hepp:1966eg}; 
it simplifies recursively 
the singularities of 
Feynman parameterizations  and allows  a straightforward expansion 
in $\epsilon$. The method of sector decomposition has been very succesfully 
employed in the
computation of several multi-leg integrals at one, two and three loops
\cite{heinrich_oneloop,sector_loop,Binoth:2003ak}.  
This method has been introduced, recently, for 
the purely numerical evaluation of multi-loop amplitudes~\cite{muon}. 
However, it is perplexing how to apply it for  loop-integrals in 
non-Euclidean regions.

In 1999, Smirnov~\cite{smirnov} and, soon later, 
Tausk~\cite{tausk}, introduced a new method for the evaluation of loop 
integrals. In their pioneering papers, Smirnov and Tausk~\cite{smirnov,tausk} 
computed analytically the first infrared divergent 
double box integrals. In ~\cite{veretin,tausk1} new 
two-loop integrals  for $2 \to 2$ massless processes were computed. 
In ~\cite{smirnovB} the method was applied to double-box integrals 
with one additional mass-scale. The method was 
spectacularly applied in the computation of three-loop 
amplitudes~\cite{smirnovC,smirnovD}  in ${\cal N}=4$ supersymmetric 
Yang-Mills theory, shedding light to novel cross-order perturbative 
relations of the theory~\cite{smirnovD,msym}. 

The Smirnov-Tausk method is based on a few 
simple ideas. Starting from the Feynman parameterization of a loop integral
we can derive a new representation in terms of 
a multiple complex contour integral. Such, Mellin-Barnes (MB),  
parameterizations have enabled compicated loop calculations by 
using powerful methods for complex integration~\cite{usyukina}.  
Smirnov and Tausk exploited a novel property of these representations.  
Infrared divergences localize on simple  
poles  inside the complex integration volume. 
We can isolate the divergent pieces of the integral at 
$\epsilon =0$, by using the Cauchy theorem. 
After subtracting the divergent residues, we can perform a Taylor 
expansion in  $\epsilon$, and  sum up the remaining infinite 
series of residues.  Finally, we can work to derive analytic expressions 
for the infinite sums in the coefficients of the expansion
in terms of logarithms, 
generalized polylogarithms~\cite{remiddi_vermaseren}, and 
more complicated functions.

The Smirnov-Tausk method is very powerful; however, it is laborious
and intricate. The isolation of the divergent residues in multiple 
Mellin-Barnes integrals is convoluted. In addition, it is difficult to 
identify infinite sums in terms of  
polylogarithm functions with known analyticity properties.  
As a consequence, the analytic continuation in physical 
kinematic regions is also involved. 
Due to these complications, the method has been 
applied to a few master integrals with a small number of kinematic scales.  
In this paper, we generalize the method to a broader spectrum of 
applications.

As a first task, we automate the procedure for the isolation of the 
divergent residues at $\epsilon \to 0$. Smirnov~\cite{smirnov} and 
Tausk~\cite{tausk} use different techniques for finding these residues. 
The approach of Smirnov is  very intuitive, but daedal. We have found that 
the technique described by Tausk in Ref.~\cite{tausk} resembles closely to a 
programmable algorithm. We have used it as a guide and we  
have written computer programs which subtract  the $1/\epsilon$  poles
in arbitrary Mellin-Barnes integrals.

In our method, we avoid entirely the painstaking tasks of finding analytic 
expressions for infinite sums in terms of polylogarithms, 
and performing the analytic continuation in the arguments of 
polylogarithms.
Mellin-Barnes representations are valid in  
kinematic regions where loop integrals may be complex-valued. 
We have found that, in a broad spectrum of applications,  
it is simple to calculate the representations numerically. 
The only analytic continuation that is ever required is that of logarithms 
with a single kinematic scale as an argument.

An important goal of our method is to calculate loop 
amplitudes in realistic gauge theories. 
We have found that tensor integrals and, furthermore,  diagrams 
which belong to the same topology can be calculated collectively. 
As we will show, the integrand of a representation 
for a scalar integral will be a product of Gamma functions and powers 
of kinematic invariants, while that of a generic tensor will be the 
same integrand as in the scalar integral multiplied by a
polynomial in the integration variables.  
The evaluation of polynomials is fast in a numerical program;
the computational cost for evaluating tensor 
integrals or loop diagrams is not  significantly larger than evaluating  
scalar integrals. The only practical issue that we need to address, 
is the book-keeping of the various terms that contribute to the 
polynomial; we present an efficient solution of this problem here.

In this paper, we apply our method to a number of examples. 
We, first, test our method in scalar and tensor integrals of 
the one-loop massless 
hexagon topology. The purpose of this computation is to introduce our method 
for tensor integrals and to demonstrate  that we can tackle problems which are 
relevant in computations of physical amplitudes. We present here results for
tensors through rank six, in both the Euclidean and 
the physical region for $2 \to 4$ processes. The numerical programs 
that we have constructed are suitable for the evaluation of the QCD 
amplitudes in four-jet production at hadron colliders.

In the second set of examples, we 
compute  scalar two and three-loop integrals which are known analytically 
in all kinematic regions: the massless planar~\cite{smirnov}  
and cross~\cite{tausk} double-box, the massless double-box with one off-shell 
leg~\cite{thomas_int,harmpol1,harmpol2,harmpol_ac,smirnovB}, and the 
massless planar triple-box~\cite{smirnovC}. They serve to 
cross-check our algorithms and to demonstrate that we can easily 
reproduce state-of-the-art computations. Our numerical 
results are in excellent agreement with the analytic expressions.
We also present a number of new 
results that would require significant efforts for their computation with 
traditional approaches. We present, for the first time, 
double-box integrals with up to four kinematic parameters, and 
triple-box integrals with up to three kinematic parameters computed in all
physical regions. 

In Section \ref{sec:MBrepresentations}, we explain our technique for 
deriving Mellin-Barnes representations for loop integrals. In Section 
\ref{sec:analytic}, we describe our routines for 
performing an $\epsilon$-expansion of Mellin-Barnes representations for
scalar integrals. We extend these results to loop integrals with
tensor numerators in Section \ref{sec:tensors} and, in Section 
\ref{sec:numerical}, we present our methods for the numerical 
integration. Section \ref{sec:results} is devoted to present our results
for several integrals at the one, two and three loop level. Finally
we present our conclusions.

\section{Mellin Barnes representations}\label{sec:MBrepresentations}

We start with a brief discussion on the derivation of Mellin-Barnes representations for 
loop integrals from their Feynman parameterization. The construction of 
parameterizations is not unique, and various 
representations of the  same integral may have quite different features. 
For example, they could have a different integral 
dimensionality, or they could be better or worse suited for numerical 
evaluation. A  valid parameterization, however, can always be 
found. A pedagogical introduction to
the topic of Mellin-Barnes representations is presented in 
Ref.~\cite{smirnov_book}. 

As a concrete example we consider the one-loop box integral with two 
adjacent external legs off-shell and massless propagators, 
\begin{equation}
\label{eq:box_2m_def}
{\cal I}_{4}^{2m} = \int \frac{d^dk}{i\pi^{\frac{d}{2}}}
\frac{1}{A_1^{\nu_1} A_2^{\nu_2} A_3^{\nu_3} A_4^{\nu_4} }, 
\end{equation}
with 
\begin{eqnarray}
A_1 &=& k^2 +i0, \nonumber \\
A_2 &=& (k+p_1)^2 +i0, \nonumber \\
A_3 &=& (k+p_1+p_2)^2 +i0, \nonumber \\
A_4 &=& (k+p_1+p_2+p_3)^2 +i0, \nonumber 
\end{eqnarray}
and $p_1^2=p_2^2=0$, $(p_1+p_2)^2=s$, $(p_2+p_3)^2=t$, $p_3^2=M_1^2$, 
$(p_1+p_2+p_3)^2=M_2^2$.  
The powers  of the propagators $\{\nu_i\}$ and the dimension $d=4-2\epsilon$ 
are kept arbitrary and we will derive a Mellin-Barnes 
parameterization of the box-integral in this general case. In this  section, 
we will seek values for $\{\nu_i, d\}$ where the representation is well 
defined and the integral is finite. In the next Section, we will explain the 
technique for the  analytic continuation to the values of the parameters, 
such as $\{\nu_i=1, \epsilon=0 \}$, where the integral develops divergences. 

Our starting point is the Feynman parameterization of the box integral: 
\begin{eqnarray}
\label{eq:box_2m_fp}
{\cal I}_{4}^{2m} &=& (-1)^N \frac{\Gamma(N-\frac{d}{2})}{\Gamma(\nu_1)
\Gamma(\nu_2)\Gamma(\nu_3)\Gamma(\nu_4)} \nonumber \\
&& \times \int_0^1 
\frac{dx_1 dx_2 dx_3 dx_4 x_1^{\nu_1-1} x_2^{\nu_2-1} x_3^{\nu_3-1} x_4^{\nu_4-1}
\delta\left( x_1+x_2+x_3+x_4-1\right)
}
{ \left(-sx_1x_3 -tx_2x_4 -M_1^2 x_3x_4 -M_2^2 x_4 x_1 -i0\right)^{N-\frac{d}{2}}},
\end{eqnarray}
with $N=\nu_1+\nu_2+\nu_3+\nu_4$. 
The main tool for obtaining the Mellin-Barnes representation 
of the above integral is the formula: 
\begin{equation}
\label{eq:mbrep}
\frac{1}{\left(A_1+A_2\right)^\alpha} = \frac{1}{2\pi i} \int_{c-i\infty}^{c+i\infty}
dw A_1^w A_2^{-\alpha-w} \frac{\Gamma(-w)\Gamma(\alpha+w)}{\Gamma(\alpha)},
\end{equation}
where the following conditions are satisfied:
\begin{enumerate}
\item $A_{1,2}$ are complex numbers with $\left| args(A_1)-args(A_2)\right| < 
\pi$,
\item the contour of integration is a straight line parallel to the imaginary 
axis separating the poles of $\Gamma(-w)$  and $\Gamma(\alpha+w)$, i.e. 
$-{\rm Re}(\alpha) < c < 0$. 
\end{enumerate}
We can use the identity of Eq.~\ref{eq:mbrep} to simplify the entangled denominator of 
Eq.~\ref{eq:box_2m_fp}, with the cost of introducing new integrations in the complex 
plane. Before we do so, we would like to point out an attractive feature of 
Eq.~\ref{eq:mbrep}. The above two 
conditions guarantee that the integrand in the right hand 
side of Eq.~\ref{eq:mbrep} vanishes at infinity. We can then close 
the contour of integration at infinity and calculate the integral using the 
Cauchy theorem. If we close 
the contour to the side of the positive real-axis we obtain 
\begin{equation}
\mbox{r.h.s.} = \frac{1}{A_2^\alpha}\sum_{n=0}^{\infty} \frac{\Gamma(\alpha+n)}{\Gamma(\alpha)
n!} \left(-\frac{A_1}{A_2} \right)^{n}
\end{equation}
This is the Taylor expansion of the left hand side of Eq.~\ref{eq:mbrep} 
in the region $\left| A_1/A_2\right| < 1$. Equivalently, by closing the 
contour of integration to the side of the negative real axis, we obtain a 
Taylor expansion in the complementary region 
$\left| A_2/A_1\right| < 1$. This is a particularly useful property: 
representations of Feynman integrals which are derived by using the 
Mellin-Barnes decomposition are valid in all kinematic regions. 
In addition, $A_{1,2}$ can be complex; therefore, we can account for the 
infinitesimal imaginary part $i0$ that is assigned to invariant masses 
and kinematic parameters. Eq. 3 may be recursively applied to denominators 
with more than two terms, yielding: 
\begin{eqnarray}
\label{eq:mbrep_multi}
\frac{1}{\left(A_1+A_2+\ldots A_m\right)^\alpha} &=& \frac{1}{\left(2\pi i\right)^{m-1}} 
\int_{c-i\infty}^{c+i\infty}
dw_1\ldots dw_{m-1} 
A_1^{w_1} \ldots A_{m-1}^{w_{m-1}} 
A_m^{-\alpha-w_1-\ldots-w_{m-1}}  \nonumber \\
&& \times 
\frac{\Gamma(-w_1) \cdots \Gamma(-w_{m-1}) \Gamma(\alpha+w_1+\ldots w_m)}{\Gamma(\alpha)},
\end{eqnarray}
We are now ready to apply the decomposition of Eq.~\ref{eq:mbrep_multi} to
the denominator of Eq.~\ref{eq:box_2m_fp}, 
introducing three contour integrations:
\begin{eqnarray}
{\cal I}_4^{2m}\left( \{\nu_i,d\}\right) &=& 
(-1)^N
\int_0^1 \left(\prod_{l=1}^4 dx_l \frac{x_l^{\nu_l-1}}{\Gamma(\nu_l)} \right)
\delta\left(x_1+x_2+x_3+x_4-1\right)
\nonumber \\
&& \times \frac{1}{(2\pi i)^3} \int dw_1 dw_2 dw_3 \Gamma(-w_1)\Gamma(-w_2)
\Gamma(-w_3) \Gamma(N-\frac{d}{2}+w_1+w_2+w_3) \nonumber \\ 
&& \times \left(-x_3x_4 M_1^2\right)^{w_1} \left(-x_4x_1 M_2^2\right)^{w_2}
\left(-x_2x_4 t\right)^{w_3} 
\left(-x_1x_3 s\right)^{\frac{d}{2}-N-w_1-w_2-w_3}\,,   
\end{eqnarray}
where the contours must satisfy condition 2 above, i.e. they must be chosen
in such a way that the arguments of the Gamma functions are all positive.
It is now straightforward to integrate out the Feynman parameters, using
\begin{equation}
\label{eq:int_feynman}
\int_0^1\left(\prod_{i=1}^{n} dx_i x_i^{\alpha_i-1} \right) 
\delta\left( 1-\sum x_i\right) = 
\frac{\Gamma(\alpha_1) \cdots \Gamma(\alpha_n)}
{\Gamma(\alpha_1+\ldots +\alpha_n)}.
\end{equation}
To exchange the order of integrations, we must assume that the 
exponents of the Feynman parameters satisfy
\begin{equation}
%x_i^{\alpha_i-1} : 
{\rm Re }(\alpha_i) > 0.
\end{equation}
With these conditions the Gamma functions on the right hand side of
Eq.\ref{eq:int_feynman} have arguments with positive real part as well.
The final result is: 
\begin{eqnarray}
\label{eq:box_2m_mb}
{\cal I}_4^{2m}\left( \{\nu_i,d\}\right) &=& 
 \frac{(-1)^N}{(2\pi i )^3 }
\int dw_1 dw_2 dw_3 \Gamma(-w_1) \Gamma(-w_2) \Gamma(-w_3)
\nonumber \\
&& \hspace{1cm} \times
f(\nu_1, \nu_2, \nu_3, \nu_4, w_1, w_2, w_3)
\nonumber \\
&& \times \left(-M_1^2-i0\right)^{w_1} \left(-M_2^2-i0\right)^{w_2}
\left(-t-i0\right)^{w_3} \left(-s-i0\right)^{\frac{d}{2}-N-w_{123}},
\end{eqnarray}
where 
\begin{eqnarray}
\label{eq:auxF}
&& f(\nu_1, \nu_2, \nu_3, \nu_4, w_1, w_2, w_3) =
\left(\prod_{l=1}^4 \frac{1}{\Gamma(\nu_l)} \right) 
\frac{
\Gamma\left(\nu_{1234}-\frac{d}{2}+w_{123}\right)} 
{\Gamma(d-\nu_{1234})}
\nonumber \\
&& \times \Gamma\left(\nu_2+w_3\right)\Gamma\left(\nu_4+w_{123}\right)
\Gamma\left(\frac{d}{2}-\nu_{234}-w_{13}\right)
\Gamma\left(\frac{d}{2}-\nu_{124}-w_{23}\right) 
\end{eqnarray}
and we have adopted the shorthand notation 
\[
\nu_{ijk \ldots} = \nu_i + \nu_j + \nu_k + \ldots
\]
In general, Mellin-Barnes representations for a 
loop-integral are well defined 
for number of dimensions and powers of propagators which 
render the real parts of the arguments in all Gamma functions positive. 
For example, it is impossible to satisfy positivity in all  Gamma function
arguments for $\{\nu_i=1, d=4\}$. This is in accordance with
the expectation that 
the scalar box integral is infrared divergent in four dimensions. But, for example, the 
representation of Eq.~\ref{eq:box_2m_mb} is well defined 
if we choose the contour ${\rm Re}(w_1)={\rm Re}(w_2)
={\rm Re}(w_3)=-0.2$ and the parameters  $\{\nu_i =1, d=5.4\}$. In the next 
section we will detail the analytic continuation from values of 
parameters for which the Mellin-Barnes representation is well-defined to 
values for which the integral develops divergences. That is, a procedure
which makes the divergences appear explicitly in the form of poles in the
parameter space.

From Eq.~\ref{eq:box_2m_mb} we observe that it is simple to 
implement the analytic continuation of kinematic parameters in a numerical 
evaluation, since  they always appear in a factorized form, 
e.g.
\[ 
(-t-i0)^w=e^{w\log(-t-i0)}.
\] 
We need the analytic continuation of simple logarithms; these 
are evaluated trivially in all kinematic regions: 
\[
\log(-t-i0) = \log(|t|) - \Theta(t)i\pi.
\]
It is clear from the above example that we can derive a Mellin-Barnes 
representation for any integral starting from its Feynman 
parameterization. The form of the parameterization depends on the choice of 
Feynman parameters, the order in which the $A_i$ terms appear in the right 
hand side of  Eq.~\ref{eq:mbrep_multi}, and  the implementation 
of the delta function constraint for the Feynman parameters. 
This arbitrariness is more pronounced beyond one-loop. 

In Section \ref{sec:tensors} we will derive representations for 
tensor one-loop integrals which simplify their numerical evaluation.  
We aim to find similarly simple representations for multi-loop tensor 
integrals as well. For this purpose, it is 
convenient to use the one-loop representations as building blocks,
by employing  a re-insertion technique~\cite{tausk1,usyukina}. 
We explain this technique in our second example.

We now derive a Mellin-Barnes representation for the massless planar 
double-box integral with two adjacent legs off-shell. The integral is 
defined: 
\begin{equation}
\label{eq:dbox_2m_def}
{\cal J}_{4}^{2m} = \int \frac{d^dk}{i\pi^{\frac{d}{2}}}
\frac{d^dl}{i\pi^{\frac{d}{2}}}
\frac{1}{A_1 A_2 A_3 A_4 A_5 A_6 A_7}, 
\end{equation}
with 
\begin{eqnarray}
A_1 &=& k^2 +i0, \nonumber \\
A_2 &=& (k+p_1)^2 +i0, \nonumber \\
A_3 &=& (k+p_1+p_2)^2 +i0, \nonumber \\
A_4 &=& (l+p_1+p_2)^2 +i0, \nonumber \\
A_5 &=& (l+p_1+p_2+p_3)^2 +i0, \nonumber \\
A_6 &=& l^2 +i0, \nonumber \\
A_7 &=& (k-l)^2 +i0, \nonumber 
\end{eqnarray}
and $p_1^2=p_2^2=0$, $(p_1+p_2)^2=s$, $(p_2+p_3)^2=t$, $p_3^2=M_1^2$, 
$(p_1+p_2+p_3)^2=M_2^2$.  
For simplicity, we have set the powers of propagators to one. We  
perform, first, the $k$-loop integral and derive the representation for this
integral  reading it off from the one-loop result of Eq.~\ref{eq:box_2m_mb},
\begin{eqnarray}
{\cal J}_{4}^{2m} &=& \frac{1}{(2\pi i)^3} \int \frac{d^dl}{i\pi^{\frac{d}{2}}} 
\frac{1}{A_4 A_5 A_6}
 \int dw_1 dw_2 dw_3 \Gamma(-w_1)\Gamma(-w_2)\Gamma(-w_3) \nonumber \\
&& \hspace{-0.8cm} \times 
f(1,1,1,1,w_1,w_2,w_3) (-s)^{\frac{d}{2}-4-w_{123}} (-A_4)^{w_1} (-A_6)^{w_2}
(-A_8)^{w_3};
\end{eqnarray}
the off-shell legs of the $k$-loop box correspond to the $A_4,A_6$ propagators 
while the $t$ invariant mass gives rise to a new propagator for the second 
integration: 
\begin{equation}
A_8 = (l+p_1)^2.
\end{equation}
The propagators $A_4,A_5,A_6,A_8$ form a new one-loop box with two adjacent 
legs off-shell, and we can read off, once again, the representation for 
this second integration using Eq.~\ref{eq:box_2m_mb}. The final result for 
the two-loop box with two adjacent legs off-shell is: 
\begin{eqnarray}
\label{eq:dbox_2m_mb}
{\cal J}_{4}^{2m} &=& \frac{-1}{(2\pi i)^6}
\int \left( \prod_{i=1}^6 dw_i\Gamma(-w_i) \right)
(-s)^{d-7-w_{456}} (-M_1^2)^{w_4} (M_2^2)^{w_5} (-t)^{w_6}
\nonumber \\
&& f(1,1,1,1,w_1,w_2,w_3)
f(1-w_2,-w_3,1-w_1,1,w_4,w_5,w_6)
\end{eqnarray}
We have produced a representation for a two-loop integral by embedding 
representations for one-loop integrals into other representations. 
It is obvious that we can use the re-insertion method for writing 
representations for arbitrary multi-loop integrals.

Representations for integrals with given kinematic scales can be used 
to derive, easily, results for simpler integrals where some of the scales 
are taken to zero. Let us, for illustration, derive the representation 
for the double-box with one-leg off-shell and the on-shell double-box. We 
must first take the limit $M_2^2 \to 0$ in Eq.~\ref{eq:dbox_2m_mb}. The term 
$(M_2^2)^{w_5}$ is vanishing in this limit unless $w_5 \to 0$ at the same 
time. If we use the Cauchy theorem, we find that the wanted limit 
is given by taking the residue of the integrand at $w_5=0$. Therefore, 
the double-box with one-leg off-shell is:
\begin{eqnarray}
\label{eq:dbox_1m_mb}
{\cal J}_{4}^{1m} &=& \frac{-1}{(2\pi i)^5}
\int \left( \prod_{i=1}^5 dw_i\Gamma(-w_i) \right)
(-s)^{d-7-w_{45}} (-M_1^2)^{w_4}  (-t)^{w_5}
\nonumber \\
&& f(1,1,1,1,w_1,w_2,w_3)
f(1-w_2,-w_3,1-w_1,1,w_4,0,w_5)
\end{eqnarray}
Similarly, the on-shell double-box is given by:
\begin{eqnarray}
\label{eq:dbox_0m_mb}
{\cal J}_{4}^{0m} &=& \frac{-1}{(2\pi i)^4}
\int \left( \prod_{i=1}^4 dw_i\Gamma(-w_i) \right)
(-s)^{d-7-w_{4}}   (-t)^{w_4}
\nonumber \\
&& f(1,1,1,1,w_1,w_2,w_3)
f(1-w_2,-w_3,1-w_1,1,0,0,w_4).
\end{eqnarray}
Note  that, in order  to have a finite limit for a vanishing kinematic scale,
the Mellin-Barnes representation should  have terms in the integrand 
which behave as $\Gamma(-g(w_i))M^{g(w_i)}$. 
These emerge naturally in non-trivial representations due 
to Eq.~\ref{eq:mbrep}.

\section{Analytic continuation}\label{sec:analytic}

In the previous Section we have noted  that the Mellin-Barnes 
representation of a
loop integral is valid if appropriate poles of the Gamma functions 
lay separated on the right and left of the integration contours. 
This condition guarantees the equivalence between the Mellin-Barnes 
integral and the original denominator in the loop integral. It often 
occurs that this condition can not be satisfied for values of the 
dimension parameter and the powers of the propagators which are relevant 
for a realistic application. Let us consider, as an example, the one-loop 
box representation of Eq.~\ref{eq:box_2m_mb} in the case of powers of 
propagators  set to unity and the dimension in $d=4-2\epsilon$, 
\begin{eqnarray}
\label{eq:box_2m_mb_epsilon}
{\cal I}_4^{2m} &=& 
 \frac{1}{(2\pi i )^3 }
\int dw_1 dw_2 dw_3 \Gamma(-w_1) \Gamma(-w_2) \Gamma(-w_3)
\frac{
\Gamma\left(2+\epsilon+w_{123}\right)} 
{\Gamma(-2\epsilon)}
\nonumber \\
&& \hspace{0.3cm} \times 
 \Gamma\left(1+w_3\right)\Gamma\left(1+w_{123}\right)
\Gamma\left(-1-\epsilon-w_{13}\right)
\Gamma\left(-1-\epsilon-w_{23}\right)
\nonumber \\
&& \hspace{0.3cm} \times  
\left(-M_1^2\right)^{w_1} \left(-M_2^2\right)^{w_2}
\left(-t\right)^{w_3} \left(-s\right)^{-2-\epsilon-w_{123}}.
\end{eqnarray}
This representation can only be valid for values of $\epsilon$ different 
from zero. For example, if we choose the contour  
${\cal C}=\left\{ {\rm Re}(w_1)=-0.1, {\rm Re}(w_2)=-0.2, {\rm Re}(w_3)=-0.3
\right\}$ we find that $\epsilon$ should be in the interval 
$ -1.4 < \epsilon < -0.6$. 

\begin{figure}[h]
\includegraphics[height=4cm]{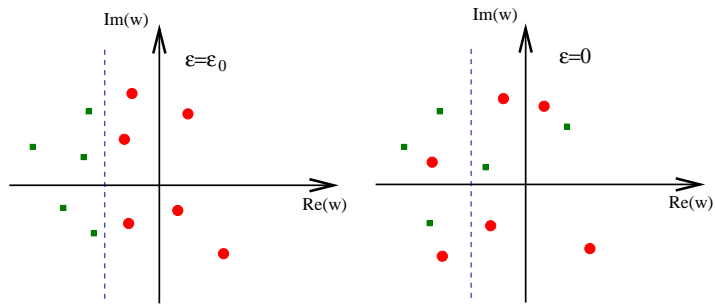}
\caption{
On the left picture, 
all poles which originate from the same Gamma function are 
positioned either to the left or to the right of a contour of integration.
On the second picture, we take $\epsilon \to 0$, and  some of the poles 
cross to the other side of the contour. To recover a valid representation 
for the loop-integral, these poles sould be isolated using the Cauchy theorem.
}\label{fig:pole_pos}
\end{figure}

We can use the Cauchy theorem to obtain a 
representation in terms of a sum of contour integrals, which is valid at 
$\epsilon=0$ and the $1/\epsilon$ poles appear explicitly. 
The key point is that if the value of 
$\epsilon$ is chosen outside the allowed region, for example $\epsilon=0$, 
some of the poles of the Gamma functions will be on the wrong side of the 
contours (see Fig.~\ref{fig:pole_pos}). To recover the original result, 
we must compare the position of the poles relative to the contours 
for values of  $\epsilon$ inside the allowed region 
and the point outside. 
Depending whether the poles crossed the
contours from left or right, we should correct the representation by 
adding or subtracting  the residue of the integrand on those poles. 
This simple observation can be easily cast into general purpose algorithms 
which extract the poles in $\epsilon$ of arbitrary Mellin-Barnes 
representations.

Let us choose a value of $\epsilon$ inside the allowed region 
and observe the position of the poles with respect to the contours 
for some $\epsilon$-dependent Gamma functions  in our example. 
The poles of $\Gamma(-1-\epsilon-w_{13})$ and 
$\Gamma(-1-\epsilon-w_{23})$, at $-1-\epsilon-
{\rm Re}(w_3)+n > {\rm Re}(w_1)$ and 
$-1-\epsilon-{\rm Re}(w_3)+n> {\rm Re}(w_2)$, are situated 
to the right of the $w_1$ and $w_2$ contours, respectively,
for all non-negative integers $n$. Let us now take $\epsilon \to 0$. 
We observe that the first poles, for $n=0$, moved to new positions,
$-1-{\rm Re}(w_3)< {\rm Re}(w_1)$ and 
$-1-{\rm Re}(w_3)< {\rm Re}(w_2)$, which are on the left side of 
the contours. The poles for $n>0$ remain on the right side. 
Therefore,  the representation is not valid at $\epsilon \simeq 0$, since the 
contours separate poles which originate from the same Gamma function. 
To obtain a valid expression for the integral at $\epsilon =0 $, we 
need to use the Cauchy theorem to isolate the crossing 
poles ($n=0$). 

\begin{figure}[h]
\scriptsize
\setlength{\unitlength}{1.8em}
%\rput[lt](0,0){
\begin{center}
\scalebox{1.20}{%
\begin{picture}(25.500000,32.500000)(-7.250000,-32.500000)
% picture environment flowchart generated by flow 0.99f
\put(3.2500,-1.2500){\oval(6.5000,2.5000)}
\put(0.0000,-2.5000){\makebox(6.5000,2.5000)[c]{\shortstack[c]{
$\displaystyle {\cal R}=K\,\Gamma(U_1)\cdots\Gamma(U_n)$\\
$\displaystyle {\cal C}=\left\{\bar{w}_1,\dots,\bar{w}_k\right\}$\\
$\displaystyle \epsilon=\epsilon_0$	
}}}
\put(3.2500,-2.5000){\vector(0,-1){1.0000}}
\put(-2.2500,-6.5000){\framebox(11.0000,3.0000)[c]{\shortstack[c]{
Find $\epsilon_1:\,|\epsilon_1|=\mbox{max}(|\tilde{\epsilon}_i|)$ with\\
$\displaystyle \tilde{\epsilon}_i: U_i({\cal C},\tilde{\epsilon}_i)=n_i$\\
${\cal P}=\{U_i: U_i \mbox{ hits a pole on ${\cal C}$ when }\epsilon=\epsilon_1\}$.
}}}
\put(-6.8500,-6.5000){\makebox(4.0000,3.0000)[r]{\shortstack[r]{
{\red $\displaystyle \mbox{sign}(\tilde{\epsilon}_i)=\mbox{sign}(\epsilon_0)$}\\
{\red  $n_i$: pole of $\Gamma(U_i)$ closer to ${\cal C}$.}
}}}
\put(3.2500,-6.5000){\vector(0,-1){1.0000}}
\put(1.2500,-9.5000){\line(1,1){2.0000}}
\put(1.2500,-9.5000){\line(1,-1){2.0000}}
\put(5.2500,-9.5000){\line(-1,-1){2.0000}}
\put(5.2500,-9.5000){\line(-1,1){2.0000}}
\put(1.2500,-11.5000){\makebox(4.0000,4.0000)[c]{\shortstack[c]{
Is ${\cal P}=\{\}$?
}}}
\put(1.2500,-8.9000){\makebox(0,0)[rt]{Y}}
\put(3.8500,-11.5000){\makebox(0,0)[lb]{N}}
\put(1.2500,-9.5000){\line(-1,0){1.0000}}
\put(0.2500,-9.5000){\vector(-1,0){1.0000}}
\put(-4.0000,-9.5000){\oval(6.5000,2.5000)}
\put(-7.2500,-10.7500){\makebox(6.5000,2.5000)[c]{\shortstack[c]{
Done!!
}}}
\put(3.2500,-11.5000){\vector(0,-1){1.0000}}
\put(1.2500,-14.5000){\line(1,1){2.0000}}
\put(1.2500,-14.5000){\line(1,-1){2.0000}}
\put(5.2500,-14.5000){\line(-1,-1){2.0000}}
\put(5.2500,-14.5000){\line(-1,1){2.0000}}
\put(1.2500,-16.5000){\makebox(4.0000,4.0000)[c]{\shortstack[c]{
Is $\epsilon_1=0$?
}}}
\put(5.2500,-13.9000){\makebox(0,0)[lt]{Y}}
\put(3.8500,-16.5000){\makebox(0,0)[lb]{N}}
\put(5.2500,-14.5000){\vector(1,0){4.5000}}
\put(3.2500,-16.5000){\vector(0,-1){1.0000}}
\put(-1.2500,-21.5000){\framebox(9.0000,4.0000)[c]{\shortstack[c]{
Solve the system of equations\\
$\left\{\displaystyle U_i=n_i,\,U_i\in {\cal P}\right\}$,\\
for the MB variables:\\
$\displaystyle w_i=w_i^R$
}}}
\put(3.2500,-21.5000){\vector(0,-1){1.0000}}
\put(1.2500,-24.5000){\line(1,1){2.0000}}
\put(1.2500,-24.5000){\line(1,-1){2.0000}}
\put(5.2500,-24.5000){\line(-1,-1){2.0000}}
\put(5.2500,-24.5000){\line(-1,1){2.0000}}
\put(1.2500,-26.5000){\makebox(4.0000,4.0000)[c]{\shortstack[c]{
Any pole  \\
nailed on $\cal C$?
}}}
\put(5.2500,-23.9000){\makebox(0,0)[lt]{Y}}
\put(3.8500,-26.5000){\makebox(0,0)[lb]{N}}
\put(5.2500,-24.5000){\line(1,0){8.0000}}
\put(13.2500,-24.5000){\line(0,1){7.5000}}
\put(13.2500,-17.0000){\vector(0,1){1.0000}}
\put(9.7500,-16.0000){\framebox(7.0000,3.0000)[c]{\shortstack[c]{
Displace the contours\\
without touching any pole\\
$\displaystyle {\cal C}=\{\bar{w}_1^\prime,\dots,\bar{w}_k^\prime\}$  
}}}
\put(13.2500,-13.0000){\vector(0,1){11.7500}}
\put(3.2500,-26.5000){\line(0,-1){1.0000}}
\put(3.2500,-27.5000){\vector(0,-1){1.0000}}
\put(-0.7500,-32.5000){\framebox(8.0000,4.0000)[c]{\shortstack[c]{
${\cal R}={\cal R}+\sum_j{\cal R}_j$\\
$\sum_j{\cal R}_j=\sum\mbox{Res}\left({\cal R},w_i=w_i^R\right)$\\
$\displaystyle \epsilon_0=\epsilon_1$
}}}
\put(7.2500,-30.5000){\line(1,0){11.0000}}
\put(18.2500,-30.5000){\line(0,1){29.2500}}
\put(18.2500,-1.2500){\vector(-1,0){11.7500}}
\end{picture}

}
\end{center}
%}
\caption{Analytic continuation algorithm}\label{fig:flow1}
\end{figure}
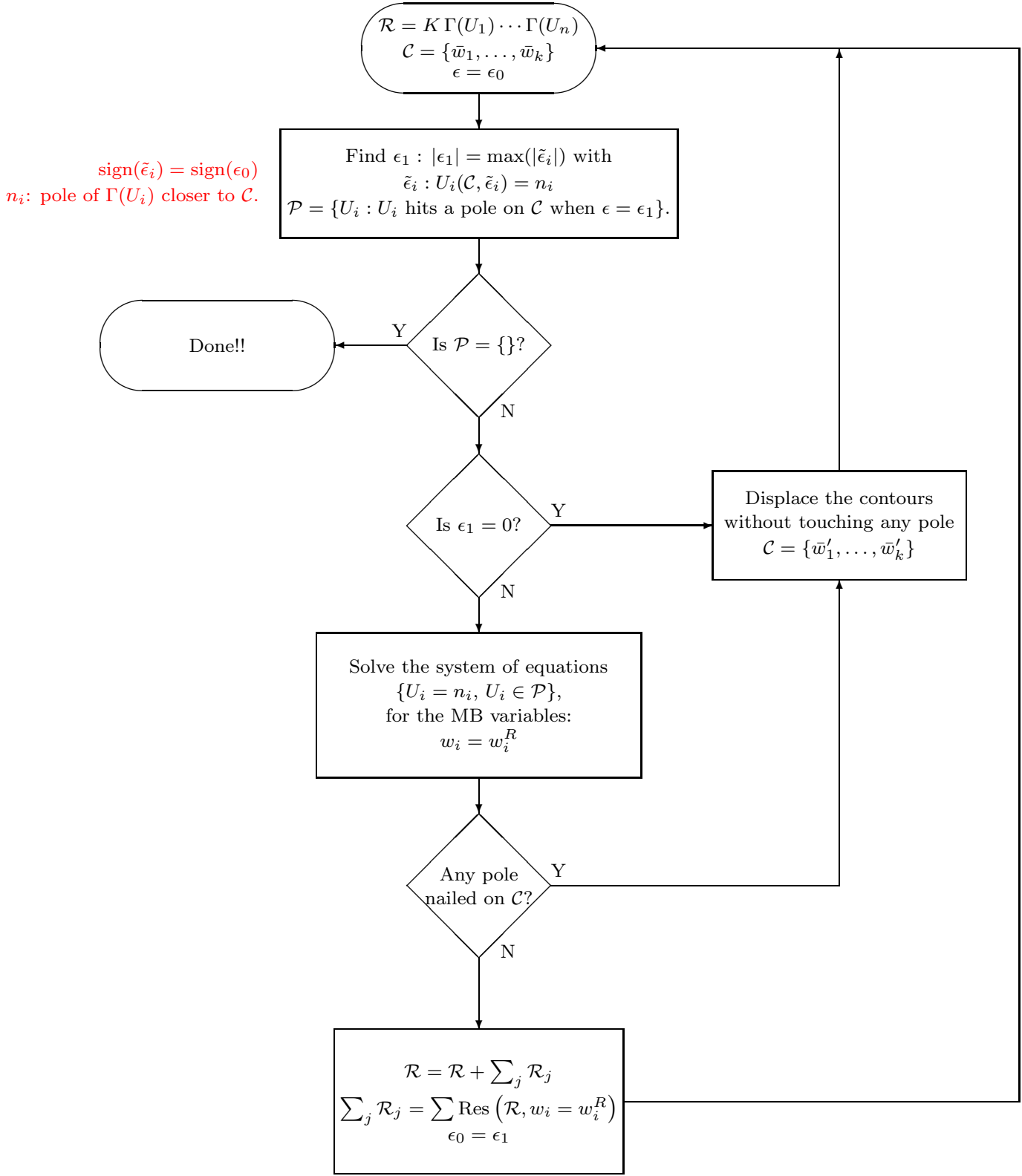

Smirnov, in Ref.~\cite{smirnov_book}, provides a number of pedagogical 
examples where this  is done. 
The approach of Smirnov is to deform the multiple contours, 
using the Cauchy theorem, so that the ``offending'' poles never cross the 
contour of integration. Tausk, in Ref.~\cite{tausk}, describes in  
detail this task as well, using as an explicit example the Mellin-Barnes 
representation for the two-loop cross box integral. 
The technique of Tausk is different from the one of Smirnov;  
the main idea is to account for the poles that end up
on the ``wrong'' side of the contour in a progressive way, as they cross the
contours when changing continuously the value of $\epsilon$. 
In this paper, we have followed the approach of Tausk. We would like 
to recommend that the interested reader studies the pedagogical example of
Ref.~\cite{tausk}.  Here we discuss our implementation of the algorithm
which is depicted in the flow diagram of Figure~\ref{fig:flow1}.

Consider the Mellin-Barnes representation ${\cal R}$ of a loop integral. 
The integrand of ${\cal R}$ depends initially on Gamma functions. 
The application of the analytic continuation algorithm requires the iterative 
evaluation of residues. These may also give rise to Psi functions, which 
have poles in the same positions the Gamma functions do, and are treated 
identically for the purposes of analytic continuation. 
The contours ${\cal C}$ of the representation are straight lines parallel to 
the imaginary axis, and are chosen so that 
the real parts of the arguments of all Gamma functions are positive. 
The same condition determines an allowed region for 
$\epsilon$, which we will assume that includes the 
point $\epsilon=\epsilon_0$. For the sake of clarity, we will consider 
the case $\epsilon_0<0$, but the method proceeds analogously when 
$\epsilon_0>0$.  

The first step is to find the maximum value   
$\epsilon=\epsilon_1$, with $\epsilon_1 \leq 0$,
for which the representation holds. 
At this value, one or some of the positions of the poles  
run into a contour. 
If the set of these poles, ${\cal P}$, is empty, 
i.e. no poles cross the contours when moving $\epsilon$
from $\epsilon_0$ to 0, then the integral can be safely expanded around 
$\epsilon=0$. 
So, let us focus on the more interesting case
in which ${\cal P}$ is non empty. 
Here we have to distinguish between $\epsilon_1=0$ and $\epsilon_1<0$.

If $\epsilon_1=0$, at least one pole lies on the contours 
for the terminating value of $\epsilon$.  Then, the representation has an 
unregulated singularity, and  it is not feasible to perform an $\epsilon$ 
expansion and evaluate the contour integrals numerically. To remedy the 
situation we must change the contour of integration; 
we use the Cauchy theorem and displace one or more of the 
contours parallel to the imaginary axis.  
Then, we attempt afresh the analytic continuation; 
the poles should now hit the contour at $\epsilon_1<0$. 
If this does not happen, due to an unfortunate choice of 
the contour we displaced, we simply iterate the procedure until we 
succeed.

When $\epsilon_1 < 0$, we have to determine the residues in the poles 
that cross the contour. If there is only one pole in ${\cal P}$ 
the situation is simple. Let us look  back at the specific example of the 
one-loop box. By applying the previous steps of the  algorithm, we find that 
the first pole to cross a contour is at 
$-1-\epsilon-w_2-w_3=0$. We then choose one of the multiple integrals, e.g. 
the $w_2$ integral, and subtract the residue of the integrand on this pole, 
at $w_2=-1-\epsilon-w_3$, to the
representation:
\begin{equation}
\label{eq:ancont}
{\cal R}(\epsilon=\epsilon_0) = {\cal R}(\epsilon_1) - \left.
{\rm Res}({\cal R}) \right|_{w_2=-1-\epsilon-w_3} 
\end{equation}
The residue  term in the above expression is a new multiple Mellin-Barnes 
integral with one integration variable less than in ${\cal R}$. 
In our example, the pole crosses the $w_2$ contour from the right side and 
we subtract its residue from  the  original representation. 
If a pole crosses the contour from the left, 
we should add its residue. 
At this point, we have achieved the analytic continuation in $\epsilon$ 
from $\epsilon_0$ to $\epsilon_1$. We now repeat the 
algorithm setting as initial value $\epsilon_0 = \epsilon_1$, and perform
the analytic continuation  to the next value of $\epsilon$ where a new 
crossing occurs. We proceed iteratively, until we 
find all poles which cross the contours of integration  
in the terms of Eq.~\ref{eq:ancont} and its descendants. 

The algorithm is more complicated if many poles cross simultaneously 
the contour at $\epsilon=\epsilon_1$. In our one-loop box example, 
if we had chosen a symmetric contour 
${\cal C}=\left\{ {\rm Re}(w_1)=-0.2, {\rm Re}(w_2)=-0.2, {\rm Re}(w_3)=-0.2
\right\} $ then we would have two poles crossing at the same time: 
${\cal P} = \left\{ -1-\epsilon-w_{13}=0, -1-\epsilon-w_{23}=0 \right\}$.
One way to deal with this situation is to displace 
the contours in such a way that the degeneracy is removed, i.e. 
introduce asymmetric contours from the beginning. 
Another alternative is to take combined residues in all the poles. 
To correctly account for all the terms in this case, we can imagine that the 
contours are only slightly displaced from their current position. 
Then the poles will hit them one at the time and we can take residues 
successively. Finally, we can restore the original 
position of the contours without affecting the position of the poles. 

There is, however, a caveat in taking multiple residues.  
Let us consider the case of the one-loop box, 
where the poles $P_1=-1-\epsilon-w_{13}$ and $P_2=-1-\epsilon-w_{23}$ 
hit the contour at the same time. 
We take residues in sequence, considering first $P_1$.  
We make the ``unfortunate'' choice to take the residue on the $w_3$ 
integration variable, and not on $w_1$. The analytic continuation 
yields:
\[
{\cal R}(\epsilon \simeq \epsilon_0) = {\cal R}(\epsilon \simeq 
\epsilon_1) - \left.
{\rm Res}({\cal R}) \right|_{w_3=-1-\epsilon-w_1} 
\]
Then we consider the crossing of the $P_2$ pole 
on each of the two terms on the right hand side of the above expression. 
We find no problems  in taking the residue of the first term. However, 
the position of the $P_2$ pole for the second term is now written: 
\[ 
P_2 = \left. 1-\epsilon-w_2-w_3 \right|_{w_3=-1-\epsilon-w_1} = w_1-w_2.
\] 
The pole $P_2$ is now independent of $\epsilon$ and gets nailed 
to the contour. 
This situation is similar to the one we encountered for $\epsilon_1=0$, 
and we can deal with it in exactly the same way by displacing the contours. 

We apply the algorithm recursively, to all the terms which are generated 
by adding the residues of crossing poles to the original parameterization. 
At the end of this procedure, we obtain a sum of contour integrals plus some 
terms that do not contain any remaining integral due to taking residues 
on all integration variables. The method assures that the sum of all 
these contributions equals the original loop integral when 
$\epsilon\simeq 0$. So we can expand in a power series of $\epsilon$.

In this Section, we have presented the algorithm assuming that only 
the dimension parameter $\epsilon$ 
is required to regulate the Mellin-Barnes 
parameterization.  
However, the method proceeds identically 
for  integrals where multiple analytic 
regulators are required. This could be the case in non-covariant 
gauges or 
in integrals with irreducible numerators~\cite{tausk1} 
or additional positive powers of propagators~\cite{veretin}. 

The actual implementation of the algorithm in a set of routines in both 
MAPLE and MATHEMATICA is very efficient, and allows to perform the analytic 
continuation in the regulator parameters in a few seconds, for all the 
Mellin-Barnes representations  that we have studied in this paper.

\section{Mellin Barnes representations for tensor integrals}
\label{sec:tensors}
Mellin-Barnes representations, with the exception of the massless two-loop 
diagonal box topology~\cite{veretin}, 
have been traditionally employed for the evaluation of master integrals.
In this Section, we introduce an efficient generalization
of the method  to tensor integrals; our method does not require any 
reduction techniques to master integrals. Common reduction methods
produce extremely large algebraic expressions which may also have spurious
singular denominators.  
Here we derive efficient representations for one-loop tensor integrals,
which are amenable to  numerical integration.  
We can derive analogous representations for multi-loop integrals by using the 
re-insertion  method of Section~\ref{sec:MBrepresentations}. 

We consider a generic one-loop tensor integral of rank $m$ with $n$ external 
legs:
\begin{equation}
I_{n,m}=
\int\frac{d^dk}{i\,\pi^{\frac{d}{2}}}\frac{k^{\mu_1}\dots k^{\mu_m}}
{
\left[ (k+q_1)^2-m_1^2 \right]
\left[ (k+q_2)^2-m_2^2 \right]
\cdots
\left[ (k+q_n)^2-m_n^2 \right]
}
\,
\end{equation}
where,  
\begin{equation}
q_1=0, \quad \quad q_j = \sum_{i=1}^{j-1} p_i,
\end{equation}
and $p_i$ are external incoming momenta. 
We first introduce Feynman parameters for the denominator. We obtain,
\begin{equation}
I_{n,m}=\Gamma(n)\,\int\frac{d^dk}{i\,\pi^{\frac{d}{2}}}\,\int
\left(\prod_{i=1}^{n}dx_i\right)\,\delta\left(1-\sum_{i=1}^n x_i\right)\,
\frac{k^{\mu_1}\dots k^{\mu_m}}{\left[(k+P)^2-\Delta\right]^n}\,,
\end{equation}
where
\begin{equation}
P=\sum_{i=1}^n x_i q_i,
\end{equation}
and 
\begin{equation}
\Delta = \sum_{i=1}^n x_i m_i^2+\sum_{j=2}^n  \sum_{i<j} x_i x_j \left[-\left(q_i -q_j\right)^2 \right]
\end{equation}
Now we perform the usual shift in the loop momentum, $k = K-P$, 
and we obtain,
\begin{equation}
I_{n,m}=
\Gamma(n)\,\,\int
\left(\prod_{i=1}^{n}dx_i\right)\,\delta\left(1-\sum_{i=1}^n x_i\right)\,
\sum_{r\le m} \int\frac{d^dK}{i\,\pi^{\frac{d}{2}}} \frac{\left\{K^r P^{m-r}\right\}^{[\mu_1,\ldots,\mu_m]}}
{(K^2-\Delta)^n}\, ,
\end{equation}
where, we denote 
\begin{eqnarray}
&&\left\{K^r P^{m-r}\right\}^{[\mu_1,\ldots,\mu_m]} =
\sum_{\{j_1,\ldots,j_m\}} K^{\mu_{j_1}}\cdots K^{\mu_{j_r}} 
P^{\mu_{j_{r+1}}}\cdots P^{\mu_{j_m}}, 
\\ 
&& \hspace{2cm} \{j_1,\ldots,j_m\} 
\in \mbox{permutations}(1,\ldots,m) \nonumber.  
\end{eqnarray}
We find the standard  
Feynman representation for the generic tensor one-loop $n$-point 
function, by integrating the loop momentum: 
\begin{eqnarray}\label{eq:tensor-feynf}
I_{n,m}&=&
(-1)^n \int
\left(\prod_{i=1}^{n}dx_i\right)\,\delta\left(1-\sum_{i=1}^n x_i\right)\,
\frac{1}{{\Delta^{n-\frac{d}{2}}}}
\nonumber \\
&& \times 
\sum_{r\le m}\,
\frac{\Gamma\left(n-\frac{d}{2}-\frac{r}{2}\right)}{2^{\frac{r}{2}}
}
\left\{{\cal A}_r P^{m-r}\right\}^{[\mu_1,\ldots,\mu_m]}
\Delta^{\frac{r}{2}},
\, 
\end{eqnarray}
 where ${\cal A}_r=0$ for $r$ odd, and 
${\cal A}_r= g^{\left[ \mu_1 \mu_2\right.}\cdots g^{\left. \mu_{r-1} \mu_r\right]}$
for $r$ even.  We observe that the sum, in the second 
line of the equation, is a polynomial in the Feynman parameters, 
with tensor coefficients: 
\begin{equation}
\label{eq:fp_polynomial}
\sum_{r\le m}\,
\frac{\Gamma\left(n-\frac{d}{2}-\frac{r}{2}\right)}{2^{\frac{r}{2}}
}
\left\{
{\cal A}_r P^{m-r}\right\}^{[\mu_1,\ldots,\mu_m]}
\Delta^{\frac{r}{2}} = \sum_{\{\nu_i >0 \}} C(\{\nu_i\}; \{p_i^{\mu_j}\})
x_1^{\nu_1-1}\cdots x_n^{\nu_n-1}
\end{equation}
For the scalar integral with unit powers of propagators, this sum is  
$\Gamma(n-\frac{d}{2})$.

We will use Eq.~\ref{eq:tensor-feynf} to derive Mellin-Barnes representations 
for the tensor loop-integral. We would like to exploit the fact that the 
Feynman representation of the scalar and tensor integrals have the same  
denominator: 
\[
\frac{1}{\Delta^{n-\frac{d}{2}}}.
\]
In Section~\ref{sec:MBrepresentations}, we have seen that we 
can obtain a Mellin-Barnes representation for a loop integral by decomposing 
the denominator of the Feynman parameterization with Eq.~\ref{eq:mbrep_multi}. 
We should anticipate that the Mellin-Barnes representations of tensors 
and  scalars are, therefore, very similar. 
We can verify this intuition, if we consider one generic term in the 
polynomial of Eq.~\ref{eq:fp_polynomial}, 
and follow the steps of 
Section~\ref{sec:MBrepresentations}. 
Decomposing the  denominator  with Eq.~\ref{eq:mbrep_multi}, we obtain  
\begin{eqnarray}
{} &&
\frac{x_1^{\nu_1-1} \cdots x_n^{\nu_n-1}}{\Delta^{n-\frac{d}{2}}}
\to  x_1^{\nu_1-1} \cdots x_n^{\nu_n-1}  
\nonumber \\
&& 
(-s_\alpha)^{\frac{d}{2}-n}\int \left( \prod_{i=1}^{\alpha-1}  
dw_i \Gamma(-w_i) (-s_i)^{w_i} (-s_\alpha)^{-w_i}\right)
\Gamma(n-\frac{d}{2}+w_{12\ldots{(\alpha-1)}})\,
x_1^{\beta_1(\vec{w})} \cdots x_n^{\beta_n(\vec{w})}  
\end{eqnarray}
where $\alpha$ is the number of kinematic scales $s_i$ 
in the integral. The factors $x_i^{\beta_i(\vec{w})}$ are universal 
for all terms in Eq.~\ref{eq:fp_polynomial}, and originate from the 
dependence of $\Delta$ on the Feynman parameters. 
We now proceed to integrate out the Feynman parameters using 
Eq.~\ref{eq:int_feynman}.  This integration yields Gamma functions 
\begin{equation}
x_i^{\beta_i(\vec{w})+\nu_i-1} \stackrel{\int dx_i}{\longrightarrow} 
\Gamma(\beta_i+\nu_i)=\Gamma(1+\beta_i) (1+\beta_i)(2+\beta_i) \cdots 
(\nu_i-1+\beta_i)
\end{equation}
Rewriting all Gamma functions as above in  
Eq.~\ref{eq:fp_polynomial}, 
we obtain, for the tensor integral, the Gamma functions 
of the scalar integral multiplied by simple factors with 
Mellin-Barnes  integration variables. 

The result for the one-loop tensor integrals 
depends on the topology and the rank of the tensors in a rather 
complicated manner. However, we  find that the 
representation is of the following form
\begin{eqnarray}
\label{eq:tensor_mbrep}
I_{n,m}&=&\frac{(-s_\alpha)^{\frac{d}{2}-n}}{(2\pi i)^{\alpha-1}} \int 
\left( 
\prod_{i=1}^{\alpha-1} dw_i \Gamma(-w_i) (-s_i)^{w_i} (-s_\alpha)^{(-w_i)} 
\right) \Gamma^{(scalar)}(w_i,\ldots,w_{\alpha-1})  
 \nonumber \\
&& \hspace{5cm} \times h^{(m)}(w_1,\ldots,w_{\alpha-1}),
\end{eqnarray}
The first 
line of the above representation contains all the  terms which already 
appear in the analogous 
representation of the scalar integral with unit powers of 
propagators. 
The function $h^{(m)}$ is a {\it polynomial in the Mellin-Barnes variables}, 
with tensor coefficients in terms of the external momenta. This is very 
important for our strategy to evaluate tensor integrals and Feynman diagrams. 
The polynomial in  the numerator is smooth, 
and it does not affect the analytic continuation in $\epsilon$. 
We can therefore perform the continuation collectively for all tensor 
integrals and diagrams of the same topology.  

We organize the evaluation of tensor integrals and diagrams of a
topology as follows.  We first find the Mellin-Barnes parameterization of 
the scalar integral of the topology, and multiply the integrand with a 
template function $h(\{ w_i\})$ which we assume depends on all 
Mellin-Barnes variables. Then, we apply the  analytic continuation algorithm 
of Section~\ref{sec:analytic} keeping $h$ general. We expand in $\epsilon$, 
and  create numerical programs for the evaluation of the expansion for 
any smooth function $h$. This part of the evaluation needs to be performed 
only once for a given topology.  Then we must identify the polynomial for 
the  evaluation of the  specific tensors or diagrams that we are 
interested in. We have written FORM~\cite{form} 
and MATHEMATICA programs which perform 
the steps that we described in this Section, and derive the explicit 
functional form of the polynomials in Eq.~\ref{eq:tensor_mbrep}. 
We then write numerical routines for their
evaluation, and link them to the general routines for the $\epsilon$ expansion
of the integrals of the topology. 

This approach is efficient for the application of our technique to tensor 
integrals. 
It turns out, however, that the  expressions 
for the polynomials are  quite long for high rank tensors.
We have observed that we can reduce the size of the expressions 
substantially with a simple modification.  
The terms  $\Delta^{\frac{r}{2}}$ in Eq.~\ref{eq:tensor-feynf} are lengthy, and 
when we integrate out the Feynman parameters they give rise to large 
expressions in $h^{(m)}$ of Eq.~\ref{eq:tensor_mbrep}. 
We rewrite Eq.~\ref{eq:tensor-feynf} in a different way, 
\begin{eqnarray}\label{eq:tensor-feynf-new}
I_{n,m}&=&
(-1)^n \int
\left(\prod_{i=1}^{n}dx_i\right)\,\delta\left(1-\sum_{i=1}^n x_i\right)\,
\nonumber \\
&& \times 
\sum_{r\le m}\,
\frac{\Gamma\left(n-\frac{d}{2}-\frac{r}{2}\right)}{2^{\frac{r}{2}}
}
%\sum_{r\le m}\,
\frac{1}{{\Delta^{n-\frac{d+r}{2}}}}
%\frac{\left(n-\frac{d}{2}\Big| -\frac{r}{2}\right)}{2^{\frac{r}{2}}}
\left\{{\cal A}_r P^{m-r}\right\}^{[\mu_1,\ldots,\mu_m]}
\,
\end{eqnarray}
Then, we introduce a Mellin-Barnes decomposition for each denominator:
\[
\frac{1}{\Delta^{n-\frac{r+d}{2}}},
\]
and integrate out the Feynman parameters. At the end of this procedure, 
we obtain a sum of $\left[ \frac{m}{2}\right]+1$ Mellin-Barnes integrals 
for a tensor of rank $m$. All integrals correspond to the Mellin-Barnes 
representation of the scalar integral with shifted dimension 
$d \to d+r$. 
Each of them has  a different 
polynomial $h^{(m)}_r$ in the numerator. 
However, these are substantially shorter than the polynomial 
$h^{(m)}$ in Eq.~\ref{eq:tensor_mbrep}. 

It is worth to note that, with our method, we never introduce 
spurious singularities. The evaluation of the higher rank tensors 
and diagrams  in gauge theories is very efficient, since we never require 
the manipulation of lengthy expressions.  The polynomials $h^{(m)}_r$ can 
be large, however, they require a minimum amount of handling before generating 
numerical codes for their evaluation. 

Here we have presented the derivation of efficient representations for 
one-loop integrals. The method may be applied to multi-loop integrals using 
the one-loop tensor integral representations as building blocks. 
The analogous functions of $h^{(m)}_r$ in multi-loop tensors 
contain Gamma functions in the denominator. However, they are  smooth 
and maintain the properties which are required for a collective evaluation of 
all tensors of a topology.  Explicit applications with multi-loop tensor integrals 
will be examined in future work.

\section{Numerical evaluation of Mellin-Barnes representations}\label{sec:numerical}

After analytic continuation, 
the contour integrals can be safely expanded in power series of $\epsilon$.
The coefficients of  the expansion are also contour integrals  
which, in general, 
contain in their integrand 
Gamma functions, Gamma function  derivatives, simple logarithms, 
and powers of the kinematic parameters.
In earlier works, these integrals were computed analytically.  
The standard approach is to evaluate them
by means of the residue theorem, i.e. summing an infinite number of 
residues and expressing the power series as logarithms, 
polylogarithms and harmonic polylogarithms. 

This  method is cumbersome for the evaluation of  
integrals with more than one Mellin-Barnes 
variable. In all cases presented in the literature, integrals 
 with more than one dimension are reduced to one-dimensional integrals 
with  a variety of clever, however not general, tricks. 
Typical procedures include direct integration
in simple situations, the application of Barnes 
lemmas~\cite{smirnov,tausk,smirnov_book},
and the expansion in Laurent series around an integer value of one of 
the variables.  In a few cases,  a careful 
inspection of the multi-dimensional integrals reveals 
exact cancellations between different components which 
yield significant simplifications. 

Many integrals which are relevant for physical applications, 
such as the massless double box with two 
external legs off-shell and one-loop tensor integrals with six external 
legs, after the analytic continuation in $\epsilon$, are expressed in terms 
of  hundreds of Mellin-Barnes 
components; some of them have a very high 
dimensionality.  The analytic methods are not suitable for such 
problems. In our method, we choose to evaluate the MB integrals
numerically by direct integration over the contours. 

The procedure we follow is, conceptually, straightforward. 
In practice, the implementation of the numerical evaluation 
requires a significant work in automating the book-keeping of the 
various terms. 
Tensor integrals require significantly more complex 
book-keeping due to large expressions that multiply the 
Gamma functions in the integrand of their 
Mellin-Barnes representation. 
Our computer programs are organized in the following way.    
We first use MAPLE, FORM~\cite{form}, 
and MATHEMATICA routines to derive Mellin-Barnes 
representations for the integrals that we need to compute.  
We then perform the  analytic continuation of Section~\ref{sec:analytic}, 
and we expand the integrands in powers of $\epsilon$ using MAPLE and MATHEMATICA. 
As the next step, we translate the integrands into FORTRAN routines 
which evaluate them as complex quantities.

There are many options to calculate numerically the sum of  
the Mellin-Barnes integrals which emerge from the routines for the 
$\epsilon$ expansion. For example, we can combine all contributions 
in a single common integrand, so that large numerical cancellations 
take place before integration. 
However, the finding of the peaks 
and  the  adaptation of the  integration routines become  less 
efficient.  Following a diametric approach, we could integrate all 
components separately and compute their sum at the end. Then, the 
adaptation of the numerical integration is optimal, however, the 
numerical precision is sensitive to cancellations. 
In practice, we follow a hybrid approach and
group together integrals with equivalent contours. 
We have achieved a reasonable efficiency and speed for all the 
computations that we present in this paper.  
Since we aim to keep our integration method general, we do not 
attempt any special simplifications by, for example, applying  Barnes lemmas 
or exploiting  symmetries of particular contours. 

For the numerical integration, we maintain the contours as straight 
lines parallel to the imaginary axis.  We map them onto a real interval
with a simple transformation, 
\begin{equation}
\label{eq:mapping}
\frac{1}{2\pi i} \int_{c-i\infty}^{c+i \infty} dw f(w) 
=\frac{1}{2\pi} \int_{0}^{1} \frac{d\lambda}{\lambda (1-\lambda)} 
f\left(c-i\ln\left( \frac{\lambda}{1-\lambda} \right) \right) 
\end{equation}
In all cases that we have considered, the integrands vanish rapidly when 
the integration variables take values away from the real line 
$(\lambda \simeq \frac{1}{2})$. The Mellin-Barnes integrals do not present 
problematic numerical instabilities, and converge rather fast. 
We perform the multidimensional integrations using the {\tt Cuhre} and 
{\tt Vegas} routines of the CUBA library~\cite{cuba}. To gain some speed,
we integrate the real and imaginary parts of the integrals concurrently,
using the same grid. The grid is adapted to the peaks of the real part, 
however, the quality of the integration for the imaginary part is only 
mildly affected. On the other hand, there is another subtle point 
concerning the convergence of the imaginary parts. As they always start 
an order later in the expansion in $\epsilon$, the analytic expressions for them would
be simpler than the corresponding real ones. When integrating numerically
this relative simplicity is reflected by a faster convergence when compared
to the real pieces at the same order.

The numerical 
results depend on the values for the kinematic scales of the integrals. 
With our approach, we use the same expressions for the evaluation of the 
integrals in all kinematic regions. As we explained earlier, the only 
analytic continuations in kinematic variables that we must perform, is 
for simple functions of the type $(-s_{ij})^z$ and $\log(-s_{ij})$. 
These are implemented trivially, 
in a branched format, in numerical programs. In this paper, we present 
multi-loop integrals for fixed kinematic parameters. However, we have 
arranged our routines so that a numerical integration over the 
phase-space can be performed simultaneously with the Mellin-Barnes 
integrations.

A problematic case for our numerical algorithms 
is  the evaluation of loop integrals  with massive internal propagators in 
non-Euclidean kinematic regions. Typically, Mellin-Barnes represenations for 
such integrals exhibit a very slow damping at $\pm i \infty$ in
physical regions. The mapping 
of  the integration region to a finite interval that we have considered 
in Eq.~\ref{eq:mapping} is not adequate, and contour deformations together 
with more sophisticated algorithms for the evaluation of oscillatory 
integrals are required.  We defer the study of such situations for 
a future project. 

We have achieved a full automatization of the code generation for the 
numerical evaluation. This has allowed us to apply our method 
to diverse problems. 
Our approach can be  improved and refined  further in the future 
by considering, for example, a clever analysis of the terms before 
integration in order to avoid numerical cancellations or 
more sophisticated mappings which can smoothen the peak structure of the 
integrands. However, as we will see in the rest of the paper, 
our first, naive, implementation is already sufficient to
obtain accurate results for complicated integrals in a reasonable time.

In what follows we present results for one,
two, and three loop integrals.  Some of the integrals are known analytically, 
and we use them to verify our numerical programs. We also present new 
loop integrals, which would be extremely tedious to calculate with 
traditional methods.

\section{Results}\label{sec:results}

In this Section we present results for one, two, and three-loop integrals 
that we  obtain  with our method. 
For each integral, we present the coefficients of its  
expansion in  $\epsilon=2-\frac{d}{2}$ through the finite term:
\begin{equation}
{\cal I}=\int \left( \prod_i \frac{d^d k_i}{i\pi^{\frac{d}{2}}}\right)
\frac{\{k_i^\mu\}}{\prod_j {{\rm propagator}}_j} 
=\sum_{i=0}^r\,c_i\,\epsilon^{-i}\,,
\end{equation}
where $r$ is the depth of the leading pole in $\epsilon$. 
We estimate the errors associated with our results for $c_i$, by 
adding, in quadrature,  the integration errors of all 
contributions from the residue decomposition.

\subsection{One loop hexagon: the scalar integral}
\begin{figure}[h]
\begin{center}
\includegraphics[width=5cm]{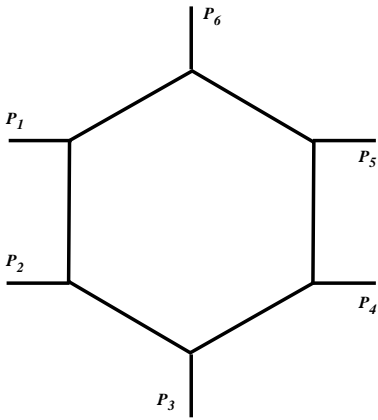}
\end{center}
\caption{The hexagon topology}
\label{fig:hexagon}
\end{figure}

The first integral we consider is the massless  one-loop hexagon,
depicted in Figure \ref{fig:hexagon}. 
For massless external legs, we 
derive an eight-dimensional Mellin-Barnes representation for this integral. 
Due to the high
dimensionality, the analytic continuation in $\epsilon$ involves 
the crossing of the contours of integration from many poles, and the 
Mellin-Barnes expression after the continuation  consists of  about 
two hundred terms. 
This example demonstrates the advantage of the automatic 
algorithm we described in Section \ref{sec:analytic}, since  
the book-keeping is done automatically and our routines 
perform the $\epsilon$ expansion in fractions of a minute.

After the analytic continuation and the expansion in series of
$\epsilon$,  only integrals with low dimensionality contribute
to the poles and finite pieces of the series. 
The integration over Feynman parameters produces
a denominator factor $1/\Gamma(-2\epsilon)$, which contributes 
factors of $\epsilon$ in the series expansion. 
The eight-dimensional component, in which no residues have
been taken, drops out from the finite part. 
In addition, most of the residues in the 
analytic continuation do not give rise to poles in $\epsilon$; 
a few components only develop a singularity that compensates the 
$\epsilon$ in the overall factor. 
For the scalar hexagon,  
we find integrals with up to three dimensions which 
contribute to the poles and the finite term. 
This mechanism, which reduces the 
dimensionality of the  representation in the leading coefficients of the 
$\epsilon$ expansion, is not specific to the hexagon topology, but is 
typical of Mellin-Barnes representations of scalar multi-leg integrals. 

In Table \ref{tb:hex-scalar.physical} we present explicit results for 
the scalar hexagon in phase space points corresponding to the 
physical region for the $2\rightarrow 4$ process.  
To check the efficiency of our  integration routines in all the phase-space, 
we have sampled hundreds of points.  
Here, we only show representative  results in the phase-space 
points of Table \ref{tb:hexagon-phasepoints}. 
These points were generated starting from
explicit values for the momenta in four dimensions, so they have a
vanishing Gram determinant.
\begin{table}[h]
{\small
\begin{tabular}{|c|c|c|c|c|c|c|c|c|c|}
\hline
point&$s_{12}$&$s_{23}$&$s_{34}$&$s_{45}$&$s_{56}$&$s_{16}$&$s_{123}$&$s_{234}$&$s_{345}$\\
\hline
$P_1$&
  1&
 -0.232033&
  0.096793&
  0.025066&
  0.465569&
 -0.015783&
  0.63356&
 -0.219996&
  0.336498\\
\hline
$P_2$&
 1&
 -0.056339&
  0.054104&
  0.111985&
  0.583564&
 -0.038557&
  0.74554&
 -0.191061&
  0.260063\\
\hline
$P_3$&
  1&
 -0.336912&
  0.099306&
  0.228826&
  0.333228&
 -0.086216&
  0.64116&
 -0.286003&
  0.580452
\\
\hline
$P_4$&
  1&
 -0.310819&
  0.012151&
  0.020687&
  0.561466&
 -0.349136&
  0.58216&
 -0.318051&
  0.114844\\
\hline
$P_5$&
  1&
 -0.07491&
  0.048452&
  0.279727&
  0.314164&
 -0.426813&
  0.73852&
 -0.449749&
  0.532752\\
\hline
$P_6$&
  1&
 -0.447378&
  0.10597&
  0.052091&
  0.277082&
 -0.484988&
  0.32942&
 -0.400468&
  0.170418\\
\hline
$P_7$&
  1&
 -0.081759&
  0.143257&
  0.139358&
  0.183411&
 -0.544412&
  0.81512&
 -0.361942&
  0.31596\\
\hline
$P_8$&
  1&
 -0.115263&
  0.054869&
  0.154172&
  0.209043&
 -0.569659&
  0.63722&
 -0.477605&
  0.410718\\
\hline
$P_9$&
  1&
 -0.331123&
  0.011523&
  0.033009&
  0.252287&
 -0.635452&
  0.38202&
 -0.441643&
  0.309497\\
\hline
$P_{10}$&
  1&
 -0.297484&
  0.010637&
  0.05663&
  0.345614&
 -0.88129&
  0.68562&
 -0.541753&
  0.083132\\
\hline
\end{tabular}
\caption{Explicit values of the invariants for some of the
phase space points used for the evaluation of the hexagon
topology. The points correspond to the physical region of the 
$2\rightarrow 4$ scattering of massless particles.}
\label{tb:hexagon-phasepoints}
}
\end{table}

As can be seen from Table \ref{tb:hex-scalar.physical}, 
our method is perfectly capable of evaluating
multi-scale integrals, such as the scalar hexagon, reliably in 
all phase-space. The results in the Table correspond to a fixed 
number of evaluations, thus
the difference in the relative errors for different points. It took 
a couple of minutes per point on a 2.8GHZ CPU to evaluate them. 

\begin{table}[h]
{\small
\begin{tabular}{|c|c|c|c|}
\hline
Point&$c_{2}$&$c_1$&$c_0$\\
\hline
$P_1$
 &
 \begin{minipage}[c]{5cm}
 \[\begin{array}{lr@{.}lc}
   -&24550&802400&(0)\\       +&i\cdot0&000000&(0)
 \end{array}\]
 \end{minipage}
 &
 \begin{minipage}[c]{5cm}
 \[\begin{array}{lr@{.}lc}
   -&129377&216040&(5)\\    -&i\cdot69329&856499&(5)
 \end{array}\]
 \end{minipage}
 &
 \begin{minipage}[c]{5cm}
 \[\begin{array}{lr@{.}lc}
        -&249435&&(40.)\\        -&i\cdot365517&9 &(3)
 \end{array}\]
 \end{minipage}
 \\
 \hline
$P_2$
 &
 \begin{minipage}[c]{5cm}
 \[\begin{array}{lr@{.}lc}
   -&10666&332700&(0)\\       +&i\cdot0&000000&(0)
 \end{array}\]
 \end{minipage}
 &
 \begin{minipage}[c]{5cm}
 \[\begin{array}{lr@{.}lc}
    -&42307&989040&(3)\\    -&i\cdot18370&333000&(3)
 \end{array}\]
 \end{minipage}
 &
 \begin{minipage}[c]{5cm}
 \[\begin{array}{lr@{.}lc}
         -&34595&&(66.)\\         -&i\cdot53139&1&(8)
 \end{array}\]
 \end{minipage}
 \\
 \hline
$P_3$
 &
 \begin{minipage}[c]{5cm}
 \[\begin{array}{lr@{.}lc}
    -&1153&682480&(0)\\       +&i\cdot0&000000&(0)
 \end{array}\]
 \end{minipage}
 &
 \begin{minipage}[c]{5cm}
 \[\begin{array}{lr@{.}lc}
     -&3091&953473&(3)\\     -&i\cdot2868&081530&(4)
 \end{array}\]
 \end{minipage}
 &
 \begin{minipage}[c]{5cm}
 \[\begin{array}{lr@{.}lc}
           &2292&&(2.)\\         -&i\cdot6531&15&(6)
 \end{array}\]
 \end{minipage}
 \\
 \hline
$P_4$
 &
 \begin{minipage}[c]{5cm}
 \[\begin{array}{lr@{.}lc}
   -&48753&897600&(0)\\       +&i\cdot0&000000&(0)
 \end{array}\]
 \end{minipage}
 &
 \begin{minipage}[c]{5cm}
 \[\begin{array}{lr@{.}lc}
   -&256890&405260&(4)\\   -&i\cdot152982&841000&(4)
 \end{array}\]
 \end{minipage}
 &
 \begin{minipage}[c]{5cm}
 \[\begin{array}{lr@{.}lc}
        -&432690&&(5.)\\       -&i\cdot806442&61&(4)
 \end{array}\]
 \end{minipage}
 \\
 \hline
$P_5$
 &
 \begin{minipage}[c]{5cm}
 \[\begin{array}{lr@{.}lc}
    -&2502&711680&(0)\\       +&i\cdot0&000000&(0)
 \end{array}\]
 \end{minipage}
 &
 \begin{minipage}[c]{5cm}
 \[\begin{array}{lr@{.}lc}
    -&10214&467200&(2)\\     -&i\cdot7425&550140&(3)
 \end{array}\]
 \end{minipage}
 &
 \begin{minipage}[c]{5cm}
 \[\begin{array}{lr@{.}lc}
          -&8934&&(16.)\\        -&i\cdot30275&25&(6)
 \end{array}\]
 \end{minipage}
 \\
 \hline
$P_6$
 &
 \begin{minipage}[c]{5cm}
 \[\begin{array}{lr@{.}lc}
    -&3857&953670&(0)\\       +&i\cdot0&000000&(0)
 \end{array}\]
 \end{minipage}
 &
 \begin{minipage}[c]{5cm}
 \[\begin{array}{lr@{.}lc}
    -&11372&775324&(2)\\    -&i\cdot12047&807600&(3)
 \end{array}\]
 \end{minipage}
 &
 \begin{minipage}[c]{5cm}
 \[\begin{array}{lr@{.}lc}
          &6715&7&(9)\\        -&i\cdot35512&41&(2)
 \end{array}\]
 \end{minipage}
 \\
 \hline

$P_7$
 &
 \begin{minipage}[c]{5cm}
 \[\begin{array}{lr@{.}lc}
    -&2078&190700&(0)\\       +&i\cdot0&000000&(0)
 \end{array}\]
 \end{minipage}
 &
 \begin{minipage}[c]{5cm}
 \[\begin{array}{lr@{.}lc}
     -&6541&063310&(1)\\     -&i\cdot6138&816950&(2)
 \end{array}\]
 \end{minipage}
 &
 \begin{minipage}[c]{5cm}
 \[\begin{array}{lr@{.}lc}
           &1413&&(2.)\\        -&i\cdot18965&48&(6)
 \end{array}\]
 \end{minipage}
 \\
 \hline
$P_8$
 &
 \begin{minipage}[c]{5cm}
 \[\begin{array}{lr@{.}lc}
    -&3356&184660&(0)\\       +&i\cdot0&000000&(0)
 \end{array}\]
 \end{minipage}
 &
 \begin{minipage}[c]{5cm}
 \[\begin{array}{lr@{.}lc}
    -&13032&698020&(3)\\    -&i\cdot10343&407301&(2)
 \end{array}\]
 \end{minipage}
 &
 \begin{minipage}[c]{5cm}
 \[\begin{array}{lr@{.}lc}
          -&8558&&(6.)\\        -&i\cdot40209&00&(3)
 \end{array}\]
 \end{minipage}
 \\
 \hline
$P_9$
&
 \begin{minipage}[c]{5cm}
 \[\begin{array}{lr@{.}lc}
   -&45864&213500&(0)\\       +&i\cdot0&000000&(0)
 \end{array}\]
 \end{minipage}
 &
 \begin{minipage}[c]{5cm}
 \[\begin{array}{lr@{.}lc}
   -&281098&336280&(4)\\   -&i\cdot144019&061999&(3)
 \end{array}\]
 \end{minipage}
 &
 \begin{minipage}[c]{5cm}
 \[\begin{array}{lr@{.}lc}
        -&649614&&(5.)\\       -&i\cdot882892&34&(2)
 \end{array}\]
 \end{minipage}
 \\
 \hline

$P_{10}$
&
 \begin{minipage}[c]{5cm}
 \[\begin{array}{lr@{.}lc}
   -&21516&252000&(0)\\       +&i\cdot0&000000&(0)
 \end{array}\]
 \end{minipage}
 &
 \begin{minipage}[c]{5cm}
 \[\begin{array}{lr@{.}lc}
    -&97739&299610&(3)\\    -&i\cdot67551&061100&(3)
 \end{array}\]
 \end{minipage}
 &
 \begin{minipage}[c]{5cm}
 \[\begin{array}{lr@{.}lc}
         -&99008&&(13.)\\       -&i\cdot306919&34&(2)
 \end{array}\]
 \end{minipage}
 \\
 \hline
\end{tabular}
}
\caption{The scalar massless hexagon evaluated in some physical 
points of the phase space. The errors of the numerical
integration are quoted in parenthesis, and they affect the last figure/s
respectively.}\label{tb:hex-scalar.physical}
\end{table}

\subsection{One-loop hexagon: tensor integrals}
The most severe problem in  evaluating 
multi-leg amplitudes, is the appearance of tensor numerators in the 
loop integrals. Our method, provides a powerful tool for their 
evaluation. The scalar hexagon, which we discussed before, 
provided a nice test-ground for the analytic 
continuation algorithm and the numerical integration strategy. Here, we 
consider tensor integrals for the same topology, in order to
demonstrate the potential of the method for computing realistic
quantities that arise in gauge theories. 

Using the procedure described in Section \ref{sec:tensors}, we 
perform the analytic continuation for an arbitrary tensor. At
variance with the scalar case, now we find components with
more than three MB variables contributing to the finite pieces in
$\epsilon$. For instance, for a typical rank three tensor,  
integrals with all eight Mellin-Barnes variables are required. 
This calculation provides  stringent tests on 
our implementation, due to the complex 
book-keeping of the tensorial terms in the 
Mellin-Barnes representations, and the 
efficiency of our routines  to evaluate 
the integrals with high dimensionality.

In Table \ref{tb:hex-tensor.check} we show results for tensors
up to rank 6 evaluated in a symmetric point $\{s_{ij}=s_{ijk}=-1\}$ 
in the Euclidean region. For these examples, we contracted 
all the loop momenta in the numerator with the same external 
momentum. All such  tensor contractions, with odd rank, should vanish;
we reproduce this result. On the table, we also
show results obtained by reducing the tensors using the program AIR~\cite{air}. 
We were not able to reduce the high rank tensors to master integrals in 
a closed analytic form.  
The reduction was only possible by 
substituting the kinematic invariants to their numerical value 
at this particular phase-space point  before solving the system 
of IBP equations. We have also made additional comparisons with 
the  reduction method, for other tensor contractions and a different 
point inside the physical region.

\begin{table}[h]
{\small
\begin{tabular}{|c|c|c|}
\hline
Tensor&MB&AIR\\
\hline
$q_2\cdot k$&
\begin{minipage}[c]{7cm}
\[\begin{array}{@{(}r@{.}lcc@{\,i\cdot}r@{.}lc@{)}r}
        0&000000&(0)&       +&0&000000&(0)&\epsilon^{-2}+\\
         0&00000&(1)&        +&0&00000&(1)&\epsilon^{-1}+\\
           0&000&(2)&        +&0&00000&(1)&
\end{array}
\]
\end{minipage}
&0\\
\hline
$q_2\cdot k\,q_2\cdot k$&
\begin{minipage}[c]{7cm}
\[\begin{array}{@{(}r@{.}lcc@{\,i\cdot}r@{.}lc@{)}r}
        0&500000&(0)&       +&0&000000&(0)&\epsilon^{-2}+\\
         0&71139&(1)&        +&0&00000&(1)&\epsilon^{-1}+\\
          0&0947&(8)&        +&0&00000&(1)&\\
\end{array}
\]
\end{minipage}
&
\begin{minipage}[c]{4cm}
\[\begin{array}{r@{.}lr}
0&5&\epsilon^{-2}+\\
0&711392&\epsilon^{-1}+\\
0&094845&
\end{array}
\]
\end{minipage}
\\
\hline
$q_2\cdot k\,q_2\cdot k\,q_2\cdot k$&
\begin{minipage}[c]{7cm}
\[\begin{array}{@{(}r@{.}lcc@{\,i\cdot}r@{.}lc@{)}r}
0&000000&(0)&       +&0&000000&(0)&\epsilon^{-2}+\\
0&0000&(1)&        +&0&00000&(1)&\epsilon^{-1}+\\
0&00&(2)&           +&0&02&(2)&
\end{array}
\]
\end{minipage}
&0\\
\hline
$q_2\cdot k\,q_2\cdot k\,q_2\cdot k\,q_2\cdot k$&
\begin{minipage}[c]{7cm}
\[\begin{array}{@{(}r@{.}lcc@{\,i\cdot}r@{.}lc@{)}r}
        0&125000&(0)&       +&0&000000&(0)&\epsilon^{-2}+\\
          0&3861&(1)&        +&0&00000&(1)&\epsilon^{-1}+\\
            0&65&(3)&           +&0&02&(3)&
\end{array}
\]
\end{minipage}
&
\begin{minipage}[c]{4cm}
\[\begin{array}{r@{.}lr}
0&125&\epsilon^{-2}-\\
0&386181&\epsilon^{-1}+\\
0&65996&
\end{array}
\]
\end{minipage}
\\
\hline
$q_2\cdot k\,q_2\cdot k\,q_2\cdot k\,q_2\cdot k\,q_2\cdot k$&
\begin{minipage}[c]{7cm}
\[\begin{array}{@{(}r@{.}lcc@{\,i\cdot}r@{.}lc@{)}r}
        0&000000&(0)&       +&0&000000&(0)&\epsilon^{-2}+\\
         0&00000&(9)&        +&0&00000&(1)&\epsilon^{-1}+\\
            0&04&(4)&           -&0&02&(4)&
\end{array}
\]
\end{minipage}
&0\\
\hline
$q_2\cdot k\,q_2\cdot k\,q_2\cdot k\,q_2\cdot k\,q_2\cdot k\,q_2\cdot k$&
\begin{minipage}[c]{7cm}
\[\begin{array}{@{(}r@{.}lcc@{\,i\cdot}r@{.}lc@{)}r}
        0&031250&(0)&       +&0&000000&(0)&\epsilon^{-2}+\\
         0&12466&(6)&        +&0&00000&(1)&\epsilon^{-1}+\\
            0&27&(1)&           -&0&00&(1)&
\end{array}
\]
\end{minipage}
&
\begin{minipage}[c]{4cm}
\[\begin{array}{r@{.}lr}
0&03125&\epsilon^{-2}-\\
0&1246703&\epsilon^{-1}+\\
0&279587&
\end{array}
\]
\end{minipage}
\\
\hline
\end{tabular}
}
\caption{Contracted tensors evaluated in a symmetric point in the Euclidean region.
We compare with results obtained  with the program AIR.}\label{tb:hex-tensor.check}
\end{table}

As in the scalar case, we sampled several phase space points in the
physical region,  calculating  a variety of contracted tensors  
at each rank. In Table
\ref{tb:hex-tensor.list} we list the cases we considered, and in Table
\ref{tb:hex-tensors.six} we show explicit results for the contraction
$q_2\cdot k\,q_2\cdot k\,q_3\cdot k\,q_4\cdot k\,q_5\cdot k\,q_6\cdot k\,$
in the same phase space points that we considered for the scalar hexagon. We
find that our routines perform extremely well, including 
the highest rank cases that we considered here.  
\begin{table}[h]
\begin{tabular}{|c|c|}
\hline
Rank&Contraction\\
\hline
1&
\begin{tabular}{c}
$q_2\cdot k$
\end{tabular}\\
\hline
\hline
2&
\begin{tabular}{c}
$q_2\cdot k\,q_4\cdot k$\\
$q_2\cdot k\,q_5\cdot k$
\end{tabular}\\
\hline
\hline
3&
\begin{tabular}{c}
$q_2\cdot k\,q_2\cdot k\,q_4\cdot k$\\
$q_2\cdot k\,q_4\cdot k\,q_5\cdot k$
\end{tabular}\\
\hline
\hline
4&
\begin{tabular}{c}
$q_2\cdot k\,q_2\cdot k\,q_3\cdot k\,q_4\cdot k$\\
$q_2\cdot k\,q_4\cdot k\,q_5\cdot k\,q_6\cdot k$\\
$q_2\cdot k\,q_3\cdot k\,q_5\cdot k\,q_6\cdot k$
\end{tabular}\\
\hline
\hline
5&
\begin{tabular}{c}
$q_2\cdot k\,q_3\cdot k\,q_4\cdot k\,q_5\cdot k\,q_6\cdot k$\\
$q_2\cdot k\,q_2\cdot k\,q_4\cdot k\,q_6\cdot k\,q_6\cdot k$\\
$k\cdot k\,q_2\cdot k\,q_3\cdot k\,q_4\cdot k$
\end{tabular}\\
\hline
\hline
6&
\begin{tabular}{c}
$q_2\cdot k\,q_2\cdot k\,q_3\cdot k\,q_4\cdot k\,q_5\cdot k\,q_6\cdot k$\\
$k\cdot k\,q_2\cdot k\,q_3\cdot k\,q_4\cdot k\,q_5\cdot k$\\
$k\cdot k\,k\cdot k\,k\cdot k$
\end{tabular}\\
\hline
\end{tabular}
\caption{Tensor contractions evaluated for the hexagon topology}\label{tb:hex-tensor.list}
\end{table}

\begin{table}[h]
{\small
\begin{tabular}{|c|c|c|c|}
\hline
Point&$c_{2}$&$c_1$&$c_0$\\
\hline
$P_1$
 &
 \begin{minipage}[c]{5cm}
 \[\begin{array}{lr@{.}lc}
     -&402&790580&(0)\\       +&i\cdot0&000000&(0)
 \end{array}\]
 \end{minipage}
 &
 \begin{minipage}[c]{5cm}
 \[\begin{array}{lr@{.}lc}
        -&1919&50&(1)\\        -&i\cdot1248&33&(2)
 \end{array}\]
 \end{minipage}
 &
 \begin{minipage}[c]{5cm}
 \[\begin{array}{lr@{.}lc}
          -&3019&&(55.)\\          -&i\cdot6068&&(50.)
 \end{array}\]
 \end{minipage}
 \\
 \hline

$P_2$
 &
 \begin{minipage}[c]{5cm}
 \[\begin{array}{lr@{.}lc}
     -&253&831615&(0)\\       +&i\cdot0&000000&(0)
 \end{array}\]
 \end{minipage}
 &
 \begin{minipage}[c]{5cm}
 \[\begin{array}{lr@{.}lc}
          -&876&7&(2)\\          -&i\cdot676&4&(1)
 \end{array}\]
 \end{minipage}
 &
 \begin{minipage}[c]{5cm}
 \[\begin{array}{lr@{.}lc}
           -&538&&(56.)\\          -&i\cdot2440&&(50.)
 \end{array}\]
 \end{minipage}
 \\
 \hline
$P_3$
 &
 \begin{minipage}[c]{5cm}
 \[\begin{array}{lr@{.}lc}
      -&47&453378&(0)\\       +&i\cdot0&000000&(0)
 \end{array}\]
 \end{minipage}
 &
 \begin{minipage}[c]{5cm}
 \[\begin{array}{lr@{.}lc}
        -&149&099&(6)\\        -&i\cdot145&128&(4)
 \end{array}\]
 \end{minipage}
 &
 \begin{minipage}[c]{5cm}
 \[\begin{array}{lr@{.}lc}
            -&79&&(2.)\\           -&i\cdot485&&(2.)
 \end{array}\]
 \end{minipage}
 \\
 \hline

$P_4$
 &
 \begin{minipage}[c]{5cm}
 \[\begin{array}{lr@{.}lc}
    -&3876&804240&(0)\\       +&i\cdot0&000000&(0)
 \end{array}\]
 \end{minipage}
 &
 \begin{minipage}[c]{5cm}
 \[\begin{array}{lr@{.}lc}
      -&23083&200&(2)\\      -&i\cdot12161&074&(2)
 \end{array}\]
 \end{minipage}
 &
 \begin{minipage}[c]{5cm}
 \[\begin{array}{lr@{.}lc}
         -&53835&&(6.)\\         -&i\cdot72507&&(5.)
 \end{array}\]
 \end{minipage}
 \\
 \hline

$P_5$
 &
 \begin{minipage}[c]{5cm}
 \[\begin{array}{lr@{.}lc}
      -&52&452711&(0)\\       +&i\cdot0&000000&(0)
 \end{array}\]
 \end{minipage}
 &
 \begin{minipage}[c]{5cm}
 \[\begin{array}{lr@{.}lc}
        -&214&000&(2)\\        -&i\cdot158&220&(2)
 \end{array}\]
 \end{minipage}
 &
 \begin{minipage}[c]{5cm}
 \[\begin{array}{lr@{.}lc}
           -&214&&(2.)\\           -&i\cdot663&&(1.)
 \end{array}\]
 \end{minipage}
 \\
 \hline

$P_6$
 &
 \begin{minipage}[c]{5cm}
 \[\begin{array}{lr@{.}lc}
     -&224&589492&(0)\\       +&i\cdot0&000000&(0)
 \end{array}\]
 \end{minipage}
 &
 \begin{minipage}[c]{5cm}
 \[\begin{array}{lr@{.}lc}
       -&887&7120&(5)\\       -&i\cdot701&8064&(5)
 \end{array}\]
 \end{minipage}
 &
 \begin{minipage}[c]{5cm}
 \[\begin{array}{lr@{.}lc}
          -&888&&(1.)\\         -&i\cdot2796&9&(9)
 \end{array}\]
 \end{minipage}
 \\
 \hline

$P_7$
 &
 \begin{minipage}[c]{5cm}
 \[\begin{array}{lr@{.}lc}
      -&54&532983&(0)\\       +&i\cdot0&000000&(0)
 \end{array}\]
 \end{minipage}
 &
 \begin{minipage}[c]{5cm}
 \[\begin{array}{lr@{.}lc}
        -&172&177&(4)\\        -&i\cdot161&007&(7)
 \end{array}\]
 \end{minipage}
 &
 \begin{minipage}[c]{5cm}
 \[\begin{array}{lr@{.}lc}
            -&26&&(2.)\\           -&i\cdot531&&(2.)
 \end{array}\]
 \end{minipage}
 \\
 \hline

$P_8$
 &
 \begin{minipage}[c]{5cm}
 \[\begin{array}{lr@{.}lc}
      -&47&254266&(0)\\       +&i\cdot0&000000&(0)
 \end{array}\]
 \end{minipage}
 &
 \begin{minipage}[c]{5cm}
 \[\begin{array}{lr@{.}lc}
        -&179&409&(1)\\        -&i\cdot143&862&(2)
 \end{array}\]
 \end{minipage}
 &
 \begin{minipage}[c]{5cm}
 \[\begin{array}{lr@{.}lc}
           -&127&&(1.)\\           -&i\cdot558&&(1.)
 \end{array}\]
 \end{minipage}
 \\
 \hline

$P_9$
 &
 \begin{minipage}[c]{5cm}
 \[\begin{array}{lr@{.}lc}
     -&865&507076&(0)\\       +&i\cdot0&000000&(0)
 \end{array}\]
 \end{minipage}
 &
 \begin{minipage}[c]{5cm}
 \[\begin{array}{lr@{.}lc}
      -&5612&7500&(2)\\      -&i\cdot2716&2898&(1)
 \end{array}\]
 \end{minipage}
 &
 \begin{minipage}[c]{5cm}
 \[\begin{array}{lr@{.}lc}
         -&15300&&(2.)\\         -&i\cdot17634&&(1.)
 \end{array}\]
 \end{minipage}
 \\
 \hline

$P_{10}$
 &
 \begin{minipage}[c]{5cm}
 \[\begin{array}{lr@{.}lc}
    -&1318&087970&(0)\\       +&i\cdot0&000000&(0)
 \end{array}\]
 \end{minipage}
 &
 \begin{minipage}[c]{5cm}
 \[\begin{array}{lr@{.}lc}
      -&6160&7460&(9)\\      -&i\cdot4137&4238&(2)
 \end{array}\]
 \end{minipage}
 &
 \begin{minipage}[c]{5cm}
 \[\begin{array}{lr@{.}lc}
          -&7611&&(2.)\\         -&i\cdot19356&&(2.)
 \end{array}\]
 \end{minipage}
 \\
 \hline

\end{tabular}
}
\caption{Results for the rank six tensor 
$q_2\cdot k\,q_2\cdot k\,q_3\cdot k\,q_4\cdot k\,q_5\cdot k\,q_6\cdot k\,$
  for the hexagon topology at one loop. The
phase space points are detailed in Table \ref{tb:hexagon-phasepoints}}\label{tb:hex-tensors.six}
\end{table}

We should note that, due to the way that we organize the 
evaluation of the tensors in Section~\ref{sec:tensors}, 
we do not anticipate that the evaluation 
of Feynman graphs, such as the ones in the one-loop 
six photon amplitude, are significantly harder than 
the six rank tensors that we  presented. In addition, our routines 
allow a combined integration over the Mellin-Barnes variables and 
the kinematic invariants.

\subsection{On-shell planar double box}
\begin{figure}[h]
\begin{center}
\includegraphics[width=6cm]{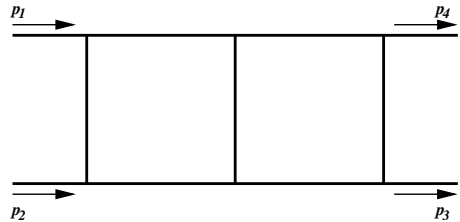}
\end{center}
\caption{The massless double box.}
\label{fig:2Bonshell}
\end{figure}
We now compute two-loop integrals, and we consider first 
the massless planar double box with on-shell legs shown in Fig.~\ref{fig:2Bonshell}.
This integral is 
known analytically  in Ref.~\cite{smirnov}. As discussed in
Section \ref{sec:MBrepresentations}, it can be expressed 
as a MB contour integral in
four complex dimensions. Performing the analytic continuation in $\epsilon$,
we get terms where residues in all the variables have been taken (i.e. terms
without any contour integral left); these contain poles through 
$1/\epsilon^4$. The terms through ${\cal O}(\epsilon^0)$ require 
Mellin-Barnes integrals with one and two integration variables. 

\begin{figure}[h]
\begin{minipage}{7.5cm}
\includegraphics[width=8cm]{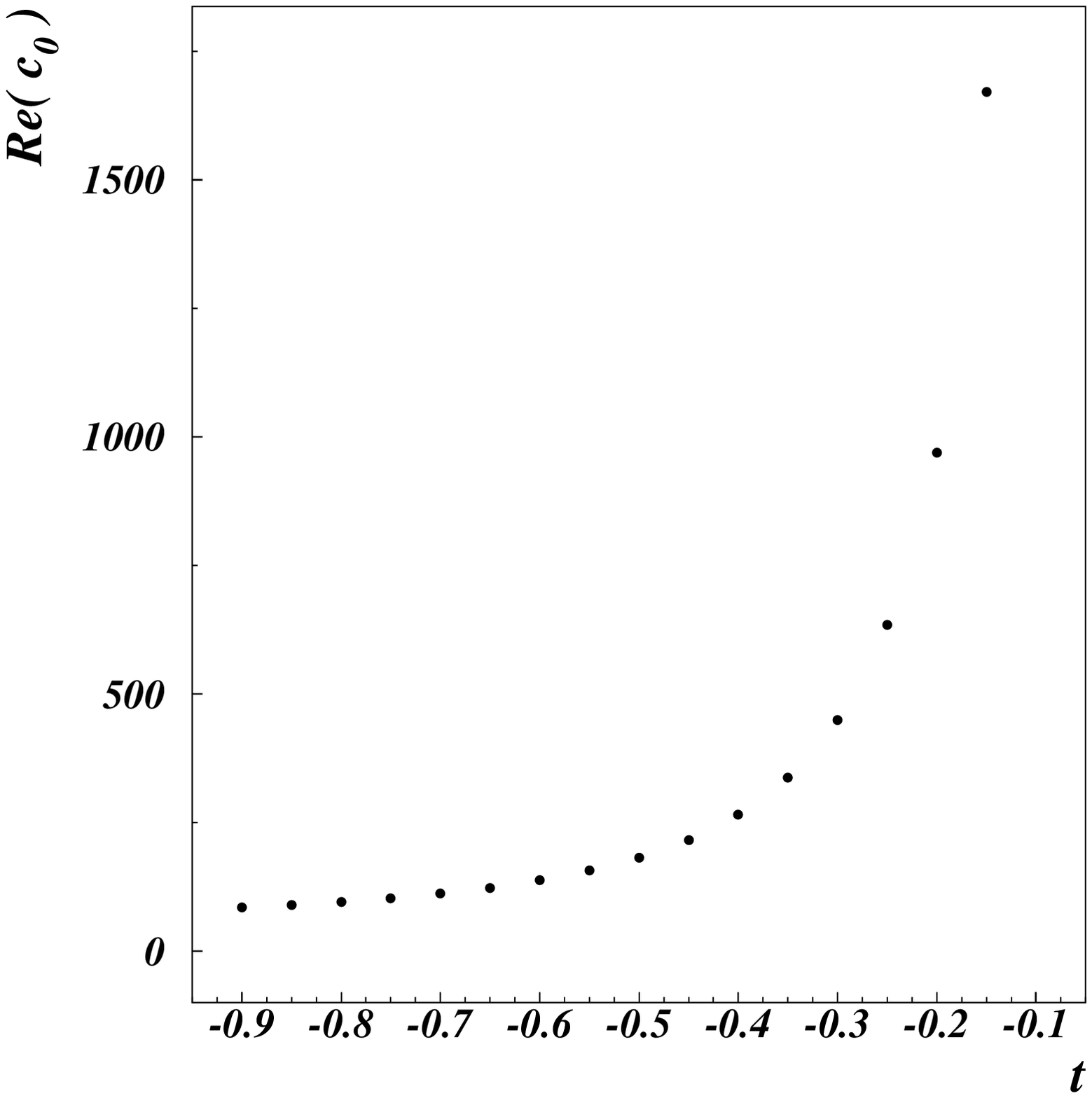}
\end{minipage}
\begin{minipage}{7.5cm}
\includegraphics[width=8cm]{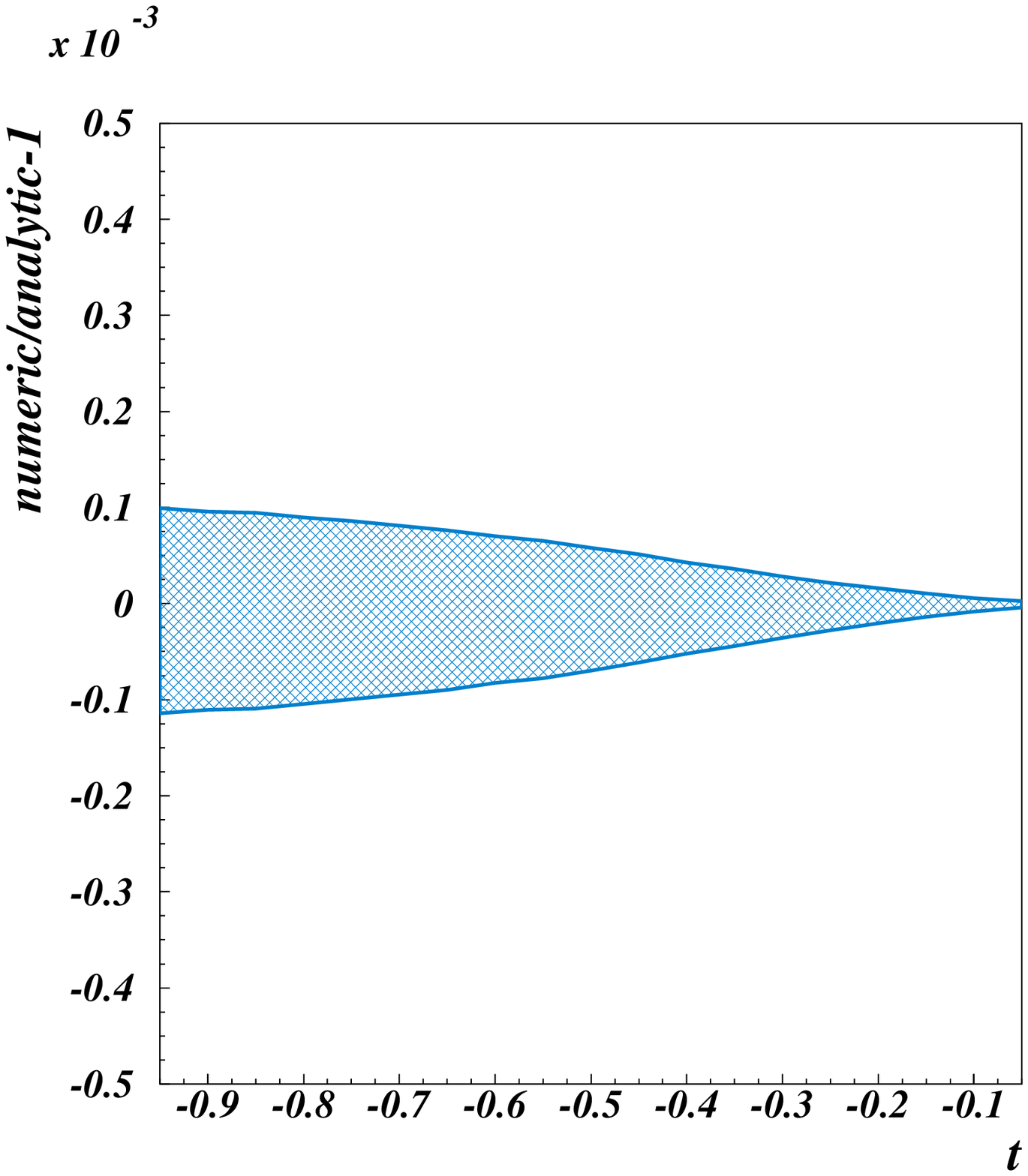}
\end{minipage}
\begin{minipage}{7.5cm}
\includegraphics[width=8cm]{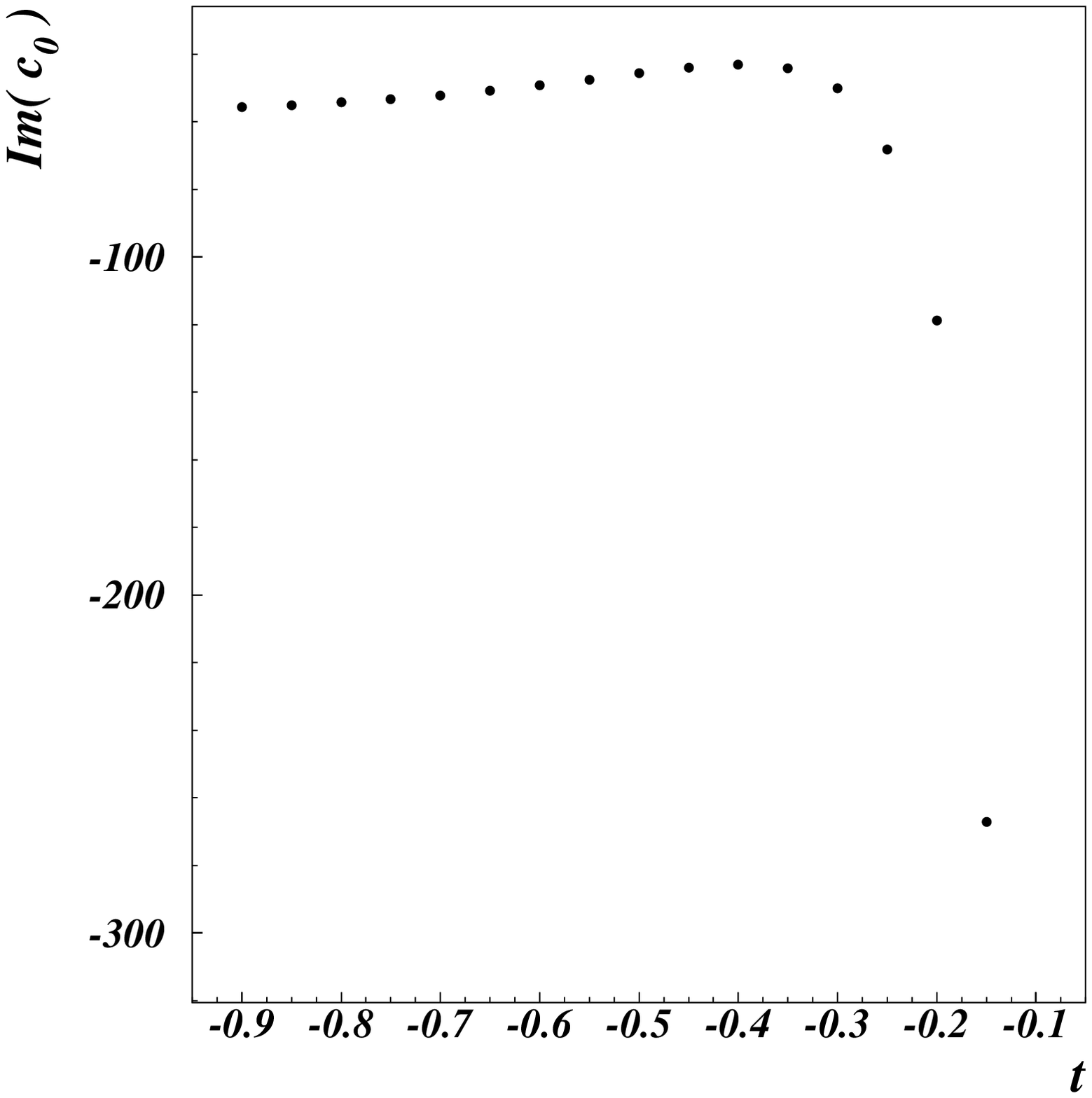}
\end{minipage}
\begin{minipage}{7.5cm}
\includegraphics[width=8cm]{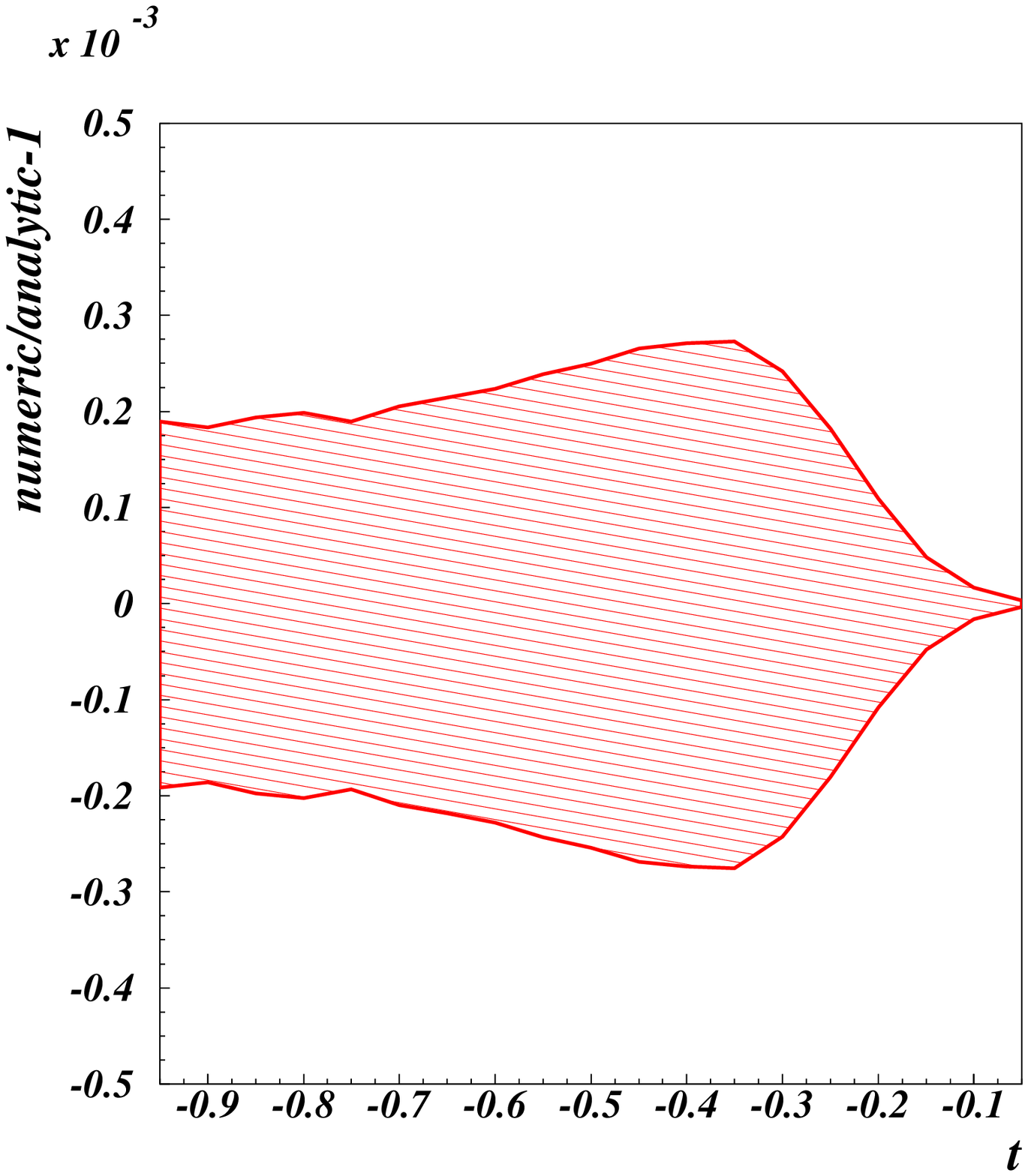}
\end{minipage}
\caption{Results for the finite part of the planar double box in the physical
region for a $2\rightarrow 2$ process. On the two left panels 
we plot the real and imaginary parts (upper and lower plots respectively) 
of the finite term as a function of $t$ for fixed value of $s=1$. 
The estimated error of the numerical integration
lies within the size of the points. On the right panel we show the
ratios of the numerical calculation to the analytic results of
\cite{smirnov} for the same kinematics, the bands in this case are given 
by the error in the numerical integrations.}
\label{fig:ON2BeM}
\end{figure}

In Figure \ref{fig:ON2BeM} we show our results for the finite pieces of
the on-shell double box in the physical region for the 
$p_1+p_2\rightarrow p_3+p_4$ process depicted in Figure \ref{fig:2Bonshell}. 
We set $s=(p_1+p_2)^2=1$, 
and plot, as functions of $t=(p_2-p_3)^2$, the finite term $c_0$ 
and the comparison of our numerical calculation to the analytic result 
of \cite{smirnov}. 
%Notice 
%that, at variance with the examples discussed in Section II, we consider
%outgoing momenta for the final state particles, which explains the difference
%in the relative sign of the two momenta in $t$.
% Diagram for the on-shell double box 

As can be seen from the figure, it is possible to achieve an 
accuracy of a few parts in ten thousand, in a few seconds per phase-space 
point. The bigger error band for the imaginary part, reflects our
naive strategy to integrate both real and imaginary part with 
the same grid. 
The coefficients for the simple and double poles in $\epsilon$, 
not shown in the figure, agree with the analytic calculation 
with even better accuracy, whereas the cubic and quartic poles are 
analytic expressions also in our case (as they do not involve any
MB integration) and coincide with the ones in \cite{smirnov}.

% Results for the on-shell double box

\subsection{On-shell non-planar double-box }
\begin{figure}[h]
\begin{center}
\includegraphics[width=6cm]{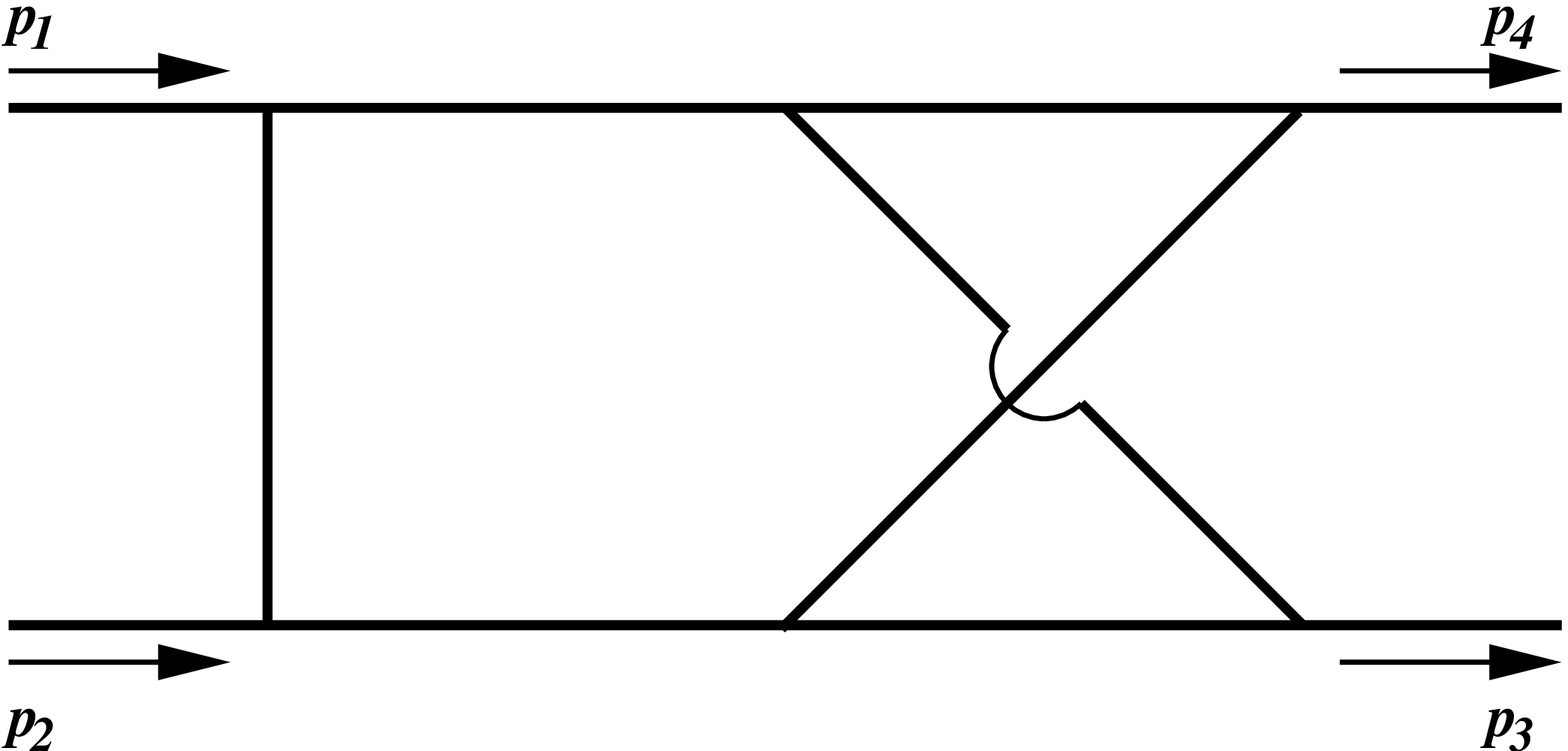}
\end{center}
\caption{The cross box.}
\label{fig:nonplanar}
\end{figure}
Now we compute the non-planar two-loop box 
with massless external legs depicted in Fig.~\ref{fig:nonplanar}. 
This integral has been computed 
analytically in Ref.~\cite{tausk} 
using a four-fold Mellin-Barnes representation. Instead of deriving 
a representation with the aid of the re-insertion method, we use the 
the representation in Ref.~\cite{tausk} for our numerical evaluation. 
This selection allows us to check our algorithms for the expansion in $\epsilon$ 
with the detailed decription in Ref.~\cite{tausk}. Furthermore, this representation
does not impose the physical constraint between the three invariants, $s+t+u=0$,
leading to a more complicated analytic structure. This is an additional test
on our programs for evaluating integrals in phase space regions which are
separated by complicated thresholds. The integral is, also,
$1/\epsilon^4$ divergent. Up to three-dimensional integrals contribute to 
the $\epsilon$ expansion through ${\cal O}(\epsilon^0)$. 
In Ref.~\cite{tausk}, it was shown that, with clever manipulations, 
it is possible to reduce all multiple integrals to integrals with 
only one dimension. Keeping our routines general,  we 
have chosen not to reduce the dimensionality of the multiple integrals 
and evaluate them numerically.

% The diagram for the non-planar double box

\begin{figure}[h]
\begin{minipage}{7.5cm}
\includegraphics[width=8cm]{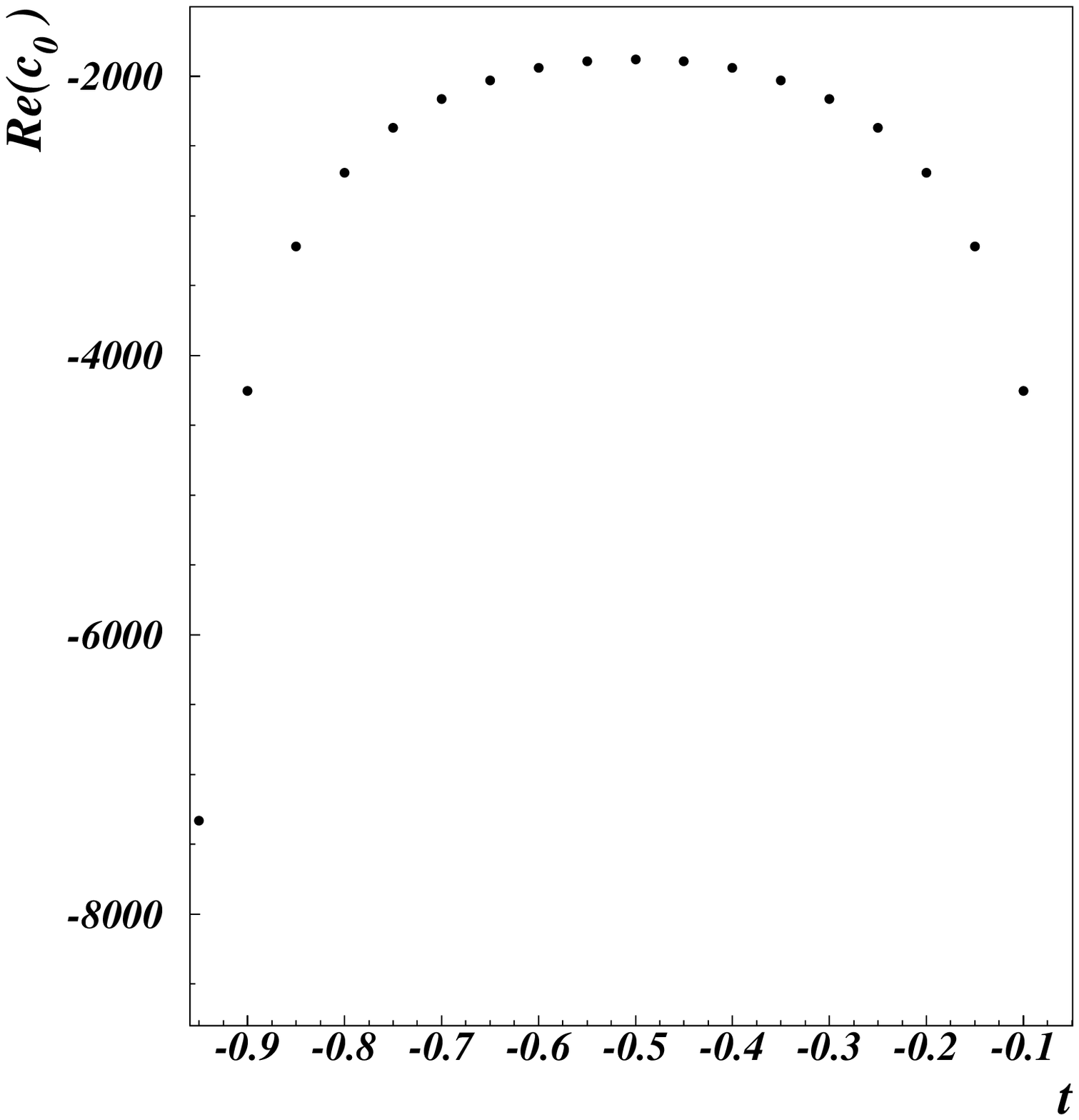}
\end{minipage}
\begin{minipage}{7.5cm}
\includegraphics[width=8cm]{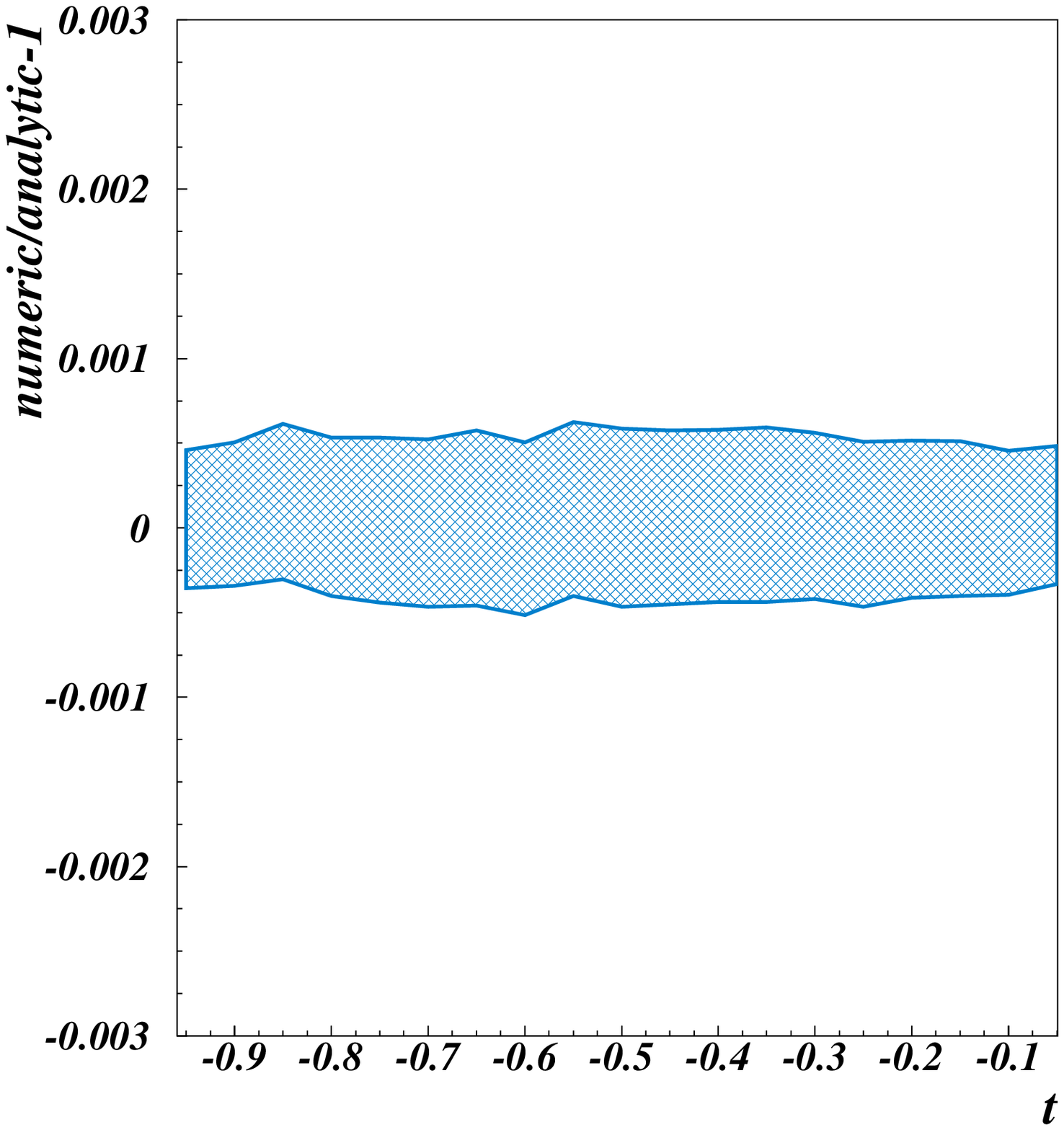}
\end{minipage}
\begin{minipage}{7.5cm}
\includegraphics[width=8cm]{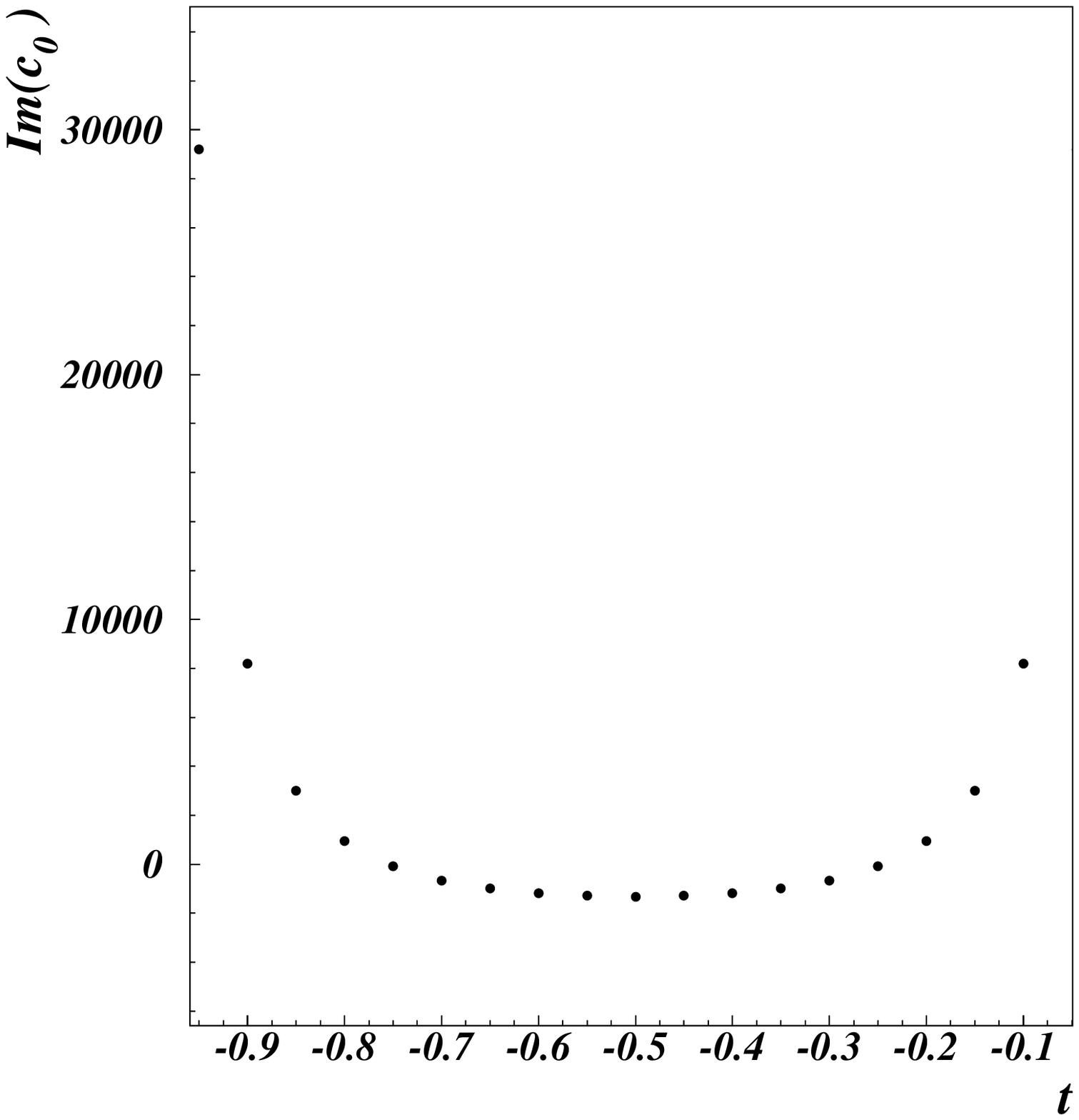}
\end{minipage}
\begin{minipage}{7.5cm}
\includegraphics[width=8cm]{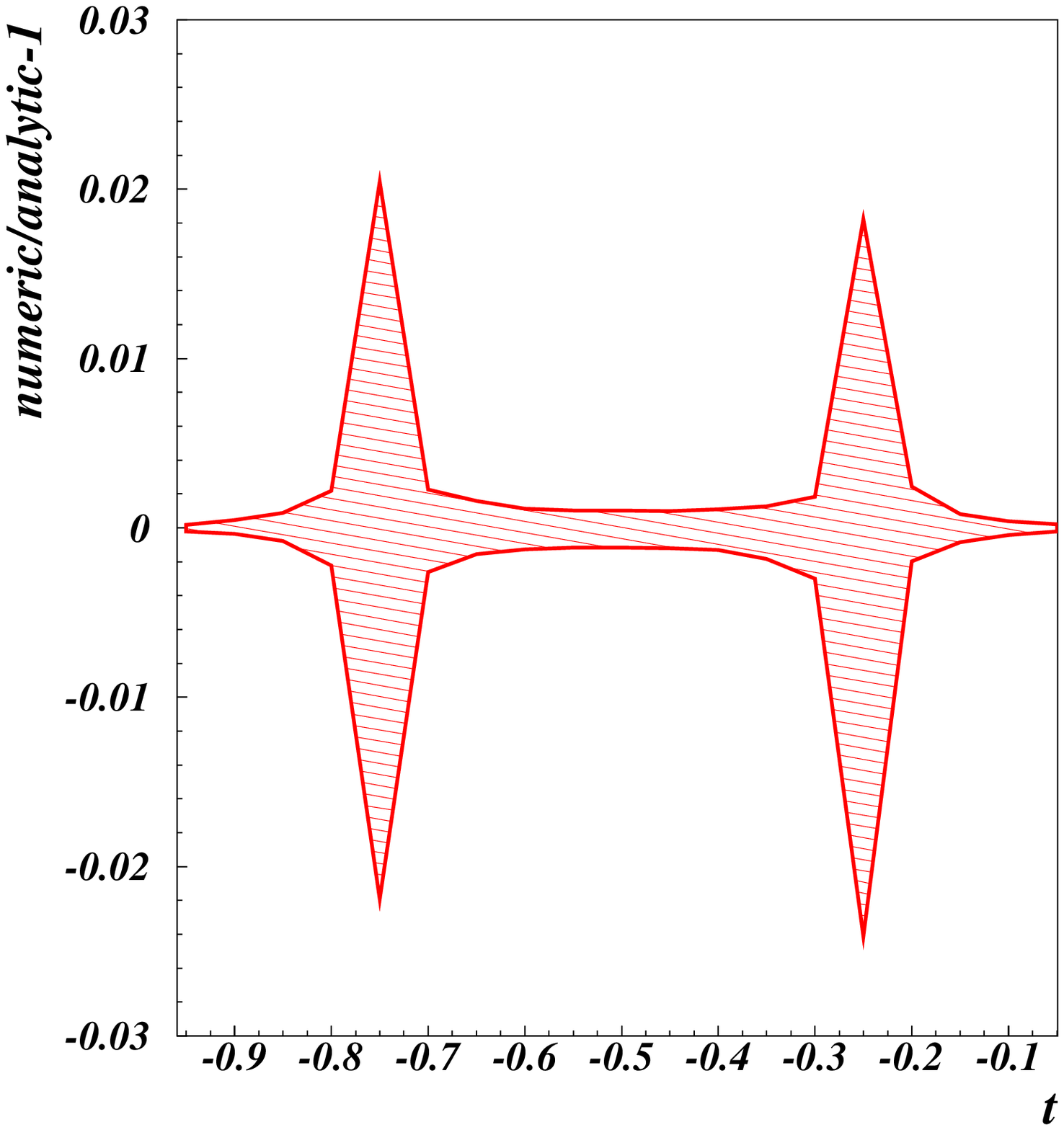}
\end{minipage}
\caption{Results for the finite piece of the non planar box with seven
propagators in the region $u,\,t<0$ and $s=-t-u$ for $s=1$ as
a function of $t$. The left panels show our numerical results for the 
real and imaginary part (upper and lower plot respectively) and the
right ones the corresponding comparisons with the analytic calculation 
of \cite{tausk}.}
\label{fig:nonplanarRI}
\end{figure}

\begin{figure}[h]
\begin{minipage}{7.5cm}
\includegraphics[width=8cm]{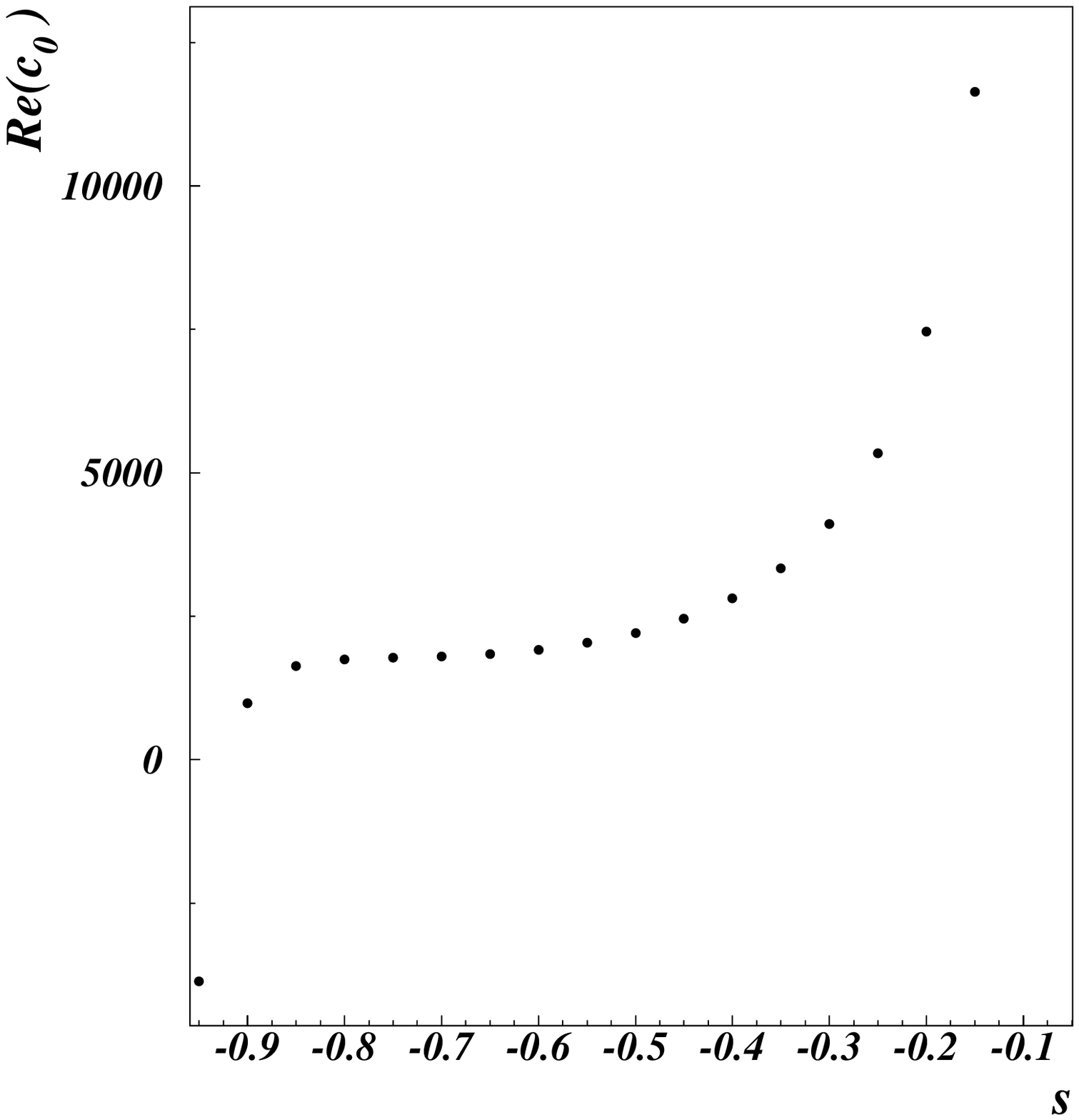}
\end{minipage}
\begin{minipage}{7.5cm}
\includegraphics[width=8cm]{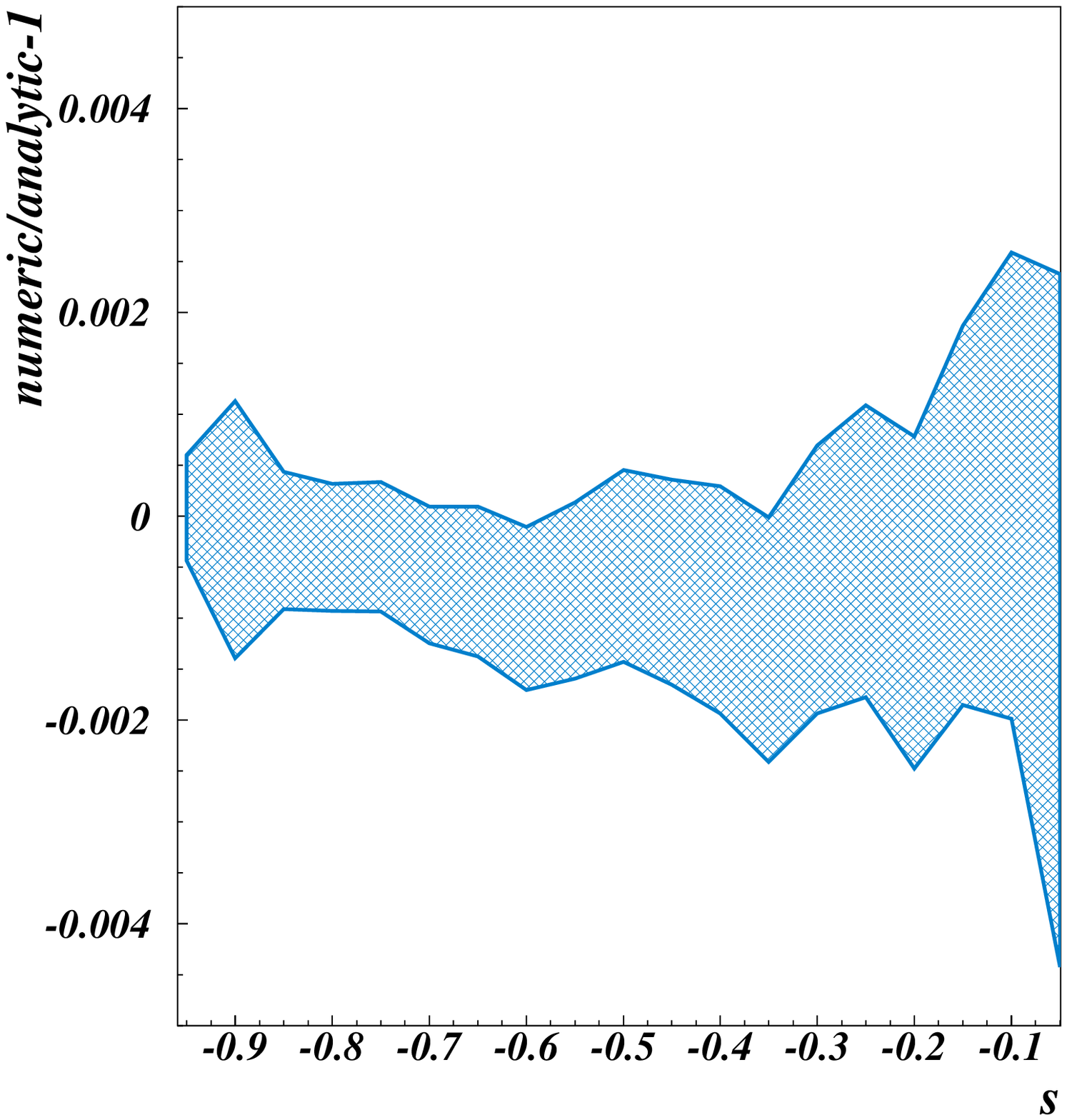}
\end{minipage}
\begin{minipage}{7.5cm}
\includegraphics[width=8cm]{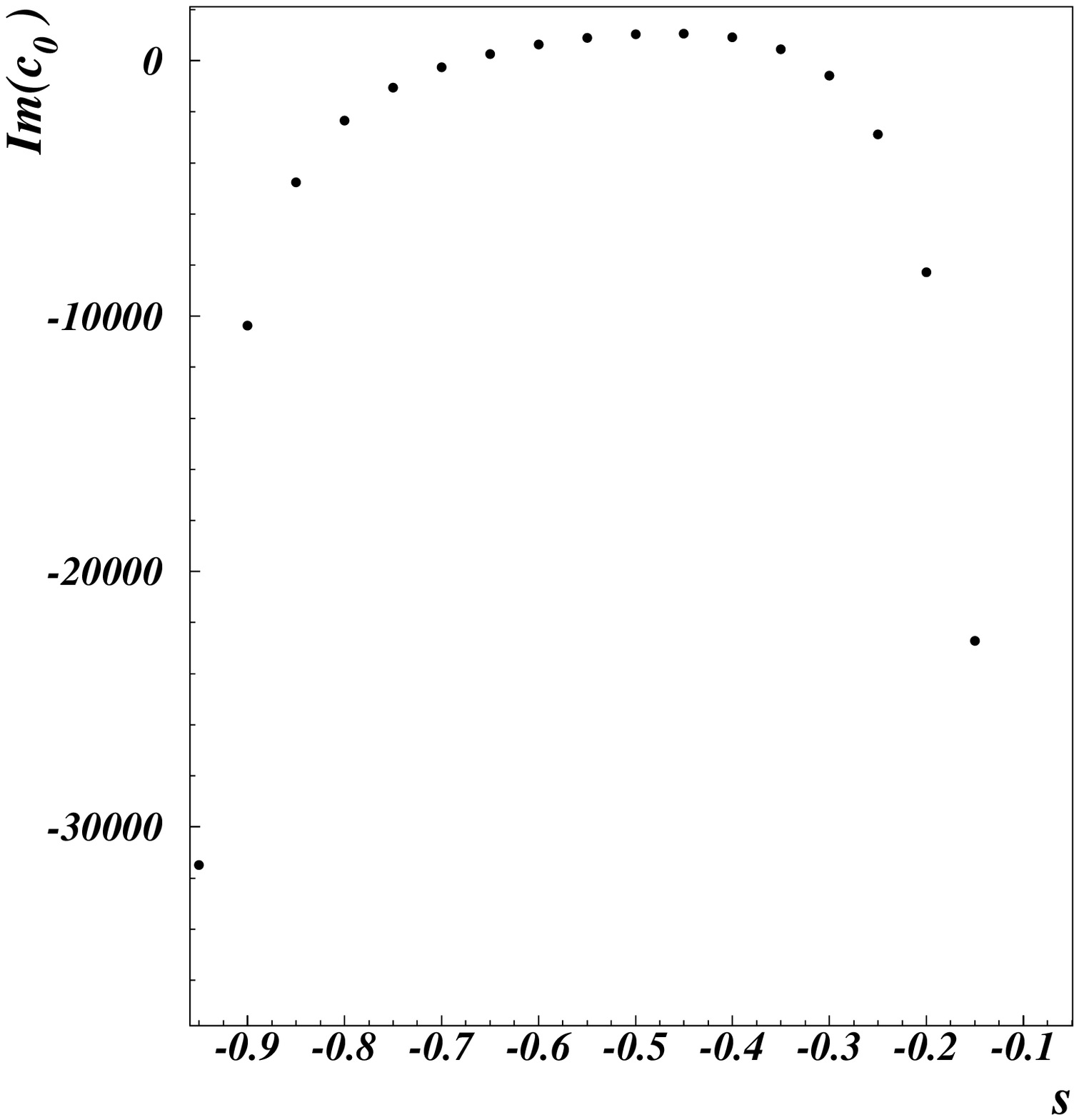}
\end{minipage}
\begin{minipage}{7.5cm}
\includegraphics[width=8cm]{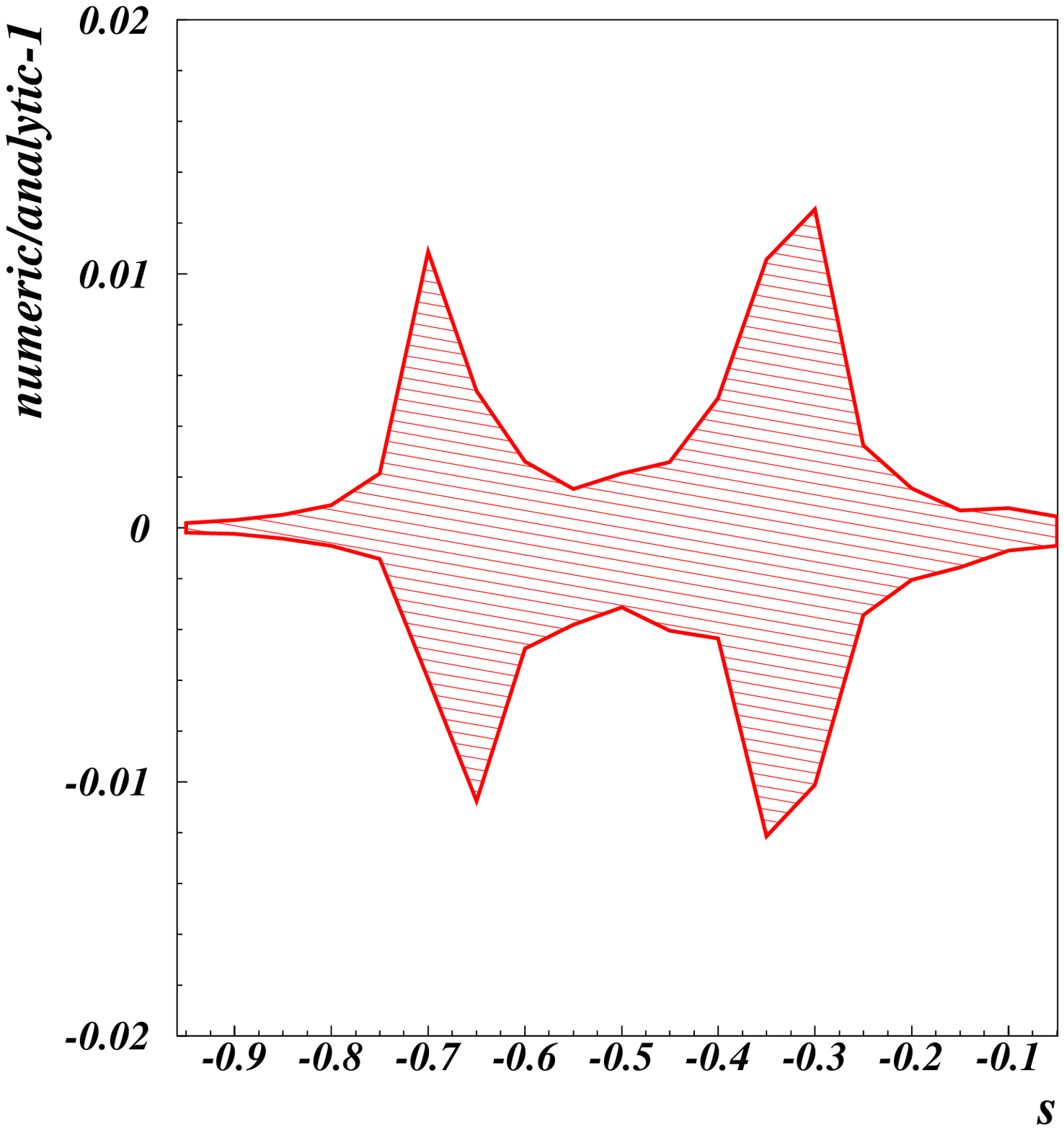}
\end{minipage}
\caption{Results for the finite piece of the non planar box with seven
propagators in the region $u,\,s<0$ and $t=-s-u$ for $t=1$ as
a function of $s$. The upper and lower left panels show our numerical 
results for the real and imaginary parts, respectively; and the
two right ones the corresponding comparisons to the analytic 
calculation of \cite{tausk}.}
\label{fig:nonplanarRII}
\end{figure}

In Figures \ref{fig:nonplanarRI} and \ref{fig:nonplanarRII} we show
our results for the finite piece of the non-planar box in the regions
(i) $u,\,t<0$ and $s=-t-u$ and (ii) $u,\,s<0$ and $t=-s-u$ respectively, 
where $s=(p_1+p_2)^2$, $t=(p_2-p_3)^2$ and $s+t+u=0$.
In case (i) we fixed $s=1$ as reference and plot the results as a function
of $t$, whereas in (ii) $t=1$ is fixed and we change $s$ within its allowed 
range. We show the comparison of our results with the 
analytic calculation of \cite{tausk}.
 
% Results for the non-planar double box

In this case, the relative errors obtained by numerical integration are 
bigger than in the the on-shell double box. This is particularly noticeable
in the imaginary part in both regions (i) and (ii), for which  
the relative error reaches 1-2\%. However,
this only happens for non-significant points of the phase-space 
in which the magnitude of the imaginary part is almost zero 
(notice that in both regions, the imaginary part of the 
finite term changes sign twice). 

The non-planar double-box has a complicated analytic structure in terms 
of  the kinematic invariants, since there is no Euclidean kinematic region
for this integral. The two kinematic regions that we study here, require 
complicated analytic continuations with traditional methods.
With our method, it is particularly simple to compute loop 
integrals in different regions of phase space.
Our results demonstrate that the numerical 
integration of the contour integrals allows for a trivial analytic 
continuation in the invariants, giving correct and accurate results 
in all phase-space.

\subsection{Planar double-box with one leg off-shell}
Our next example is the planar double-box with one 
external mass. The first results for this integral were obtained
in~\cite{sector_loop} using sector decomposition. Soon 
afterwards it was computed analytically in~\cite{smirnovB}, 
and  with differential
equation method in \cite{thomas_int}.

We derive a 5-fold MB representation for  this integral in 
Section II. The integral is $1/\epsilon^4$ divergent; after 
the expansion in $\epsilon$,  
we find that the double and simple poles and  the finite term, contain 
three-dimensional Mellin-Barnes integrals. 

% Diagrams for the double box with one mass
\begin{figure}[h]
\begin{minipage}{7.5cm}
\includegraphics[width=4cm]{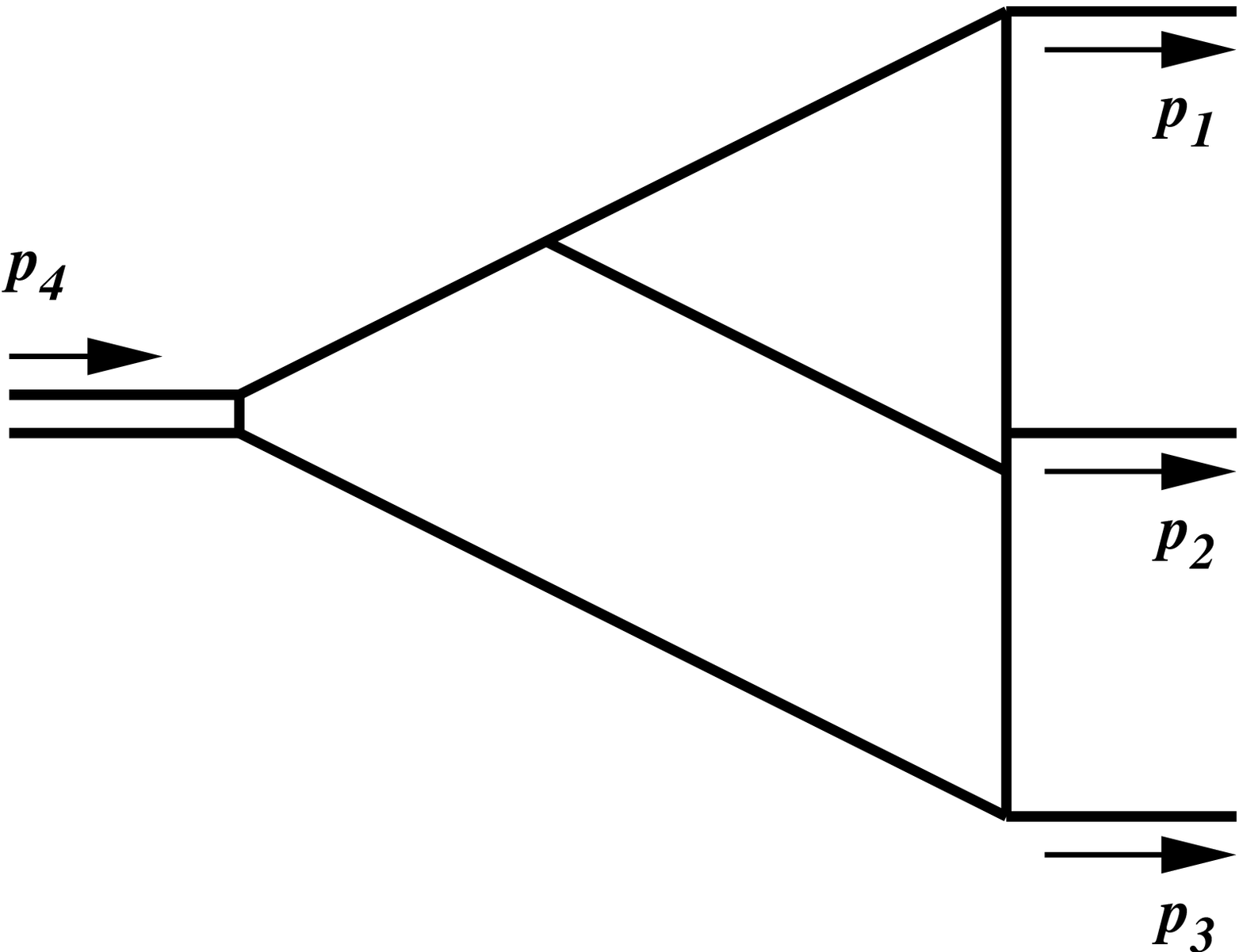}
\begin{center}
(a)
\end{center}
\end{minipage}
\begin{minipage}{7.5cm}
\includegraphics[width=6cm]{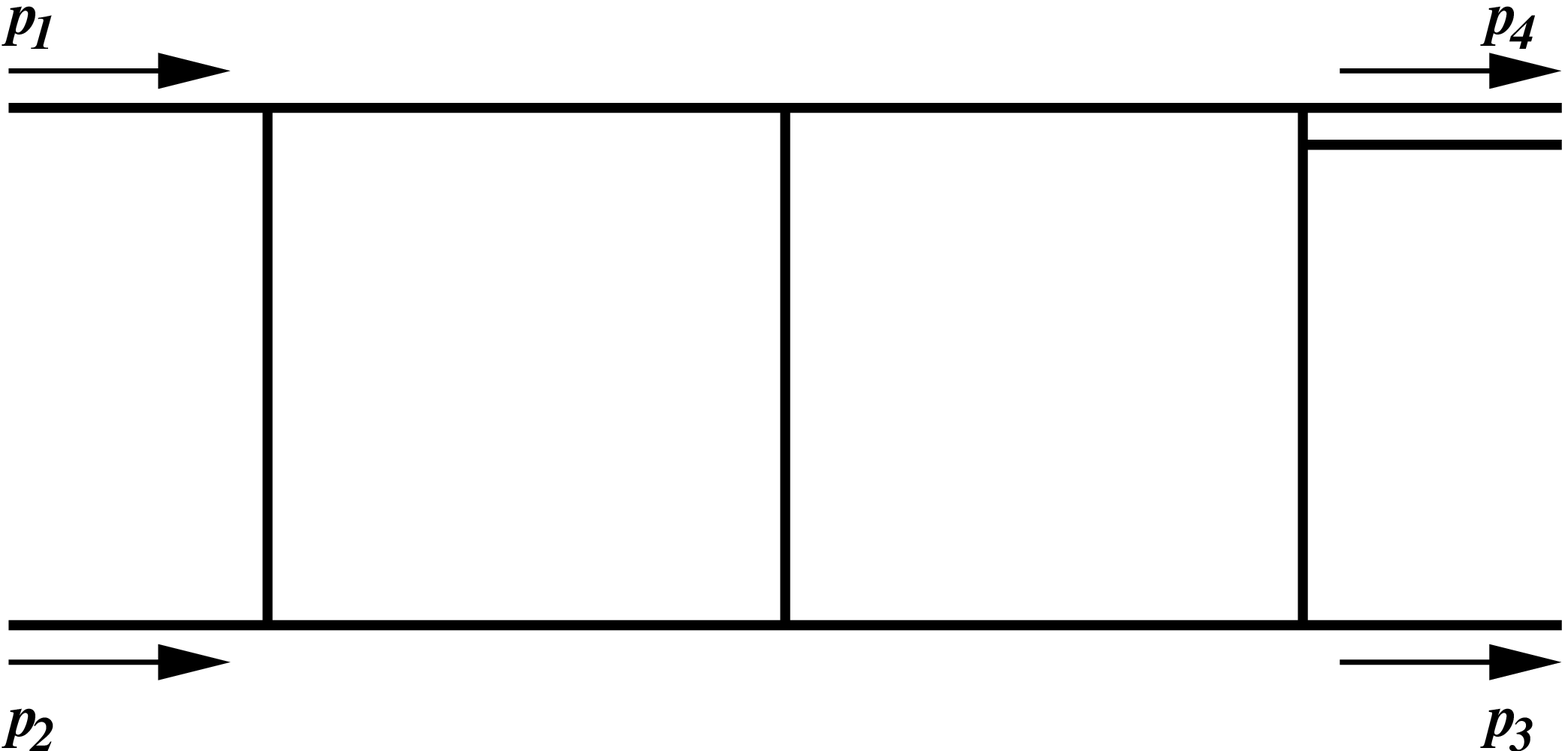}
\begin{center}
(b)
\end{center}
\end{minipage}
\caption{The two kinematical configurations considered for the double
box with one massive leg. (a) the decay process 
$p_{4}\rightarrow p_1+p_2+p_3$ and (b) the scattering
$p_1+p_2\rightarrow p_3+p_4$. The double line corresponds to 
the massive particle, $p_4$.}
\label{fig:2B1mass}
\end{figure}

\begin{figure}[h]
\begin{minipage}{7.5cm}
\includegraphics[width=8cm]{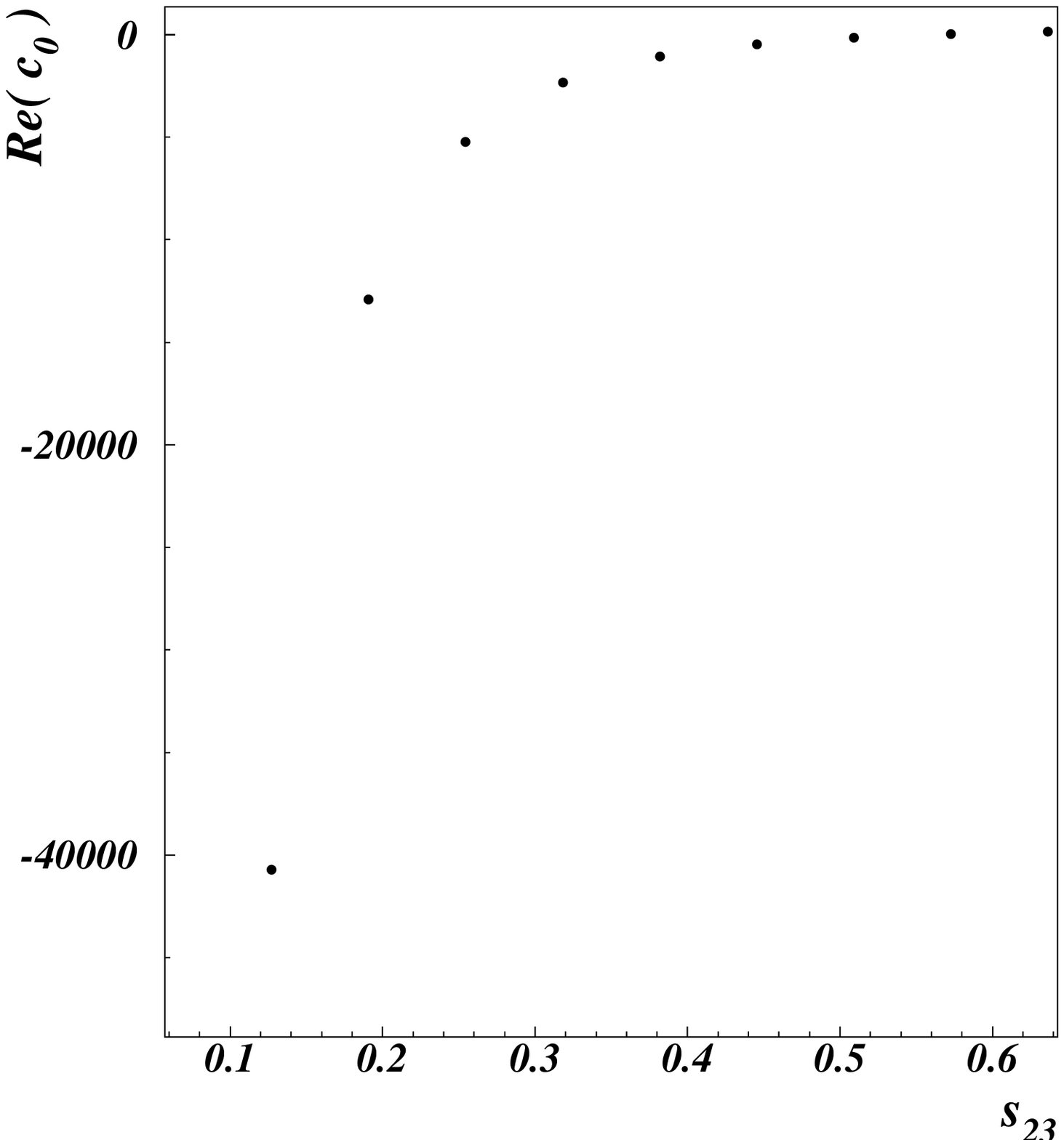}
\end{minipage}
\begin{minipage}{7.5cm}
\includegraphics[width=8cm]{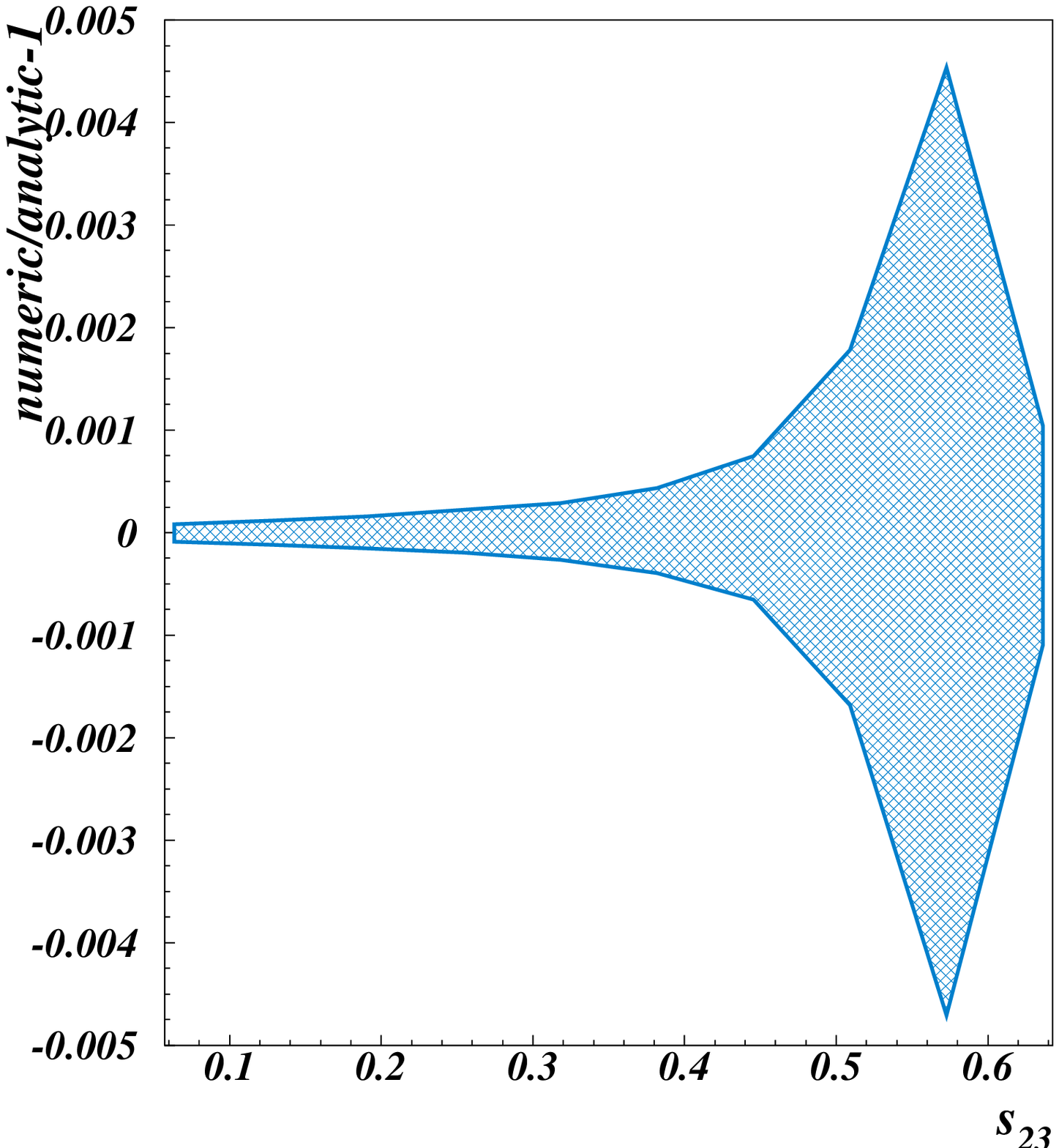}
\end{minipage}
\begin{minipage}{7.5cm}
\includegraphics[width=8cm]{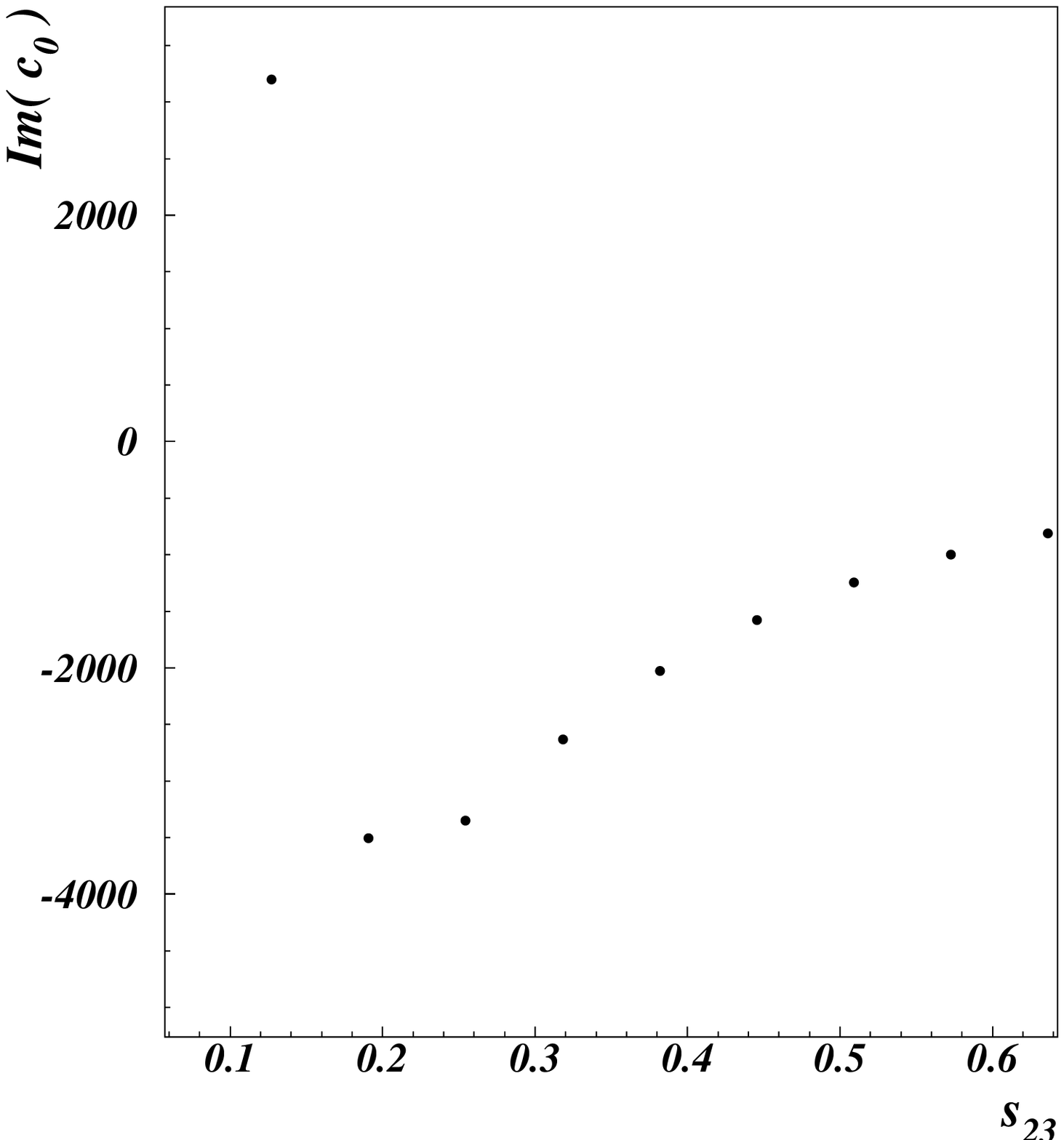}
\end{minipage}
\begin{minipage}{7.5cm}
\includegraphics[width=8cm]{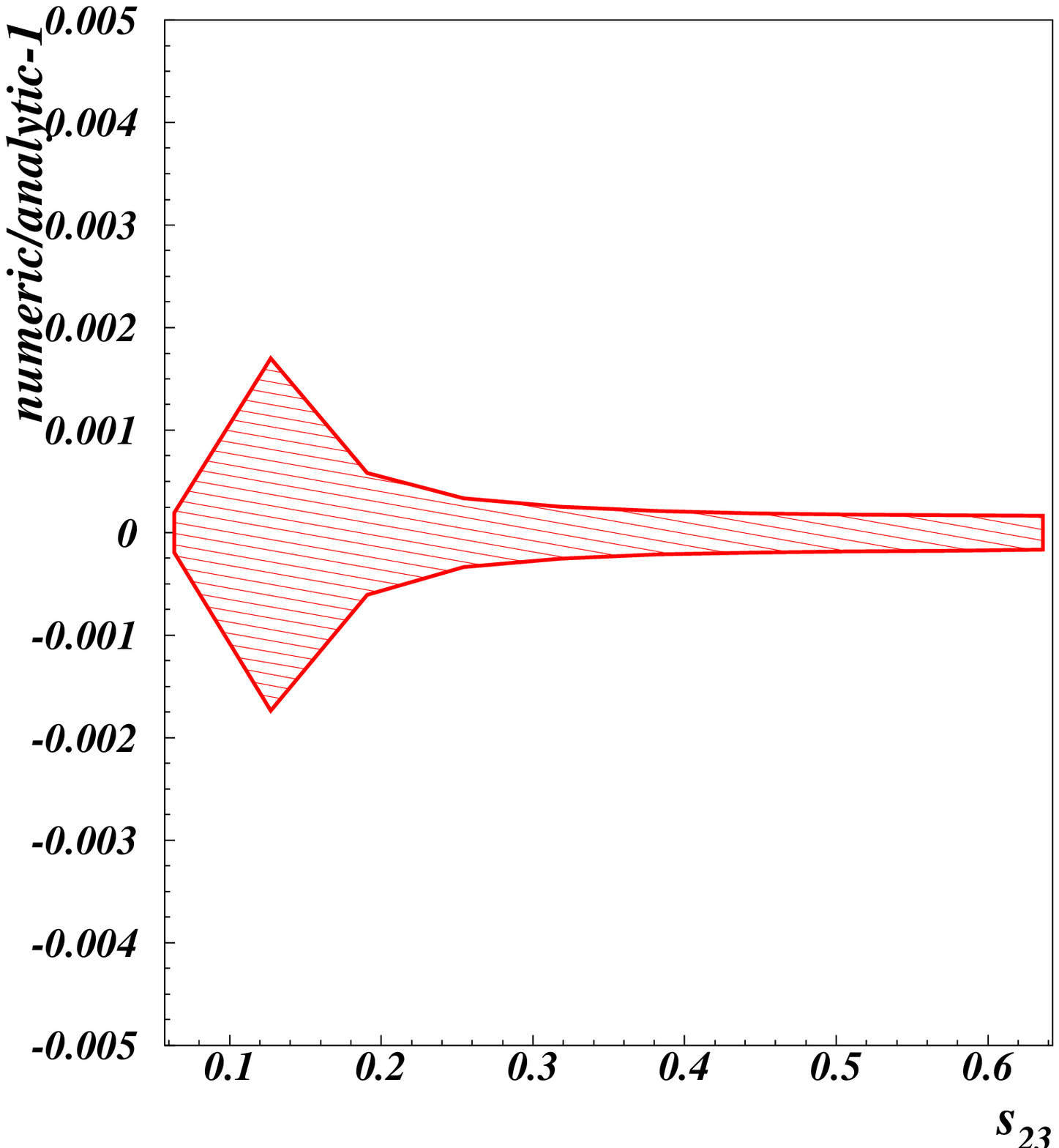}
\end{minipage}
\caption{Results for the finite term of the double box with one leg off-shell
in the physical region for the decay of a heavy particle, 
$p_{4}\rightarrow p_1+p_2+p_3$
(Figure \ref{fig:2B1mass}a). 
On the upper and lower left panel we plot the real and imaginary parts 
as a function of the invariant $s_{23}=(p_2+p_3)^2$ 
for fixed value of $p_4^2=s_{123}=1$ and $s_{13}=(p_1+p_3)^2=3/10$. 
On the right panels we show the corresponding ratios of the 
numerical calculation to the analytic result of
\cite{thomas_int} for the same kinematics, the bands are given by 
the error in the numerical integrations.}
\label{fig:1OFFdecay}
\end{figure}

In Figure \ref{fig:1OFFdecay} we show the results for this integral
in the region corresponding to the decay of a heavy particle 
 $p_{4}\rightarrow p_1+p_2+p_3$ (Figure \ref{fig:2B1mass}a), 
where we have fixed the mass of the
particle $p_4^2=s_{123}=1$ and one of the invariants 
$s_{13}=(p_1+p_3)^2=3/10$, and 
we consider values of $s_{23}=(p_2+p_3)^2$ within the
allowed range. Again, the relative errors are of the order of the few
per mille, with highest values in the points in which the absolute
value of the integral is small (in this case both the real and the imaginary
components change sign).

% results for the double box with one mass in the 1->3 process

\begin{figure}[h]
\begin{minipage}{7.5cm}
\includegraphics[width=8cm]{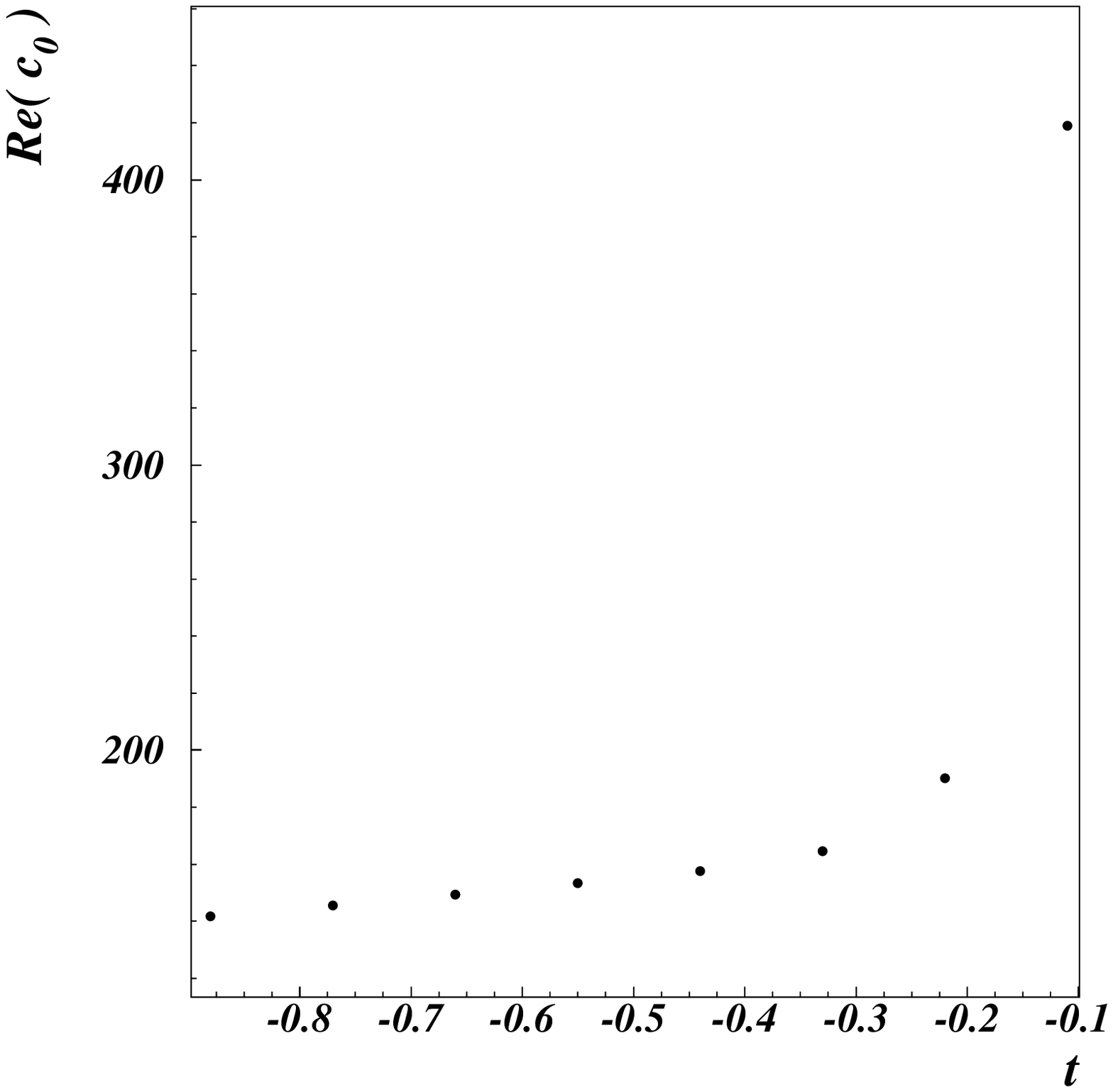}
\end{minipage}
\begin{minipage}{7.5cm}
\includegraphics[width=8cm]{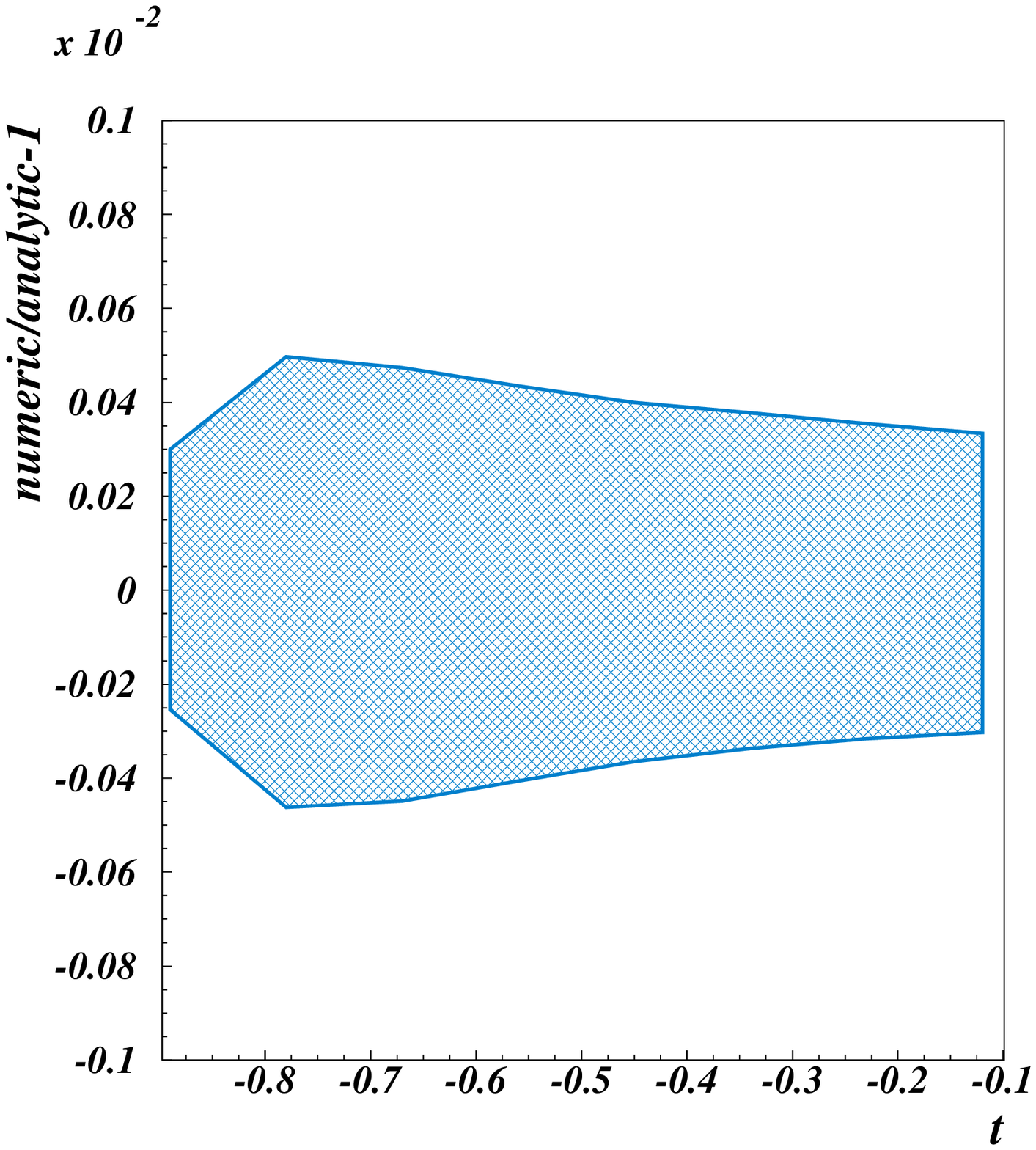}
\end{minipage}
\begin{minipage}{7.5cm}
\includegraphics[width=8cm]{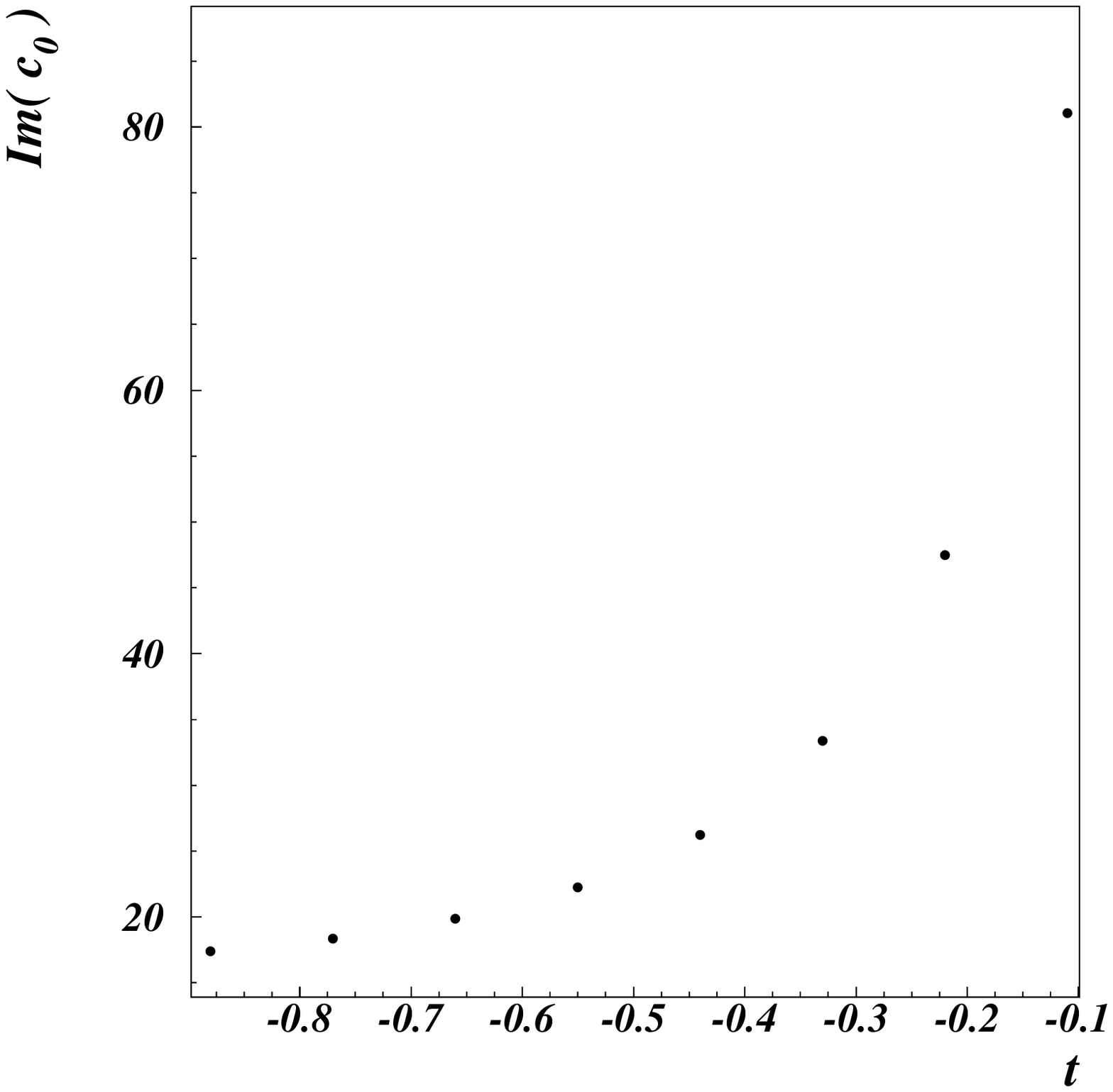}
\end{minipage}
\begin{minipage}{7.5cm}
\includegraphics[width=8cm]{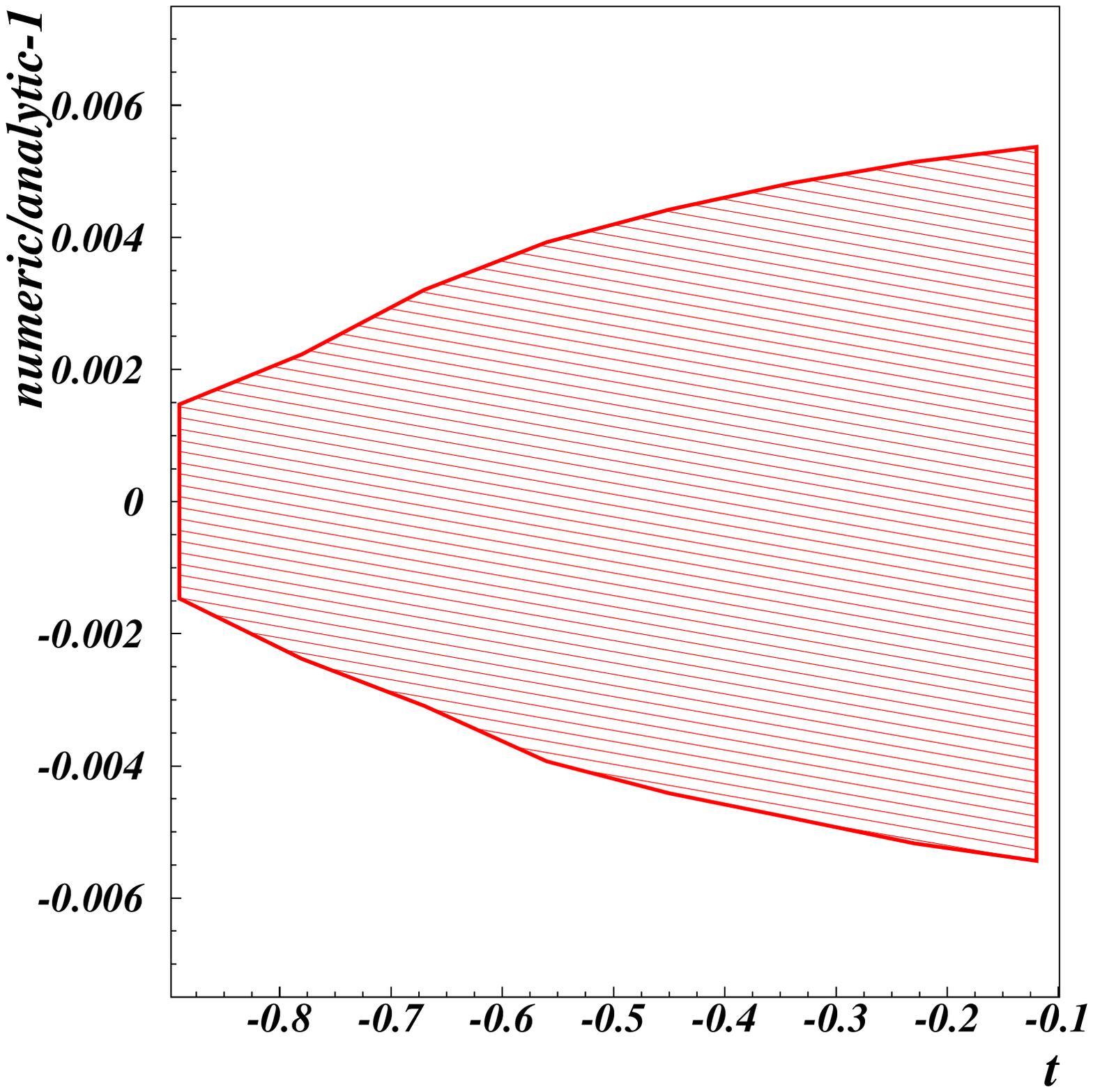}
\end{minipage}
\caption{Results for the finite part of the double box with 
one leg off-shell in the physical
region for a $2\rightarrow 2$ process with the massive leg on the final state
(Fig. \ref{fig:2B1mass}b). On the upper and lower left panels we plot the real 
and imaginary parts of $c_0$ as a function of the invariant $t=(p_2-p_3)^2$ 
for fixed value of $s=(p_1+p_2)^2=1$ and
$p_4^2=1/10$. On the right panels we show the corresponding
ratios of the numerical calculation to the analytic result of
\cite{thomas_int} for the same kinematics, the bands 
are given by the error in the numerical integrations.}
\label{fig:1OFF4a}
\end{figure}

For this topology we also considered the kinematical region corresponding 
to a $p_1+p_2\rightarrow p_3+p_4$ process being $p_4$ the massive 
momentum (see Fig. \ref{fig:2B1mass}b). The corresponding results are shown 
in Figure \ref{fig:1OFF4a} for fixed values of $s=(p_1+p_2)^2=1$ 
and $p_4^2=1/10$ as a function of $t=(p_2-p_3)^2$. Again we find 
very good agreement with the analytic results in both phase space regions.

% results for the double box with one mass in the 2->2 process

\subsection{Planar double box with two adjacent massive legs}
We now make one further step for the planar double box and consider
the case of two external adjacent masses. This integral has been 
evaluated for the first time, in some points in the Euclidean region,
in \cite{Binoth:2003ak}.

As shown in Section II, it is possible to get a MB representation for the 
double box with two adjacent massive legs 
with six MB parameters. Again, after the analytic continuation,
the effective dimensionality is reduced  by
factors containing gamma functions in the denominator. In the present 
example, only integrals in three dimensions are needed to 
compute the finite pieces in the series around $\epsilon=0$. Quartic
and cubic poles are completely determined by the pieces with no
integrals left.

\begin{figure}[h]
%\begin{minipage}{7.5cm}
\includegraphics[width=6cm]{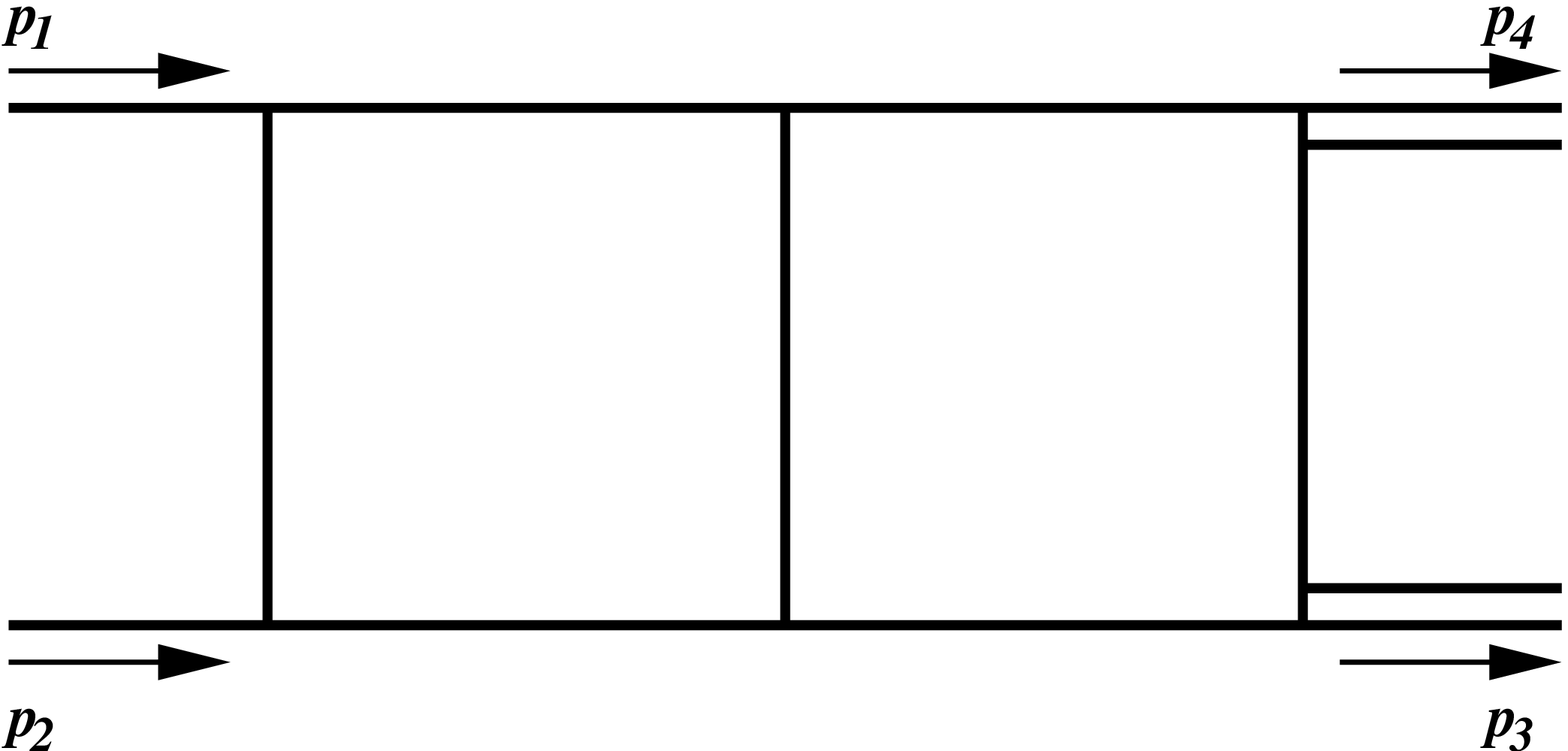}
%\end{minipage}
\caption{The double box diagram with two adjacent massive legs.}
\label{fig:2B2mass}
\end{figure}

We found complete agreement with the results quoted in \cite{Binoth:2003ak}.
We also compared our results for some values of the invariants in the
Euclidean region with an independent calculation obtained with our own 
sector decomposition code. In Table \ref{tb:2off2BE} we show the
results obtained by the two methods for two phase space points
in the mentioned region. The invariants are defined as 
$s=(p_1+p_2)^2$, $t=(p_2-p_3)^2$, $M_1^2=p_3^2$, $M_2^2=p_4^2$ 
(see Figure \ref{fig:2B2mass} for the definition of the momenta).
We find perfect agreement within the integration errors.

\begin{table}[h]
{\small
\begin{tabular}{|c|c|c|}
\hline
Point&MB&Sector Decomposition\\
\hline
$s=t=M_1^2=M_2^2=-1$&
\begin{minipage}[c]{5cm}
\[\begin{array}{r@{.}lcr}
-0&25&&\epsilon^{-4}+\\
0&288608&&\epsilon^{-3}+\\
-2&22227&(1)&\epsilon^{-2}+\\
-6&577&(6)&\epsilon^{-1}+\\
-10&15&(4)&
\end{array}
\]
\end{minipage}
&
\begin{minipage}[c]{5cm}
\[\begin{array}{r@{.}lcr}
-0&25&&\epsilon^{-4}+\\
0&2886&(2)&\epsilon^{-3}+\\
-2&222&(2)&\epsilon^{-2}+\\
-6&577&(6)&\epsilon^{-1}+\\
-10&15&(2)&
\end{array}
\]
\end{minipage}
\\
\hline
\begin{minipage}{3cm}
\[\begin{array}{r@{=}l}
s&-1\\
t&-1/2\\
M_1^2&-7/10\\
M_2^2&-2/5
\end{array}\]
\end{minipage}&
\begin{minipage}[c]{5cm}
\[\begin{array}{r@{.}lcr}
-0&5&&\epsilon^{-4}+\\
0&463887&&\epsilon^{-3}+\\
-4&47418&(1)&\epsilon^{-2}+\\
-17&46&(1)&\epsilon^{-1}+\\
-37&21&(9)&
\end{array}
\]
\end{minipage}
&
\begin{minipage}[c]{5cm}
\[\begin{array}{r@{.}lcr}
-0&5&&\epsilon^{-4}+\\
0&4639&(5)&\epsilon^{-3}+\\
-4&474&(4)&\epsilon^{-2}+\\
-17&46&(2)&\epsilon^{-1}+\\
-37&21&(9)&
\end{array}
\]
\end{minipage}
\\
\hline
\end{tabular}
}
\caption{Results for the double box with two legs off-shell for two 
points in the Euclidean region, we show results obtained with both the
Mellin-Barnes technique and wit the sector decomposition method.}
\label{tb:2off2BE}
\end{table}

\begin{figure}
\begin{center}
\begin{minipage}{6.5cm}
\includegraphics[width=6cm]{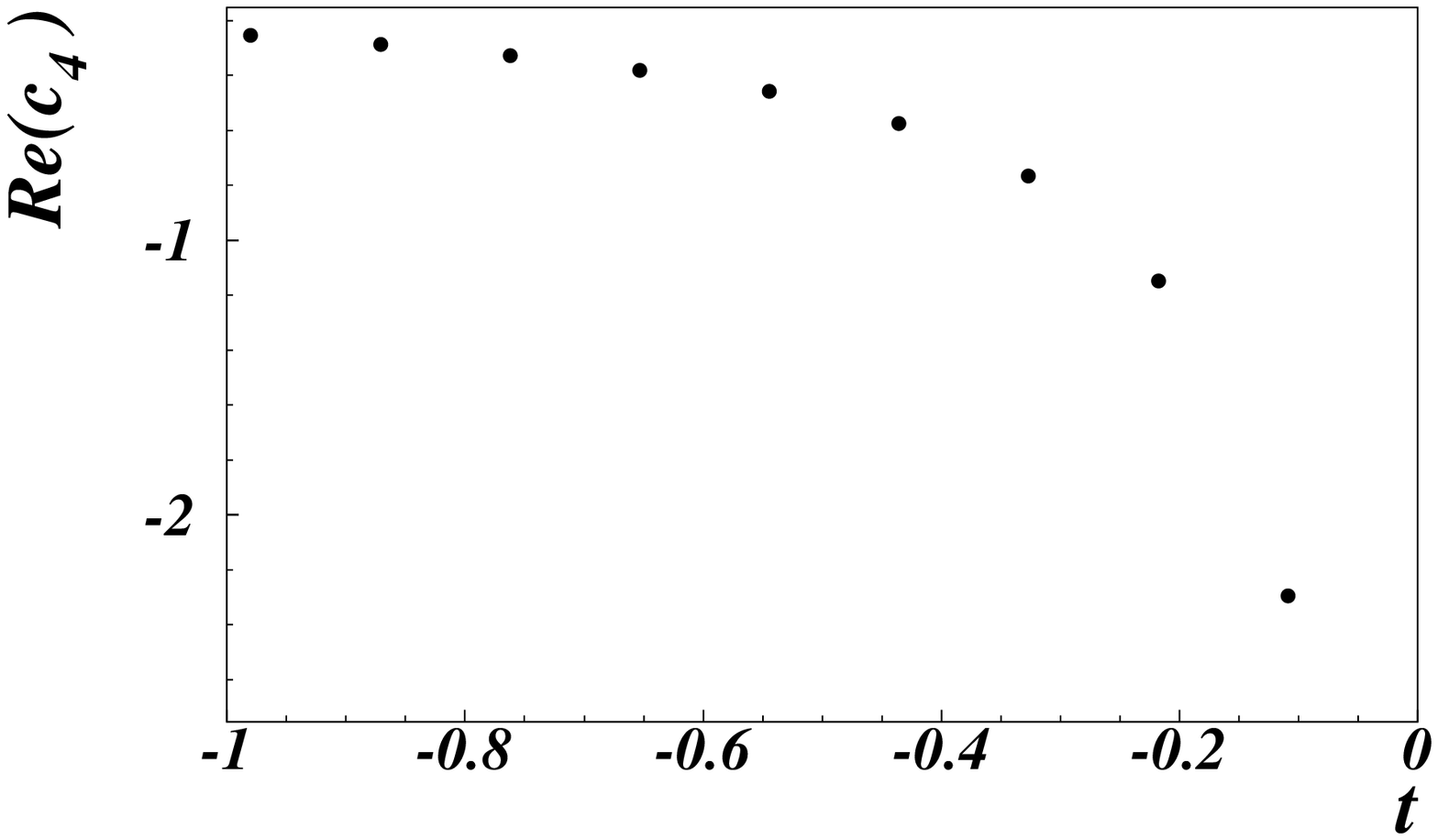}
\end{minipage}
\begin{minipage}{6.5cm}
\includegraphics[width=6cm]{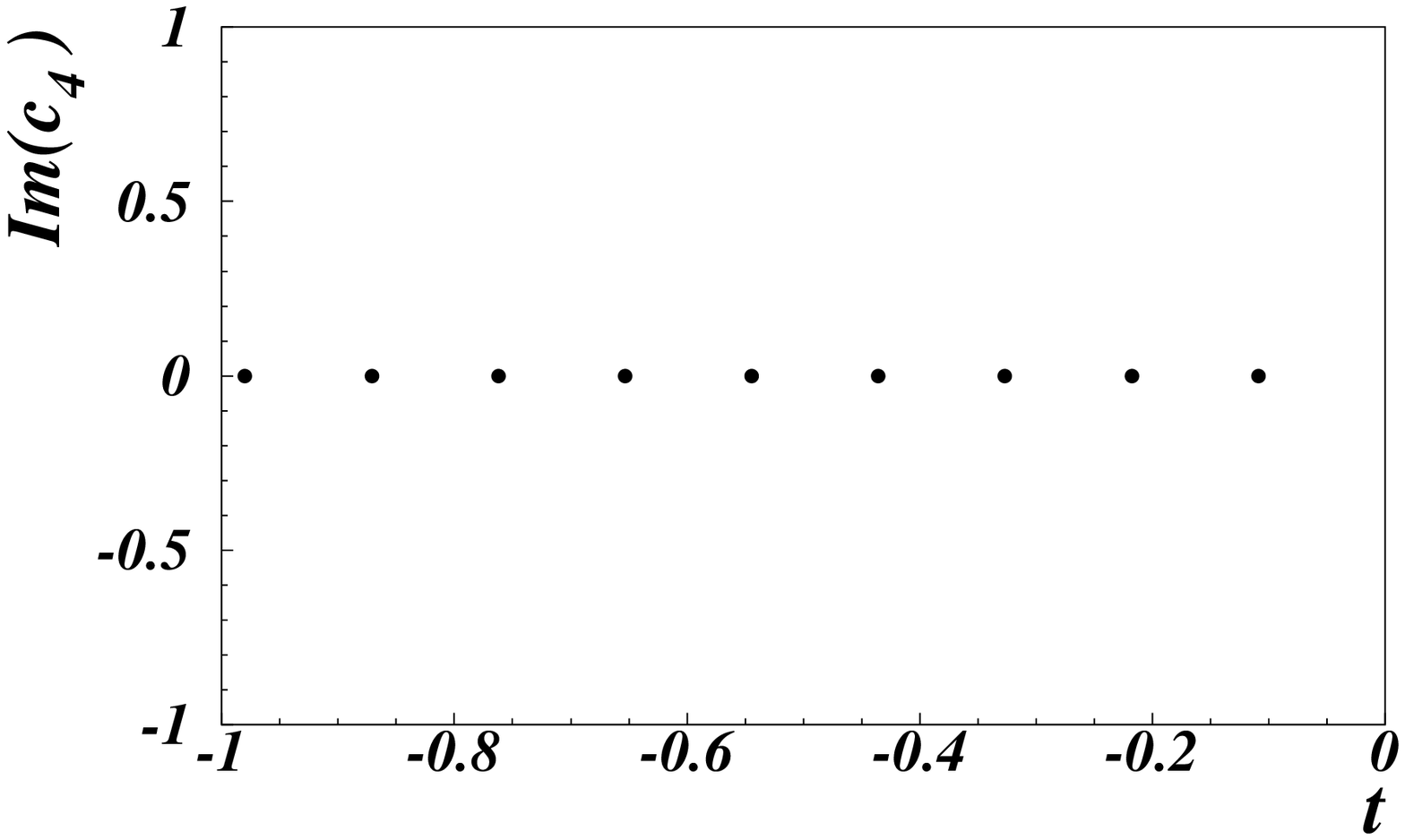}
\end{minipage}
\begin{minipage}{6.5cm}
\includegraphics[width=6cm]{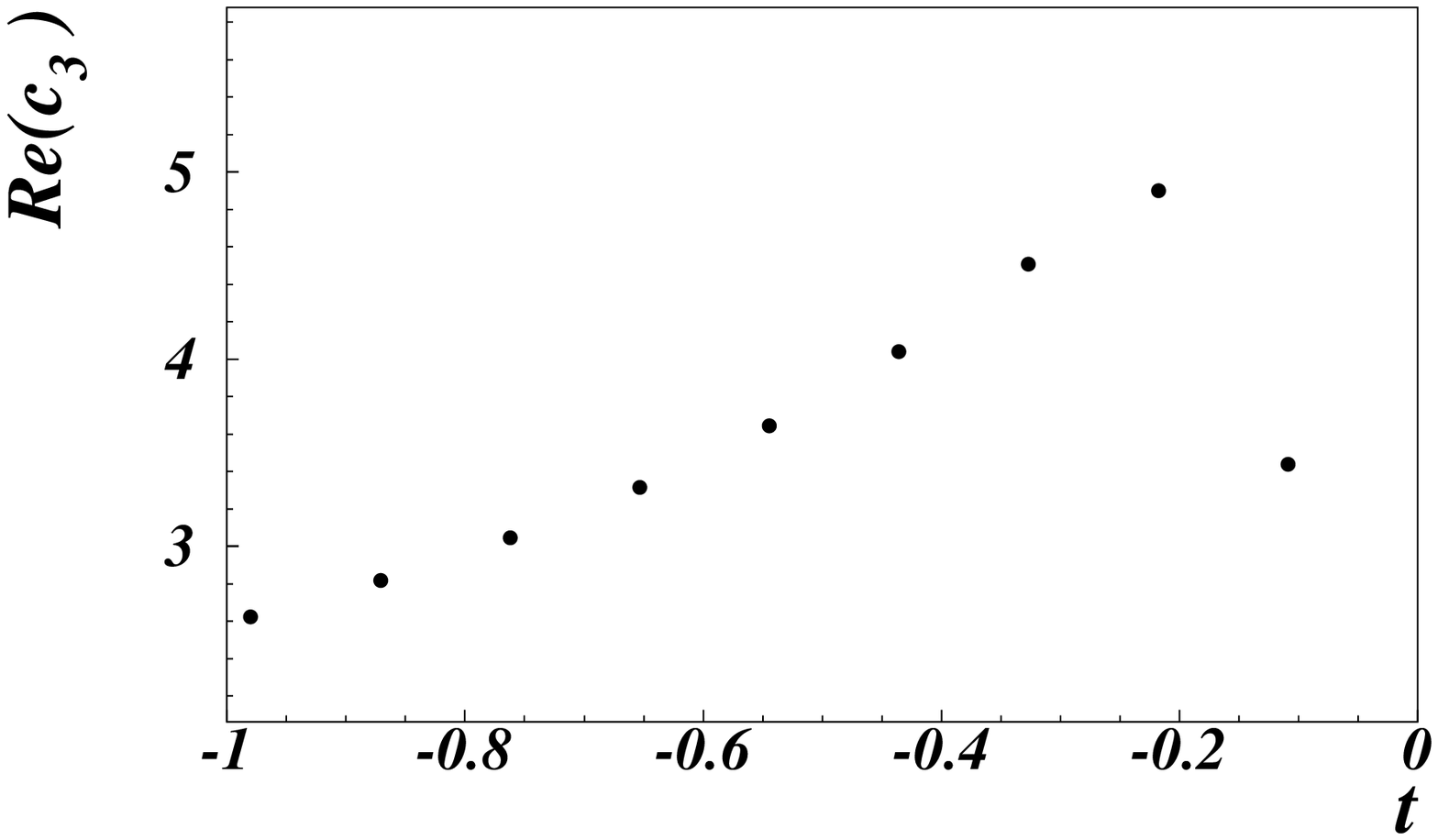}
\end{minipage}
\begin{minipage}{6.5cm}
\includegraphics[width=6cm]{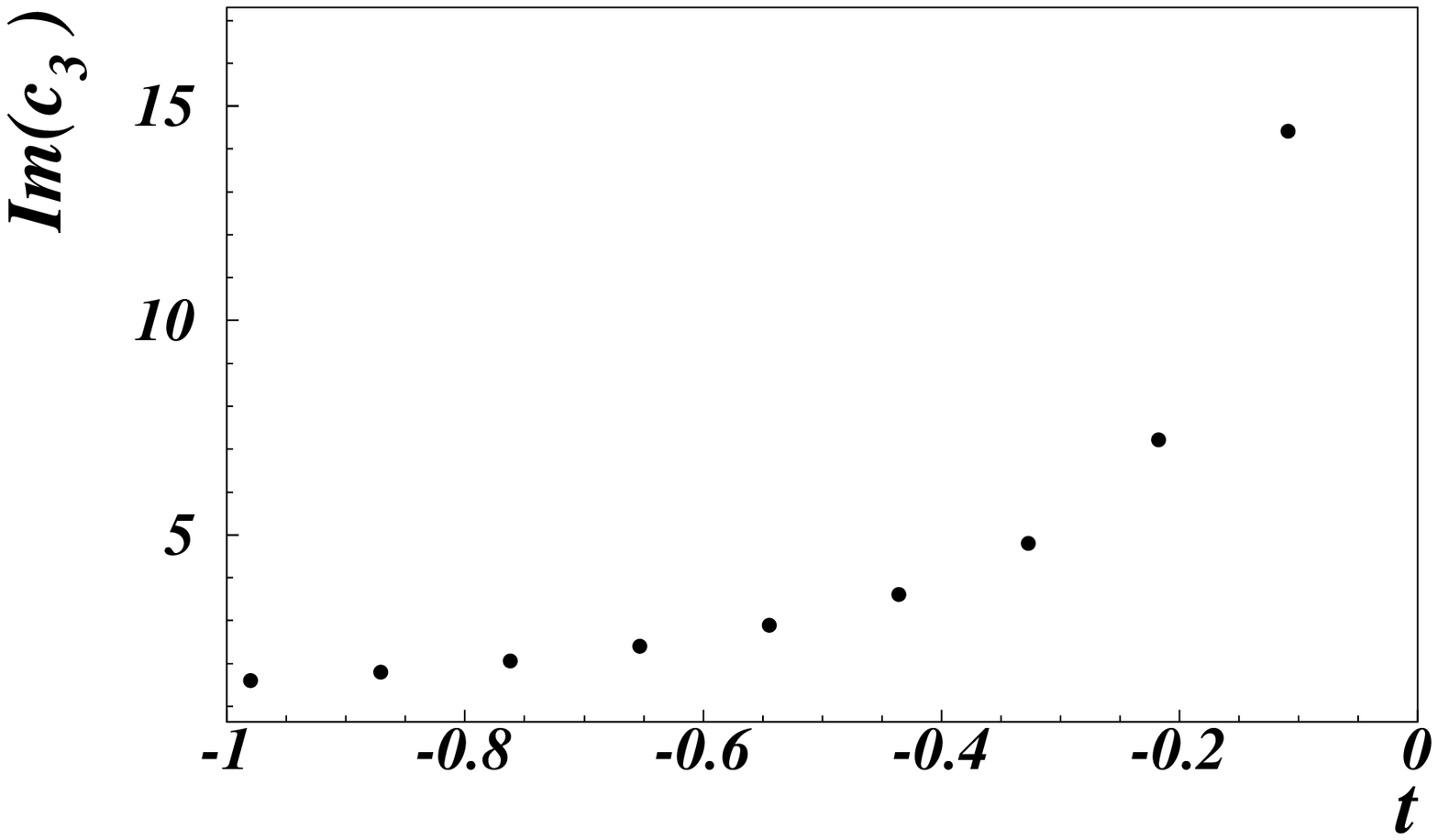}
\end{minipage}
\begin{minipage}{6.5cm}
\includegraphics[width=6cm]{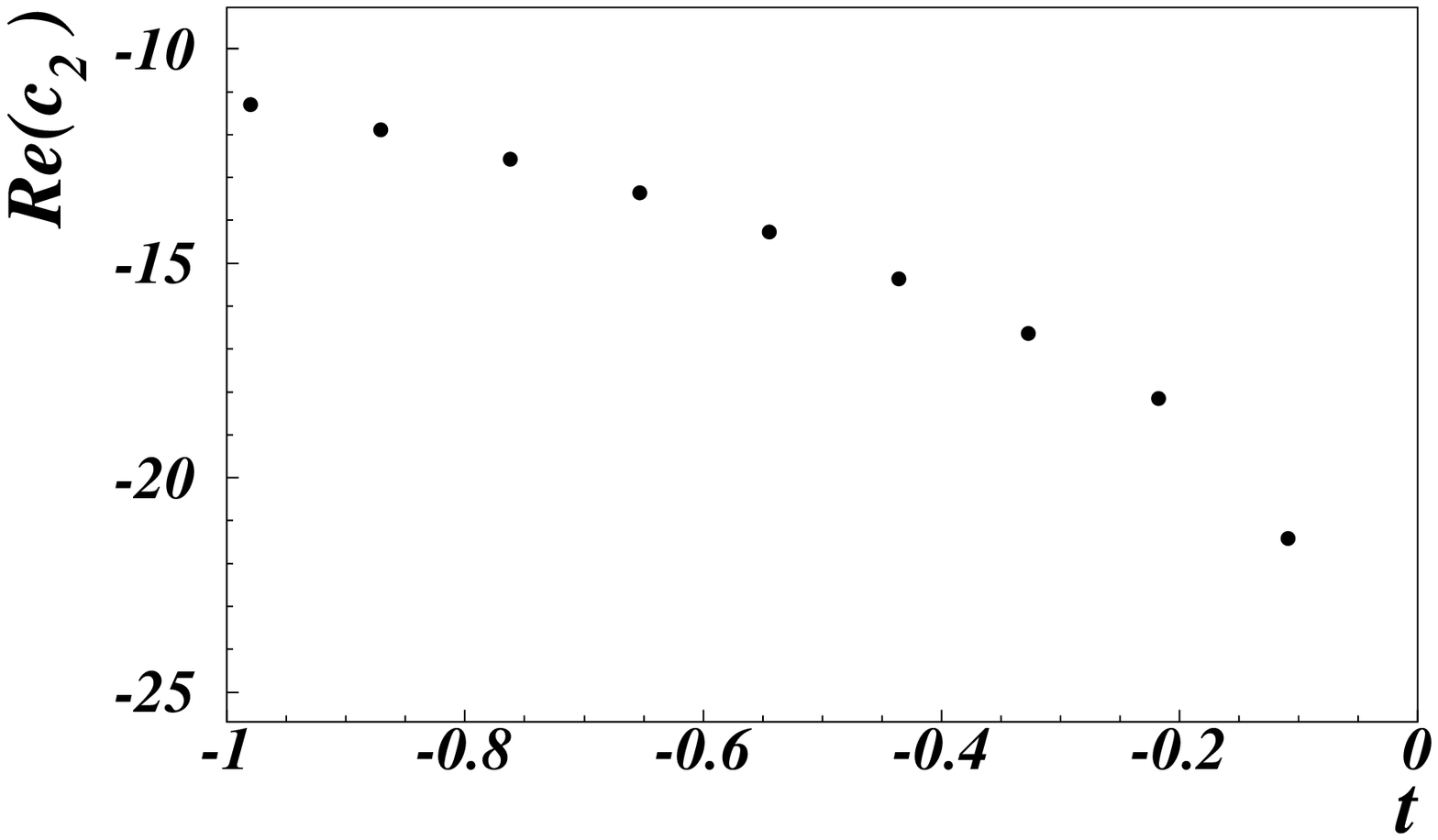}
\end{minipage}
\begin{minipage}{6.5cm}
\includegraphics[width=6cm]{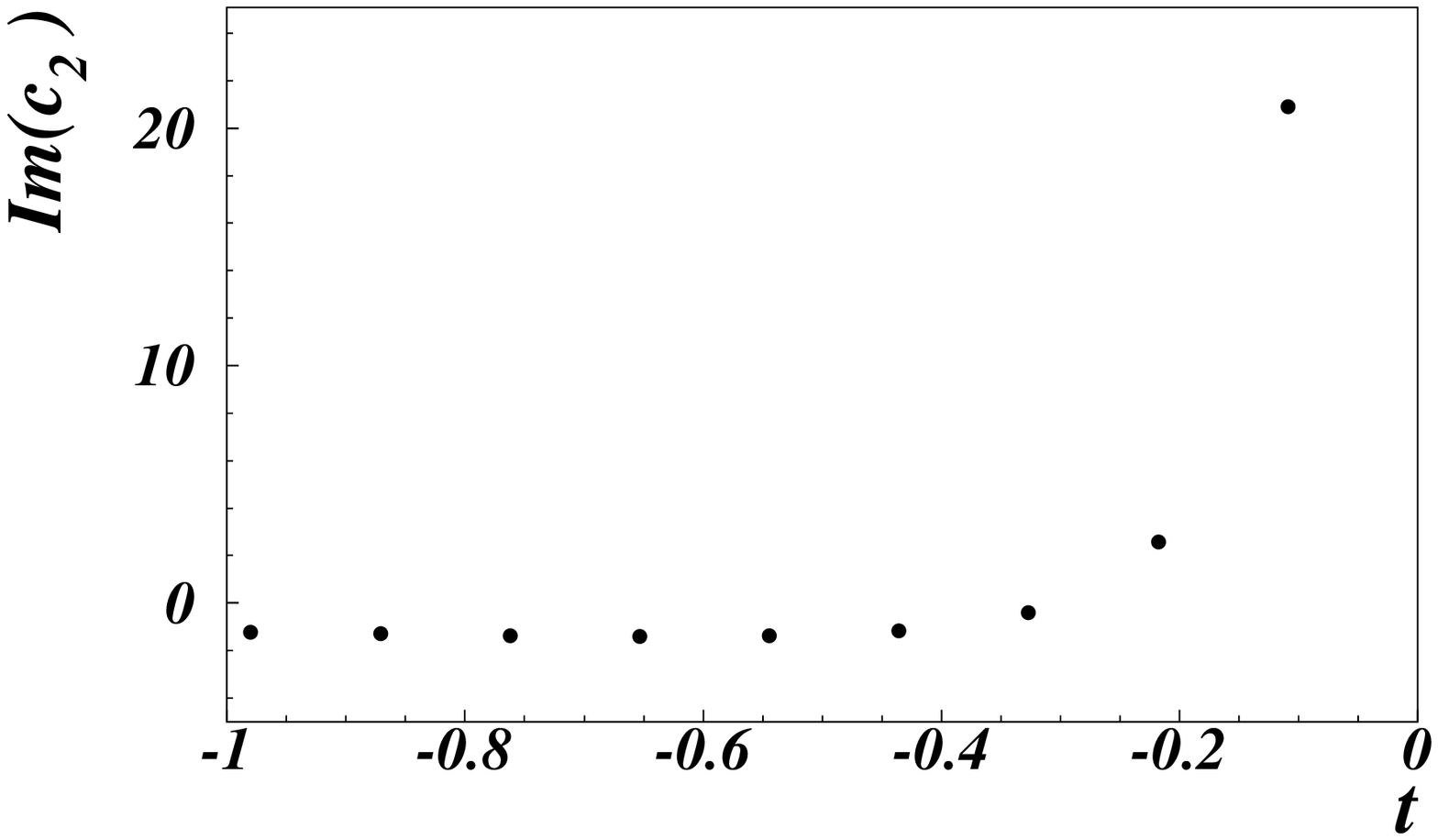}
\end{minipage}
\begin{minipage}{6.5cm}
\includegraphics[width=6cm]{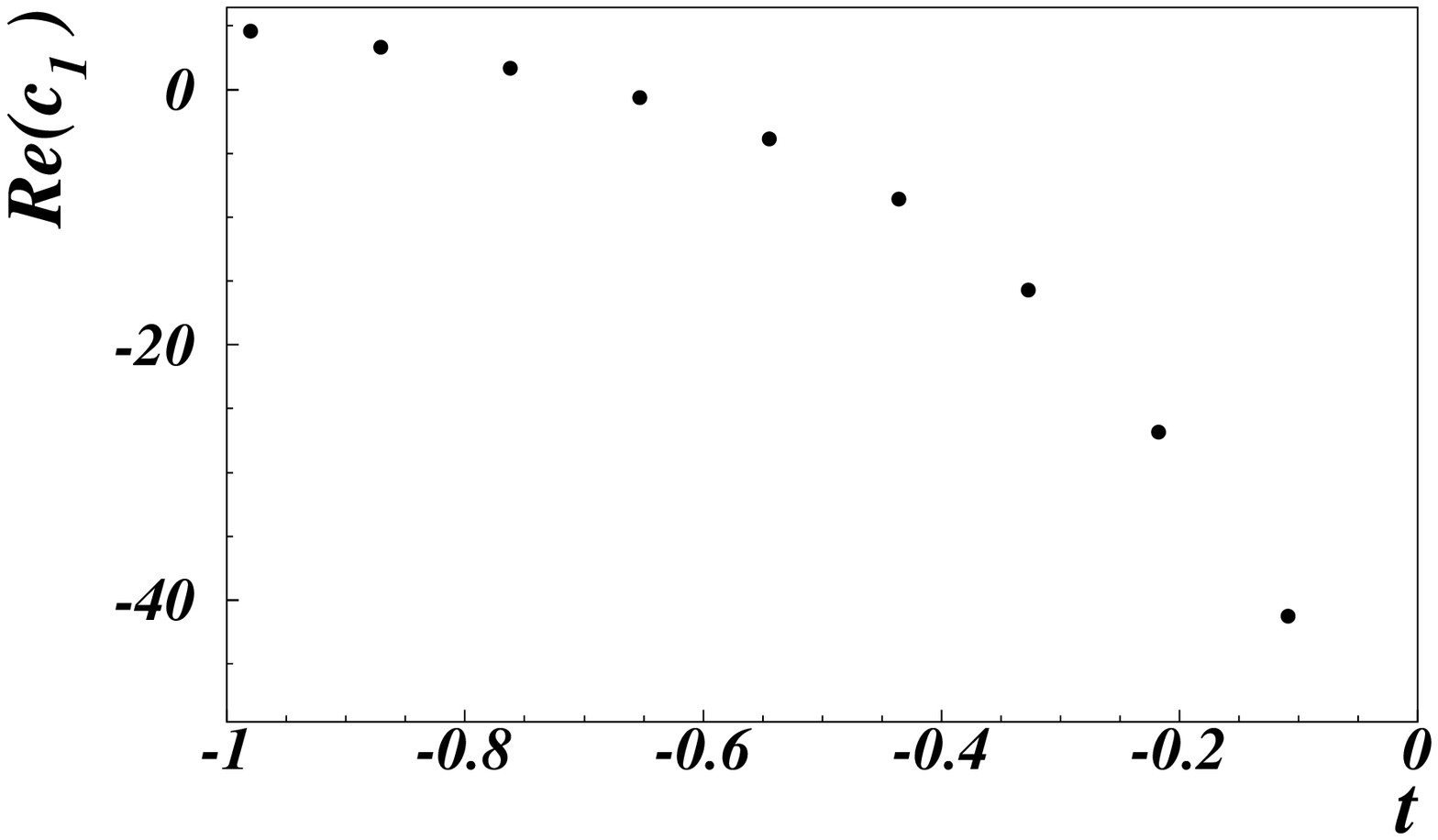}
\end{minipage}
\begin{minipage}{6.5cm}
\includegraphics[width=6cm]{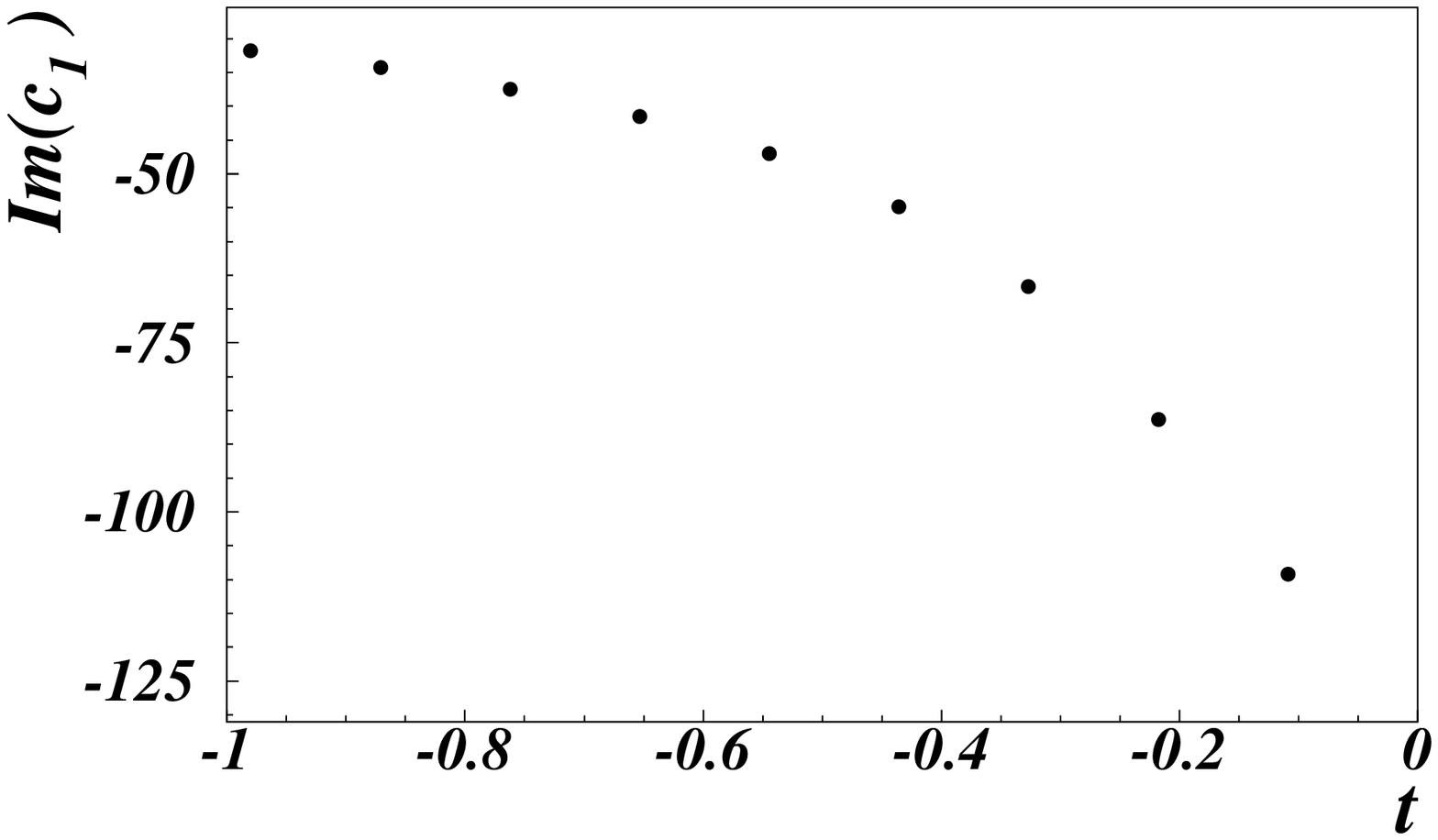}
\end{minipage}
\begin{minipage}{6.5cm}
\includegraphics[width=6cm]{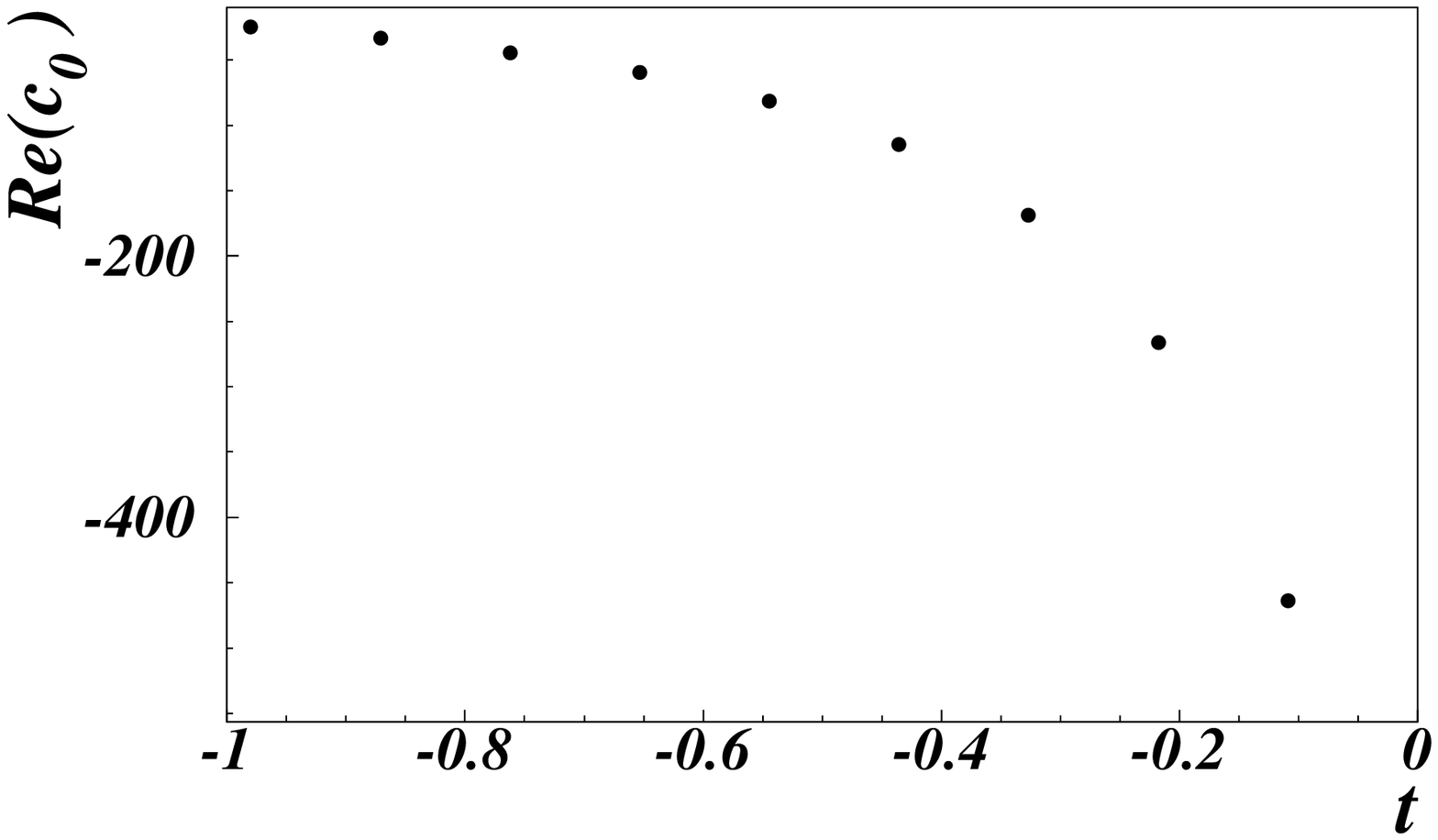}
\end{minipage}
\begin{minipage}{6.5cm}
\includegraphics[width=6cm]{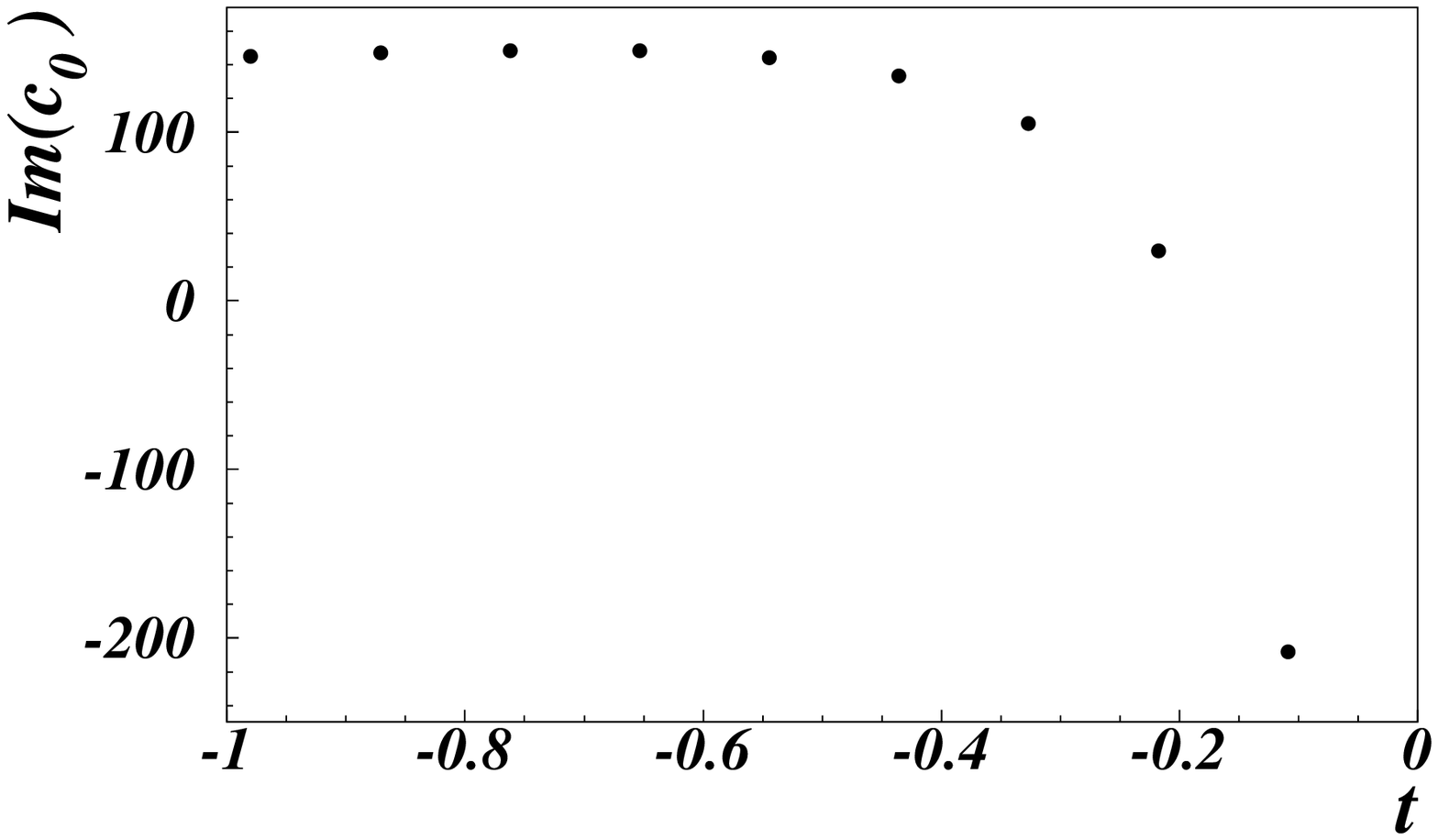}
\end{minipage}
\end{center}
\caption{Results for the double box with two adjacent masses in the 
physical region of a $2\rightarrow 2$ process where the two particles 
in the final state are massive. We plot the values of the coefficients
as function of $t$ for $s=1$, $M_1^2=M_2^2=1/10$.}
\label{fig:2off2BeM}
\end{figure}

\begin{figure}
\begin{center}
\begin{minipage}{6.5cm}
\includegraphics[width=6cm]{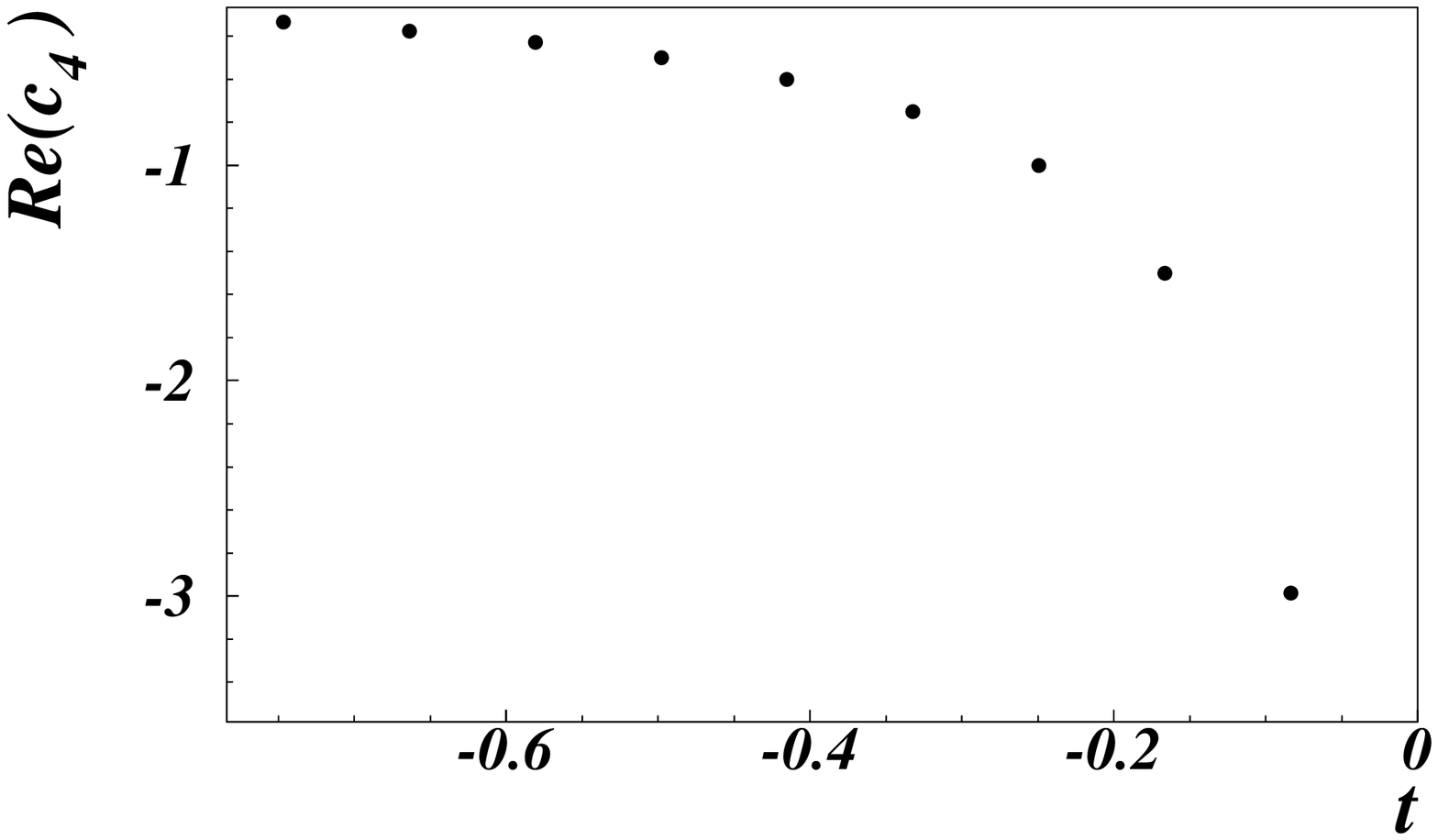}
\end{minipage}
\begin{minipage}{6.5cm}
\includegraphics[width=6cm]{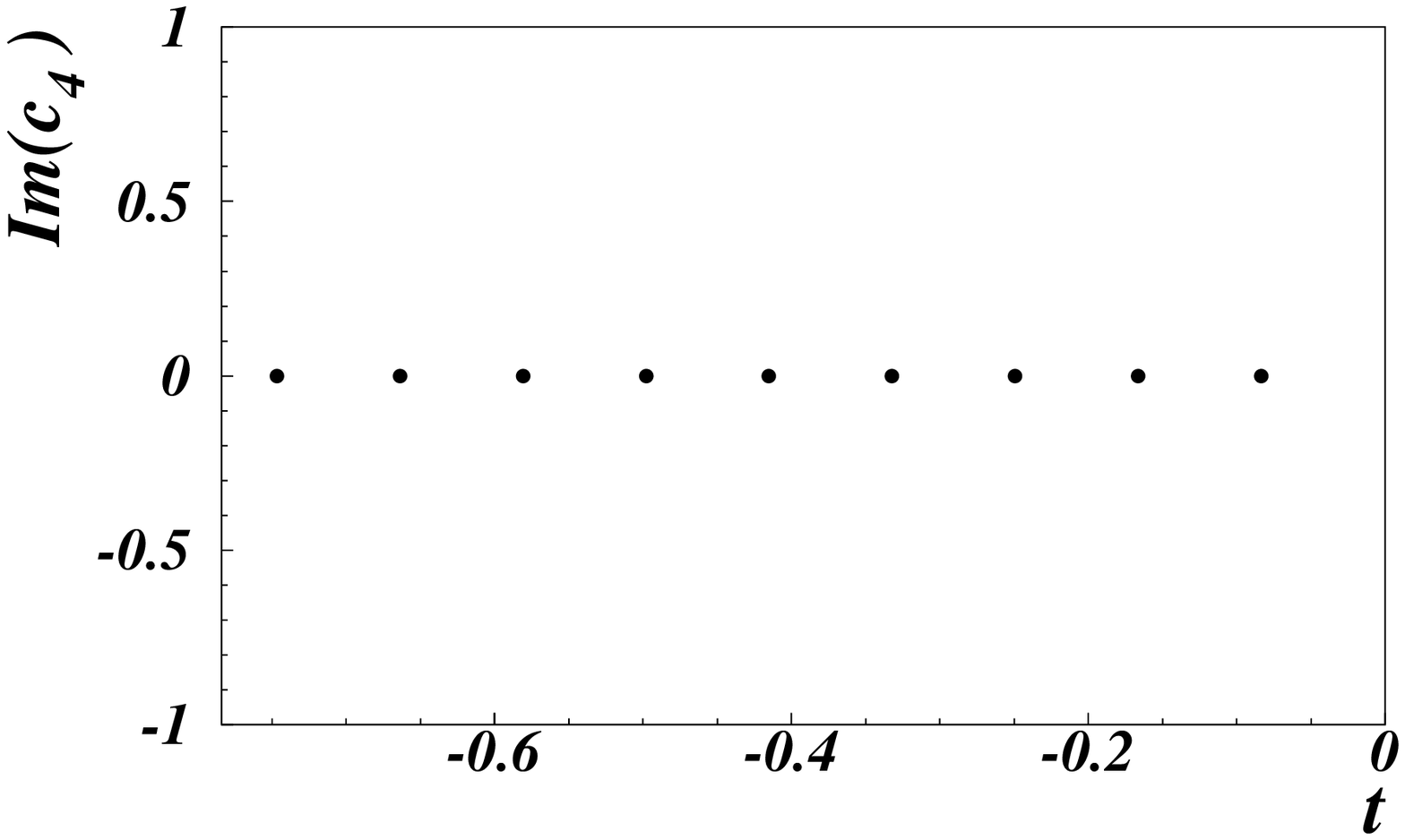}
\end{minipage}
\begin{minipage}{6.5cm}
\includegraphics[width=6cm]{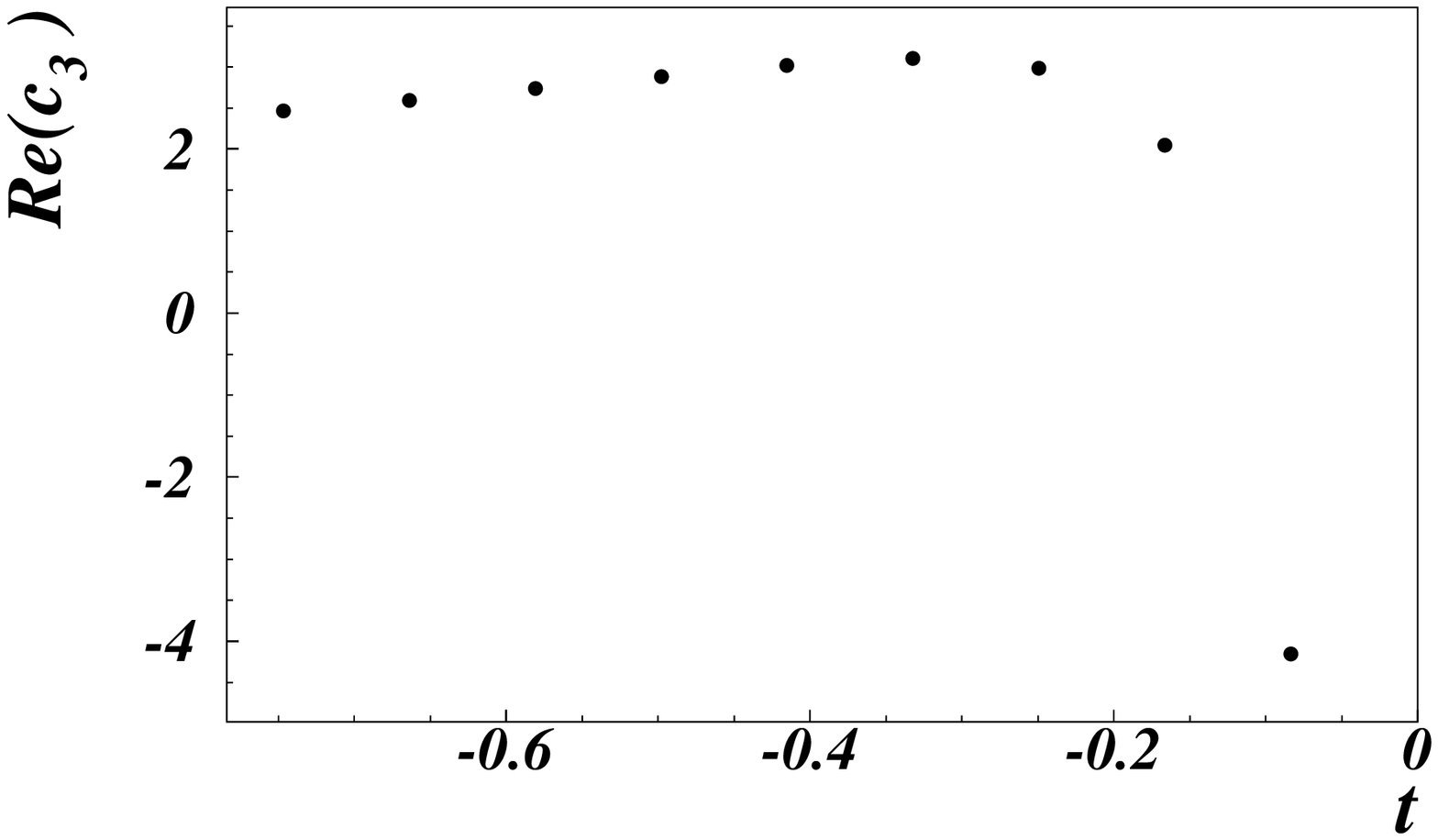}
\end{minipage}
\begin{minipage}{6.5cm}
\includegraphics[width=6cm]{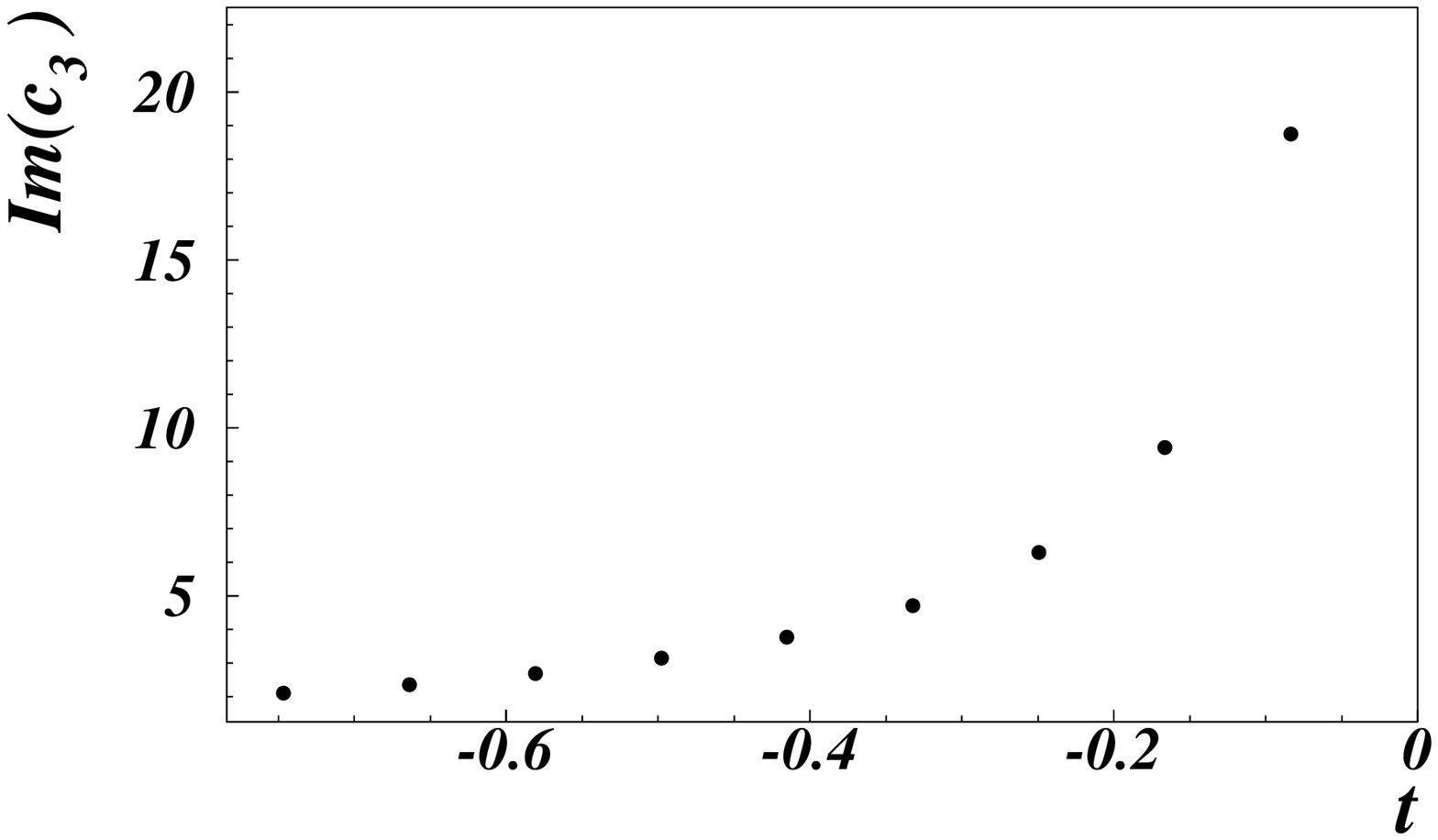}
\end{minipage}
\begin{minipage}{6.5cm}
\includegraphics[width=6cm]{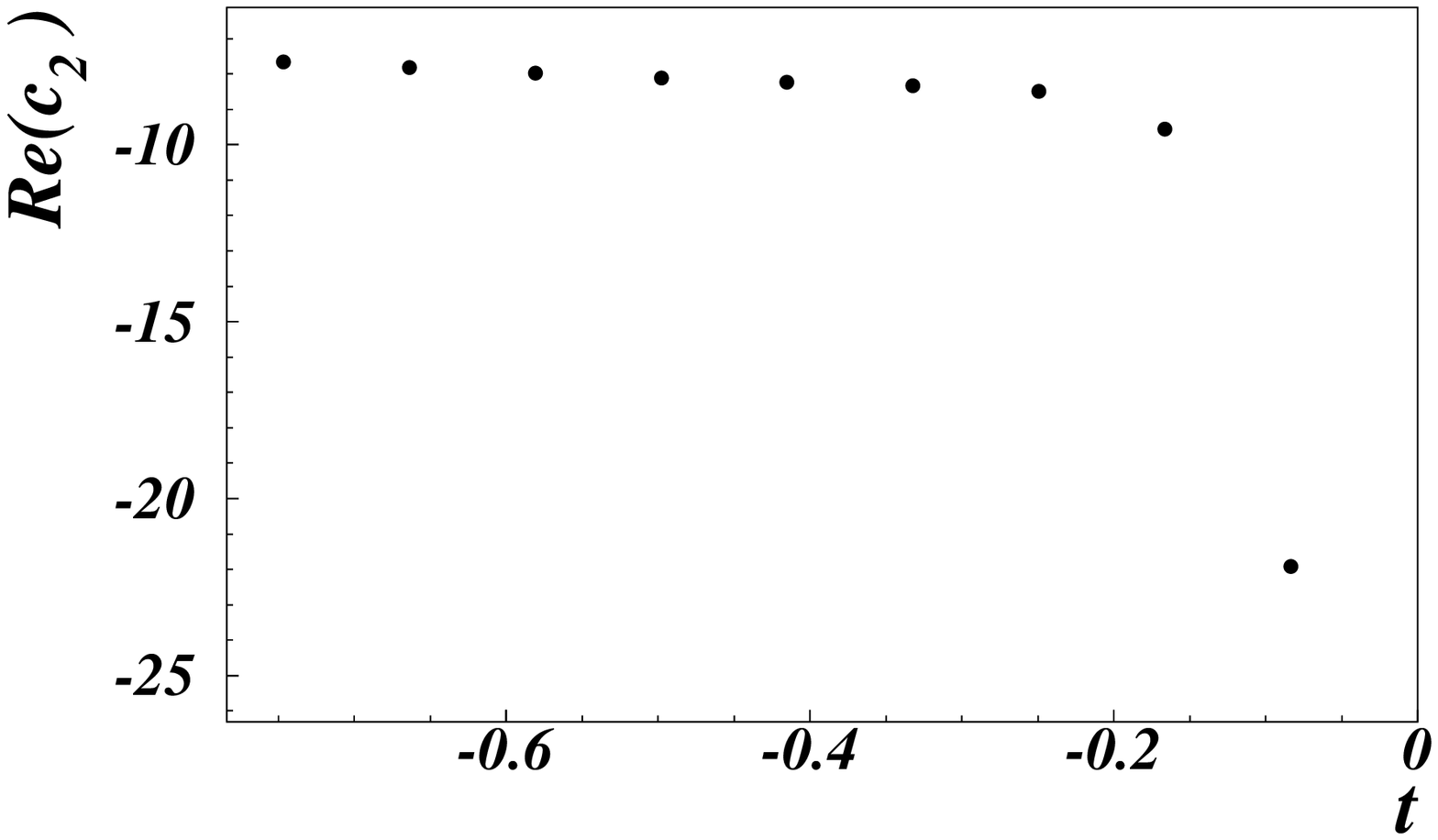}
\end{minipage}
\begin{minipage}{6.5cm}
\includegraphics[width=6cm]{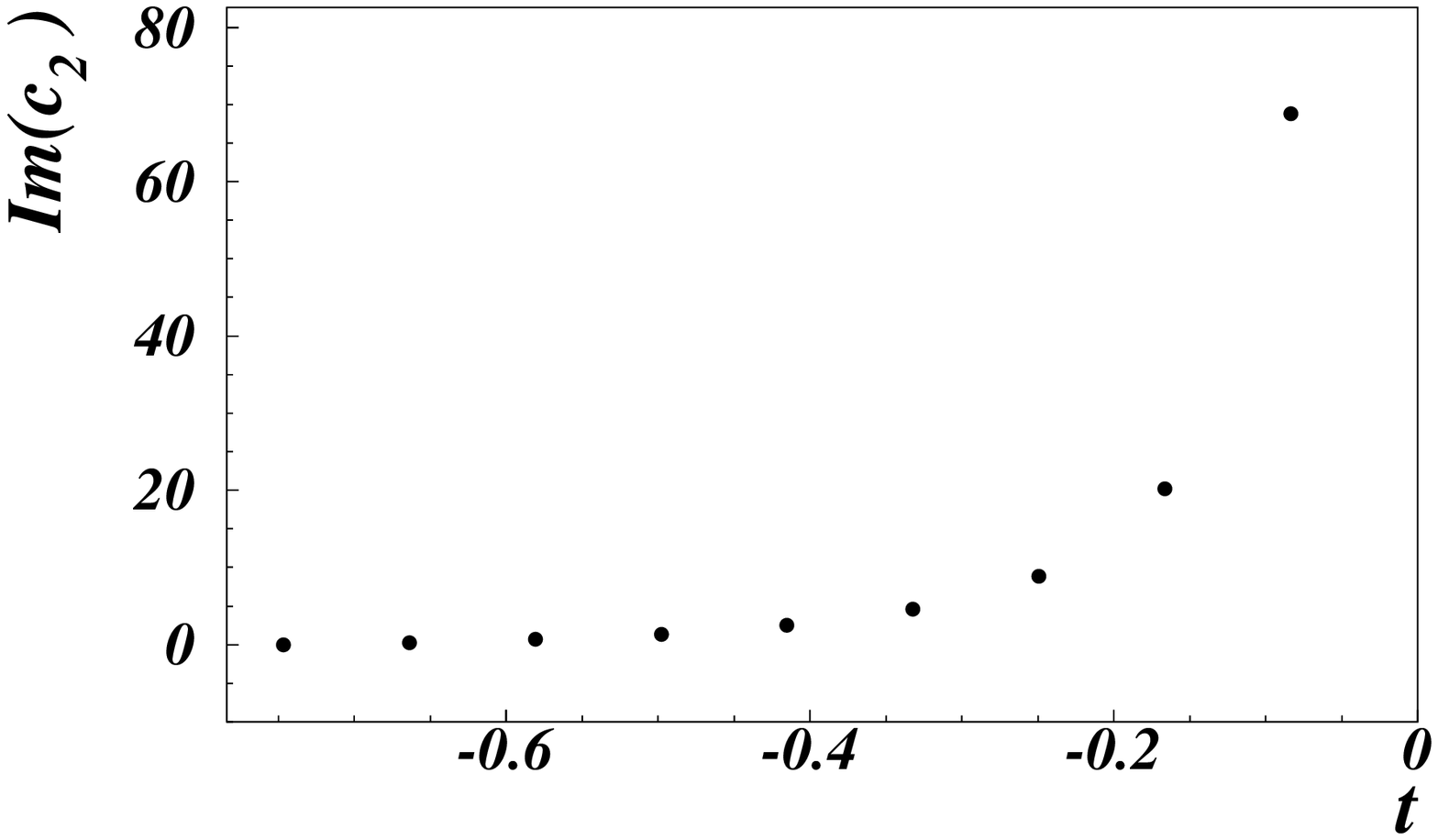}
\end{minipage}
\begin{minipage}{6.5cm}
\includegraphics[width=6cm]{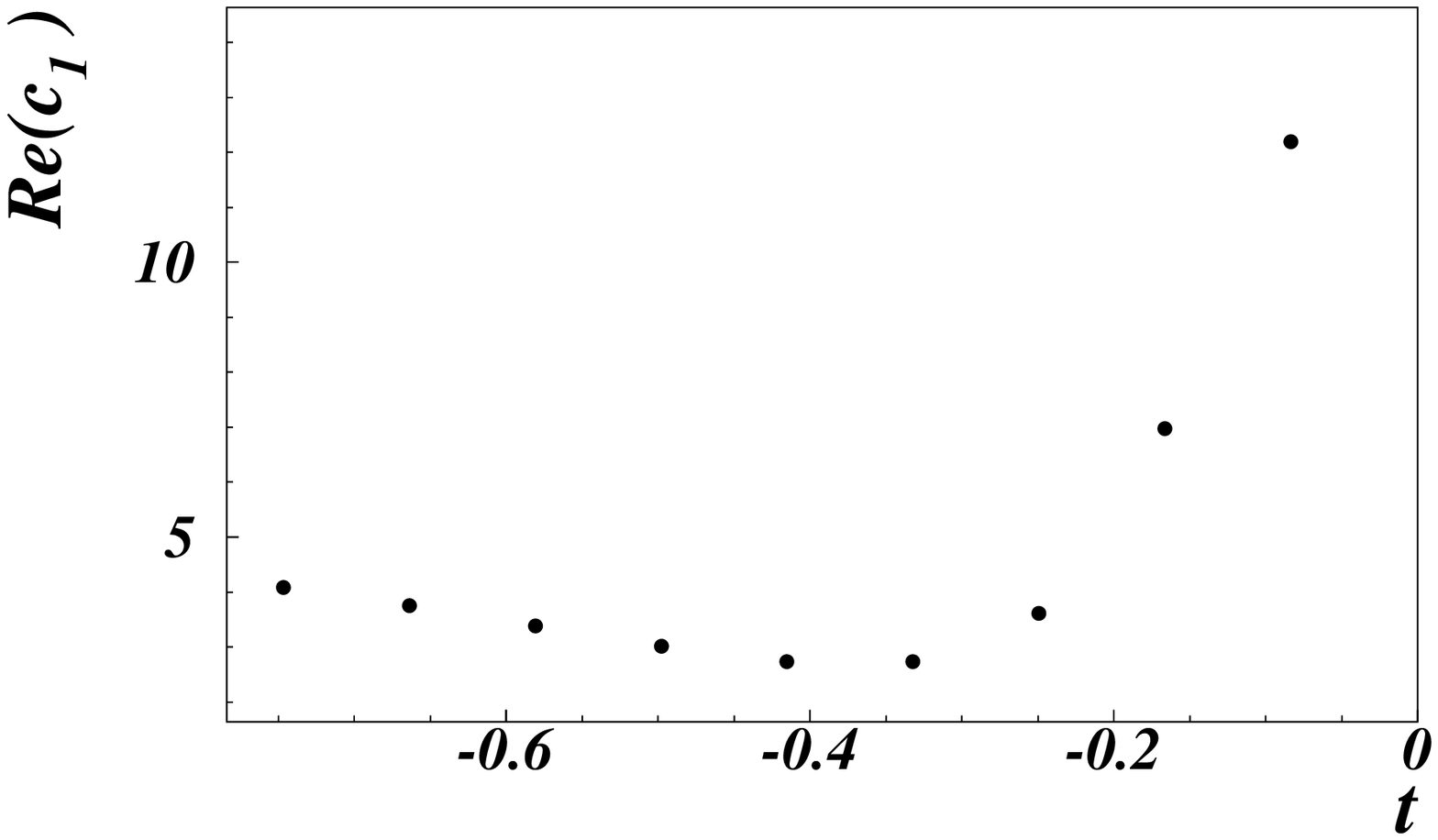}
\end{minipage}
\begin{minipage}{6.5cm}
\includegraphics[width=6cm]{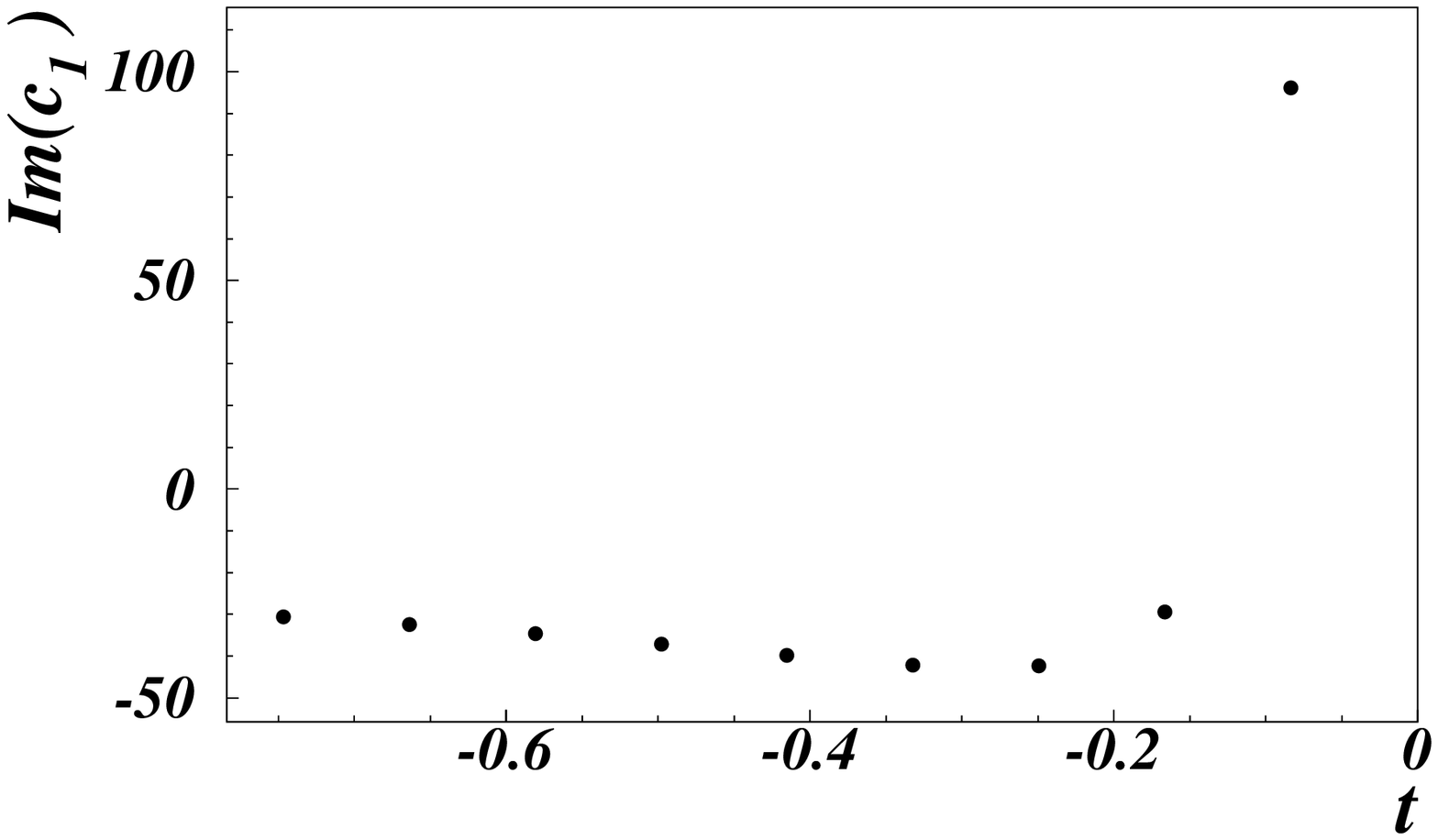}
\end{minipage}
\begin{minipage}{6.5cm}
\includegraphics[width=6cm]{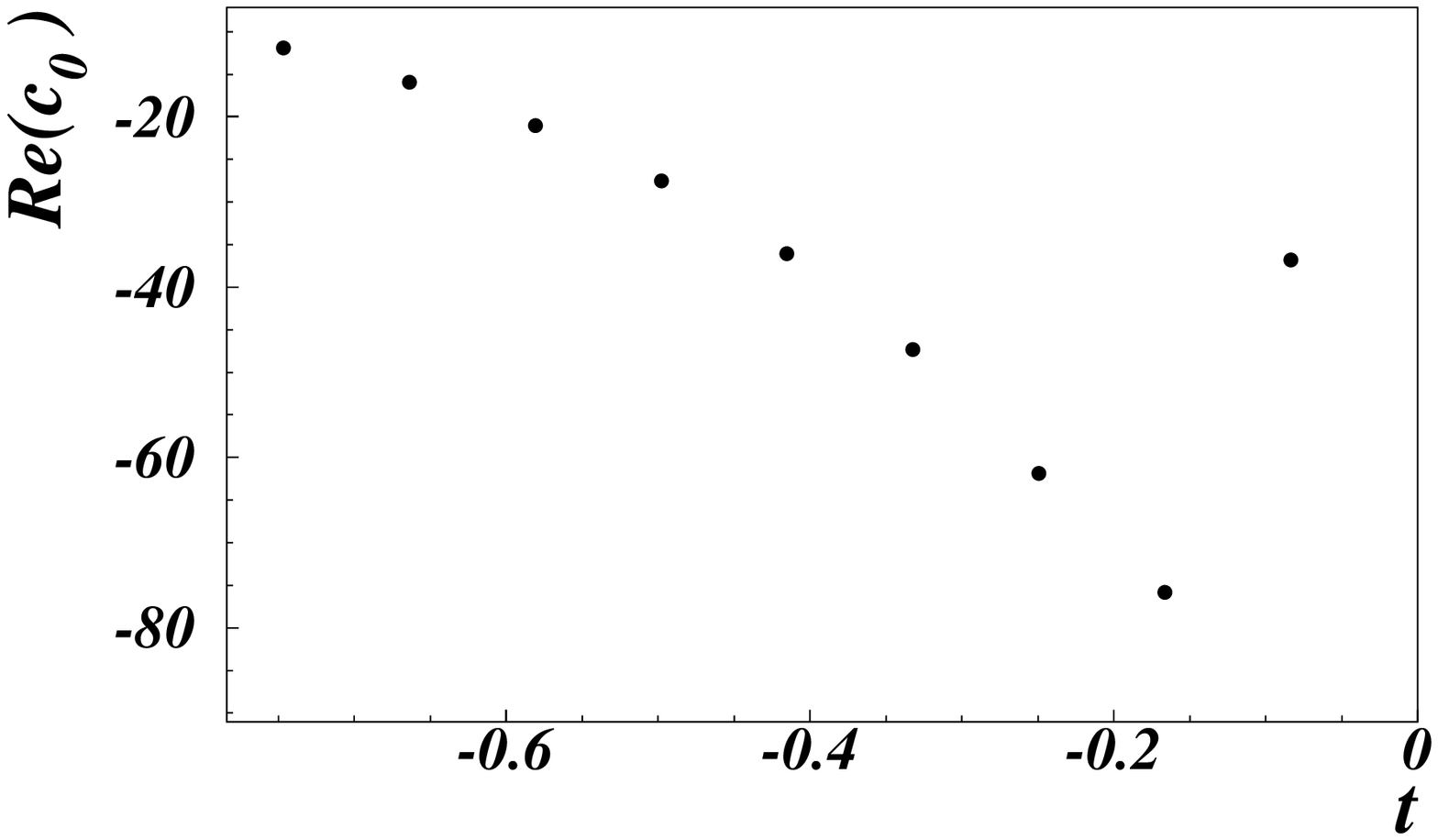}
\end{minipage}
\begin{minipage}{6.5cm}
\includegraphics[width=6cm]{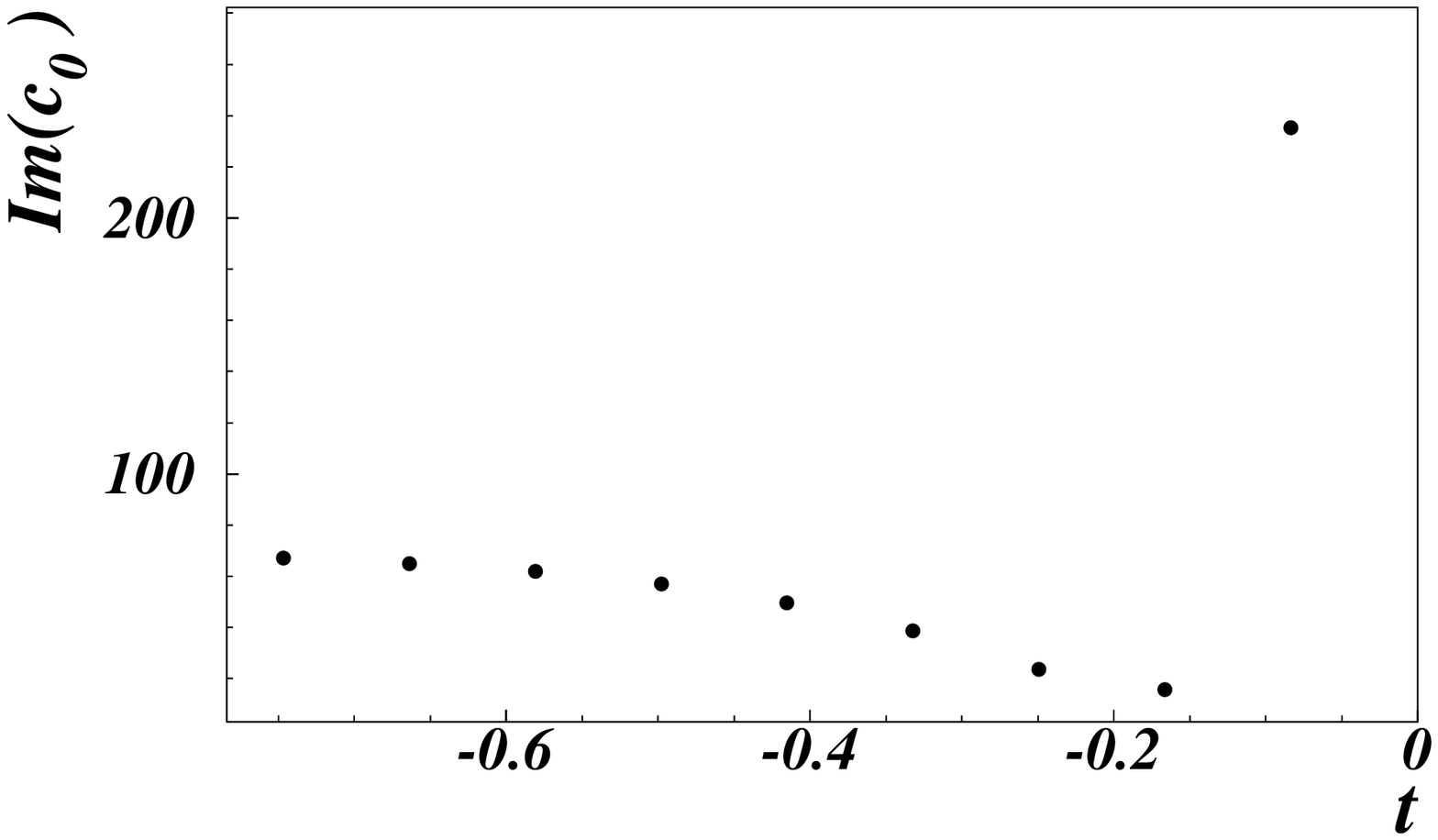}
\end{minipage}
\end{center}
\caption{Results for the double box with two adjacent masses in the 
physical region of a $2\rightarrow 2$ process where the two particles 
in the final state are massive. We plot the values of the coefficients
as function of $t$ for $s=1$, $M_1^2=1/20$ and $M_2^2=1/2$.}
\label{fig:2off2BM10}
\end{figure}

We also present our results in the kinematical region corresponding to 
the process $p_1+p_2\rightarrow p_3+p_4$ shown in Figure \ref{fig:2B2mass}, 
relevant, for instance, for the process $pp\rightarrow W^{+}W^{-}$.
In Figure \ref{fig:2off2BeM} we plot the values of the coefficients, 
for this configuration, up to
the finite terms as a function of $t$ in the case of two equal masses,
$M_1^2=M_2^2=1/10$, where the energy scale is fixed by $s=1$.
Figure \ref{fig:2off2BM10} corresponds to the same kinematical region but
for $s=1$, $M_1^2=1/20$ and $M_2^2=1/2$. The error bars again lie
within the points in the plot and are, for the finite pieces, 
typically less that 1\% after a run of a couple of minutes per 
point on a 2.8GHZ CPU.
    
\subsection{The on-shell massless triple box}
\begin{figure}[h]
\includegraphics[width=6cm]{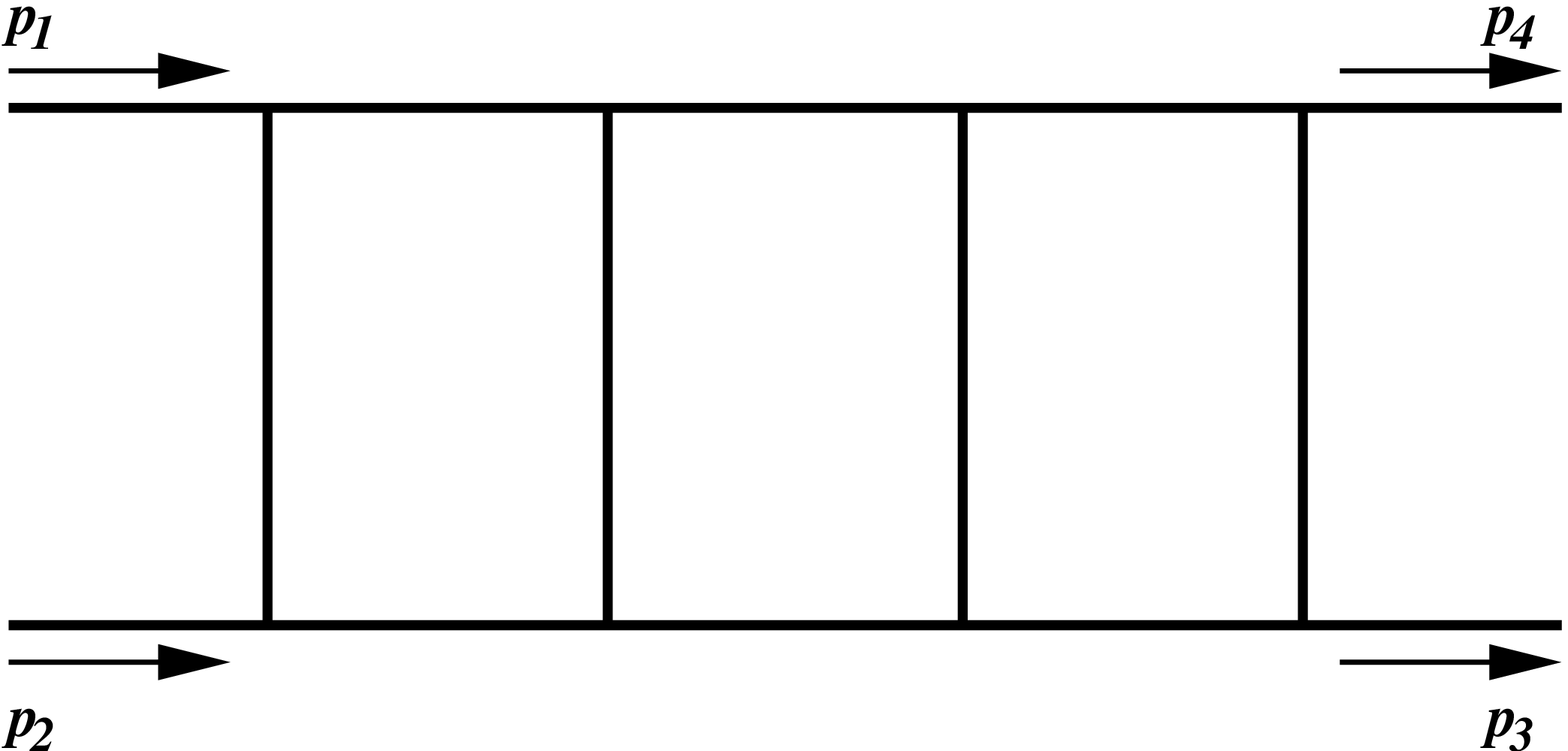}
\caption{The on-shell triple box diagram.}
\label{fig:3Bonshell}
\end{figure}

At the three loop level, we start by considering the on-shell triple box (Figure
\ref{fig:3Bonshell}), impressively
computed in an analytic way in \cite{smirnovC} using the MB technique. 

\begin{figure}[h]
\begin{minipage}{7.5cm}
\includegraphics[width=8cm]{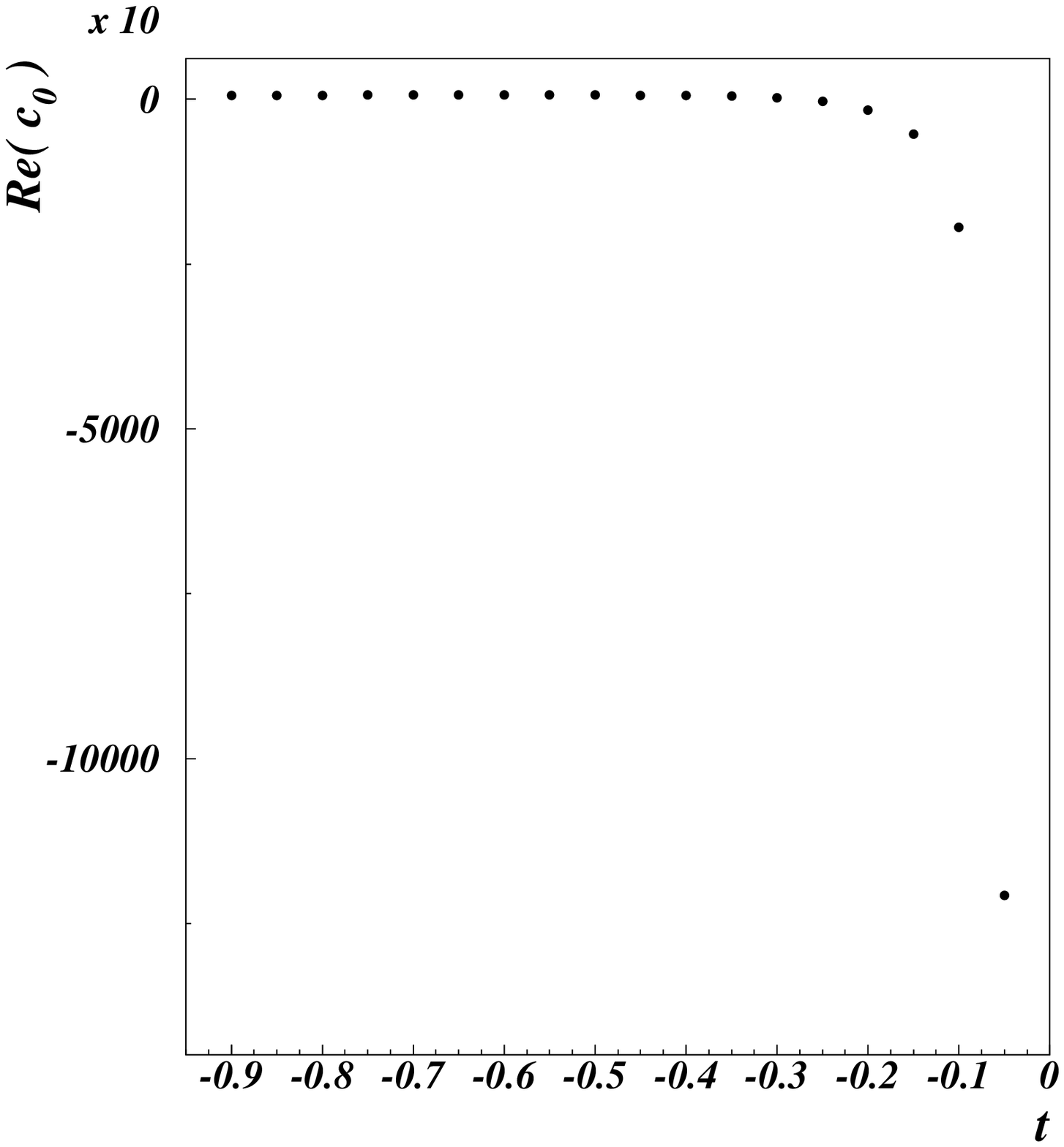}
\end{minipage}
\begin{minipage}{7.5cm}
\includegraphics[width=8cm]{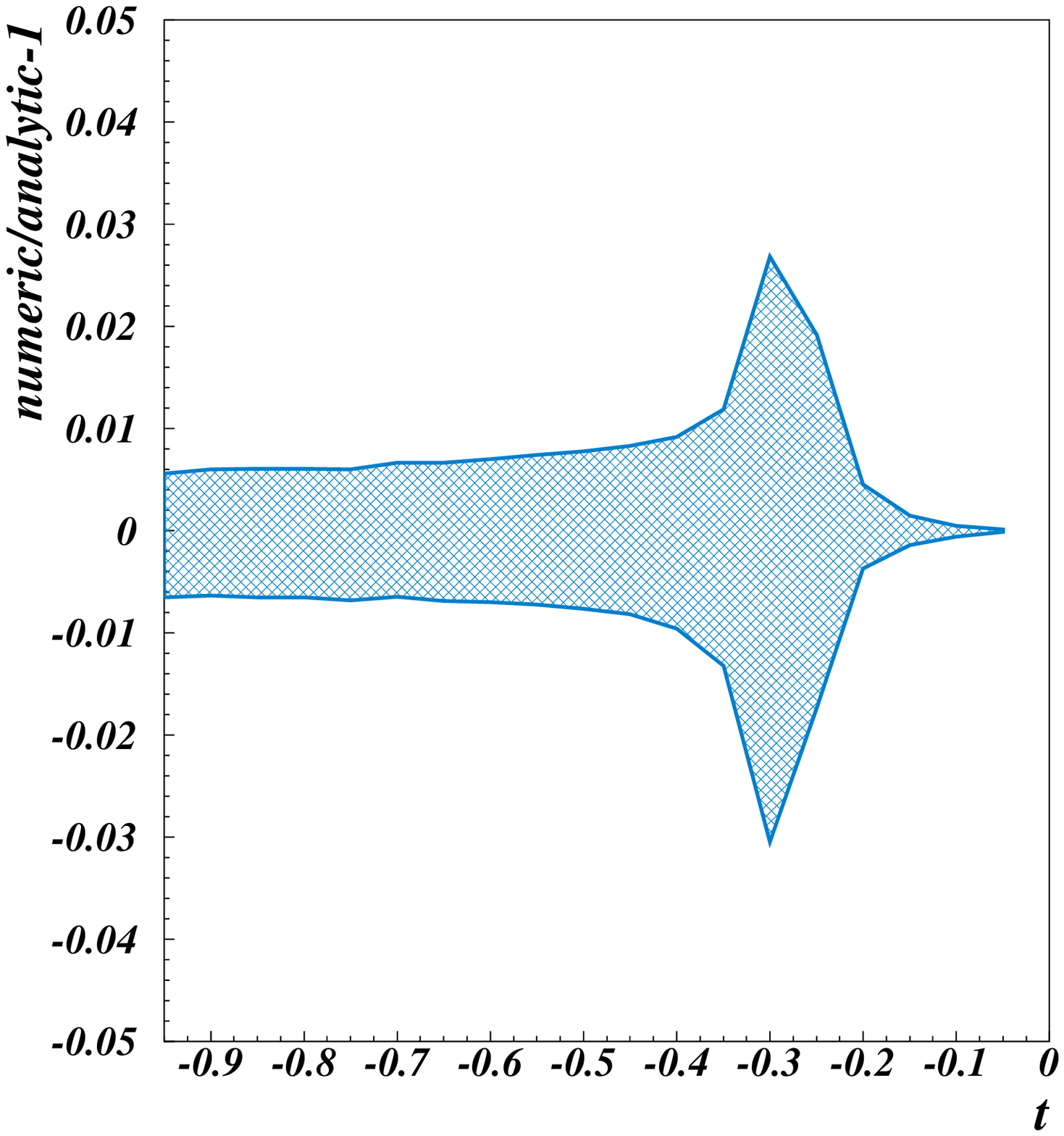}
\end{minipage}
\begin{minipage}{7.5cm}
\includegraphics[width=8cm]{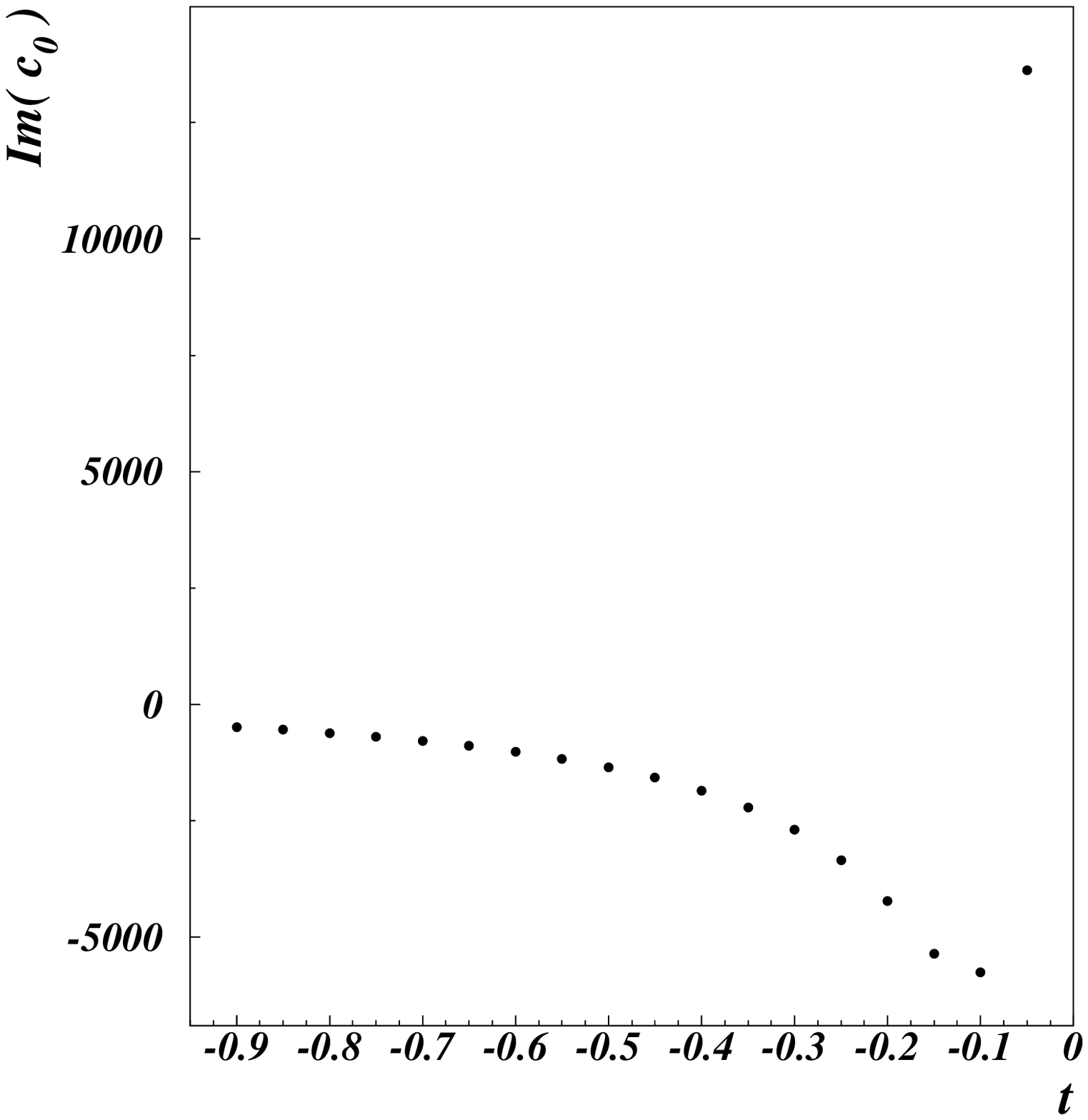}
\end{minipage}
\begin{minipage}{7.5cm}
\includegraphics[width=8cm]{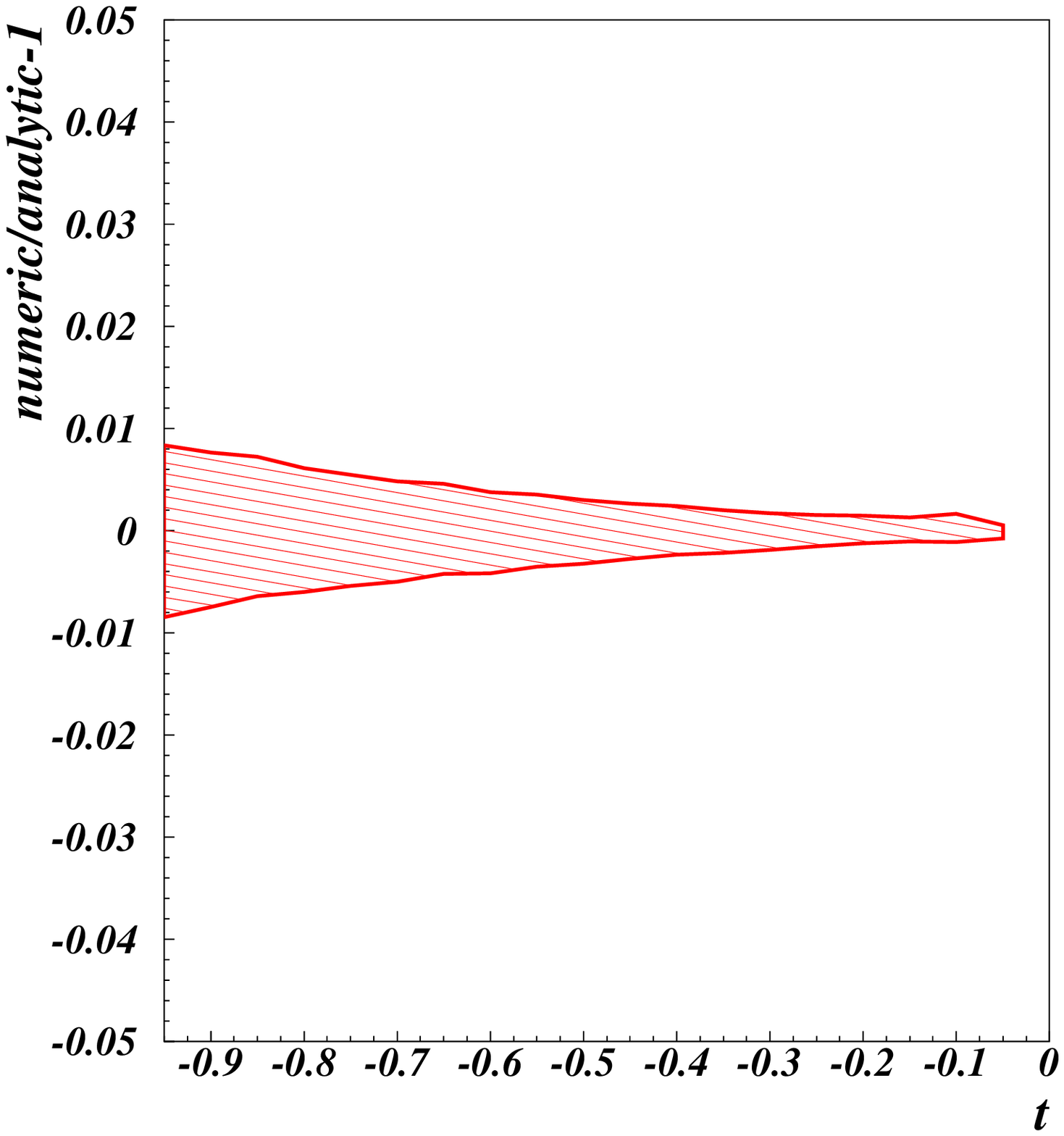}
\end{minipage}
\caption{Results for the finite part of the planar triple box in the physical
region for a $2\rightarrow 2$ process. On the upper and lower left panel 
we plot the real and imaginary parts, respectively, of $c_0$ 
as a function of $t$
for fixed value of $s=1$. On the right panels we show the corresponding 
ratios of the numerical calculation to the analytic result of
\cite{smirnovC} for the same kinematics, the bands are given 
by the error in the numerical integrations.}
\label{fig:ON3BeM}
\end{figure}

Using the re-insertion method described in Section II, we get a 7-fold
MB representation. The analytic continuation leaves us with contributions
with up to 5 contour integrals for the constant pieces in the $\epsilon$
expansion, and poles up to $\epsilon^{-6}$. Although the numerical 
integration is certainly more involved in this case, due to the depth of the
poles in $\epsilon$ which generates very complicated expressions for the
finite terms, it is relatively easy to achieve a precision of $\sim 1\%$, 
as shown in Figure \ref{fig:ON3BeM}. As in the two loop case we 
show results for the physical region
corresponding to $s>0$, $t<0$ with $s=(p_1+p_2)^2$ and $t=(p_2-p_3)^2$. We 
compare our results with the analytic computation of \cite{smirnovC}. We
used the MATHEMATICA package HPL \cite{maitre} for the evaluation of the
one-dimensional harmonic polylogarithms.

% Diagram for the on-shell triple box

%Results for the on-shell triple box

\subsection{The triple box with one external mass}
As we did for the double box topology, we consider now a further step
in complication for the triple box, adding a mass in the external 
state. This is, as far as we know, the first evaluation of 
a three-loop box with three mass scales. 

We obtained a MB representation with 8 variables. 
As usual, the dimensionality of the
problem is reduced after the analytic continuation and we found that,
up to order $\epsilon^0$, only up to six-fold integrals contribute.

\begin{figure}[h]
\includegraphics[width=4cm]{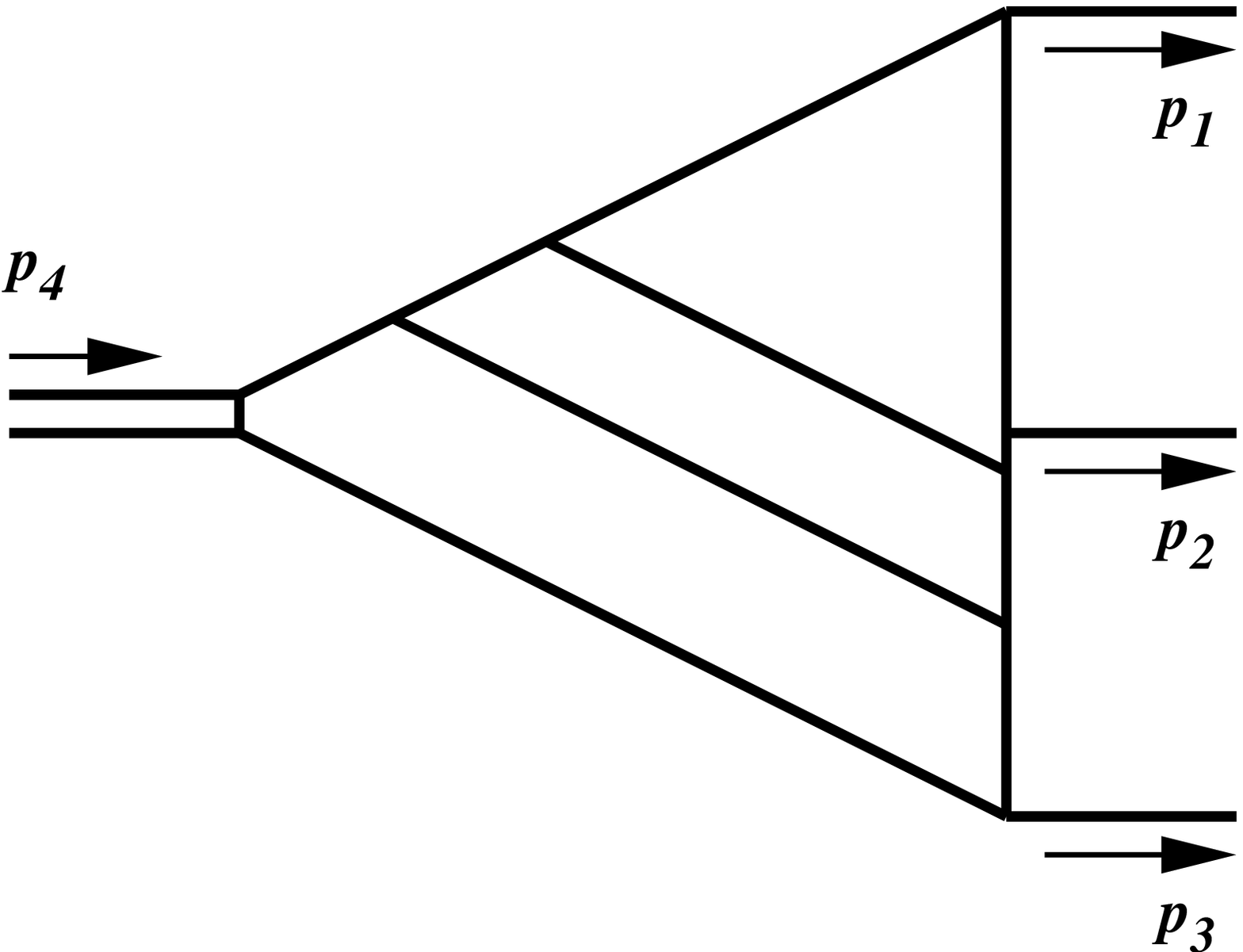}
\caption{The triple box in the decay process 
$p_{4}\rightarrow p_1+p_2+p_3$}
\label{fig:3Bdecay}
\end{figure}

\begin{figure}
\begin{center}
\begin{minipage}{6.5cm}
\includegraphics[width=6cm]{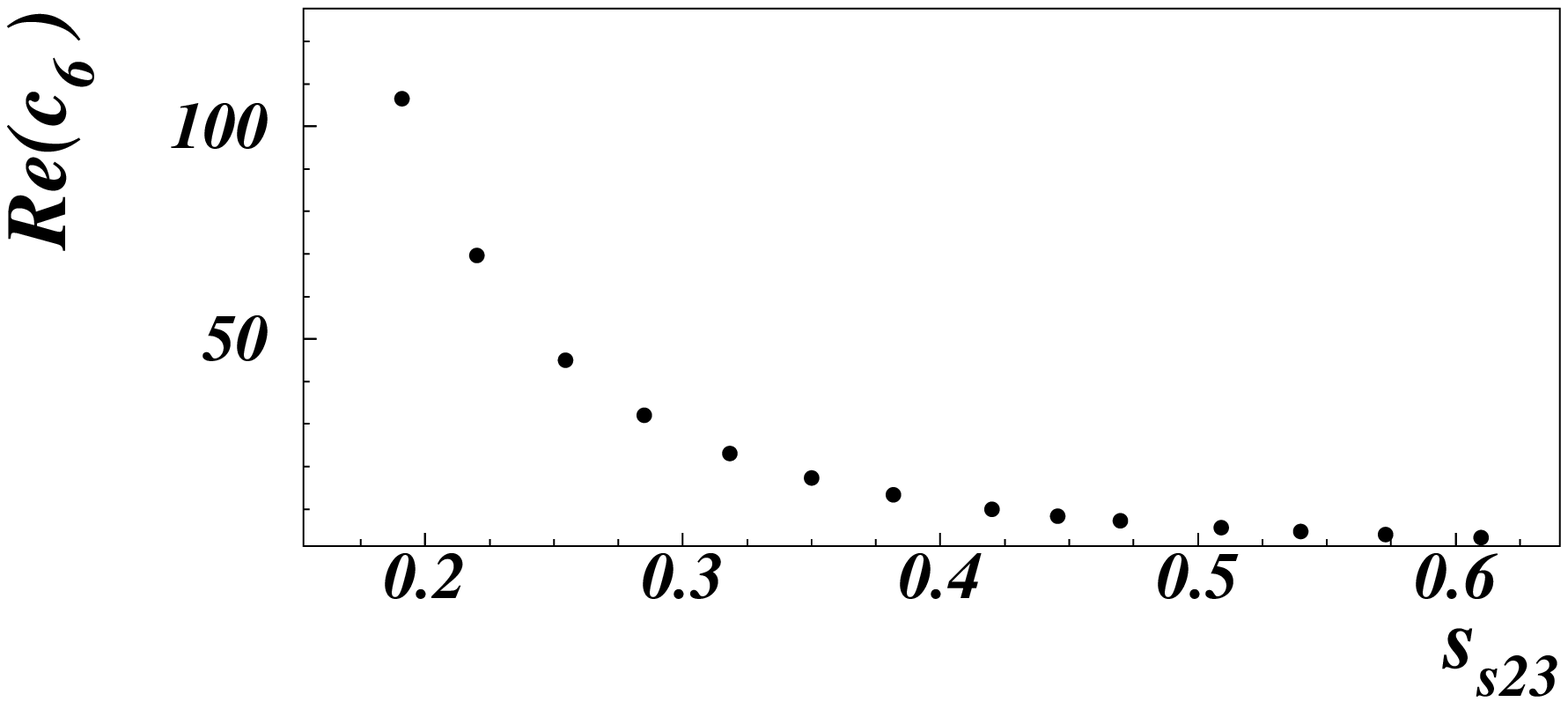}
\end{minipage}
\begin{minipage}{6.5cm}
\includegraphics[width=6cm]{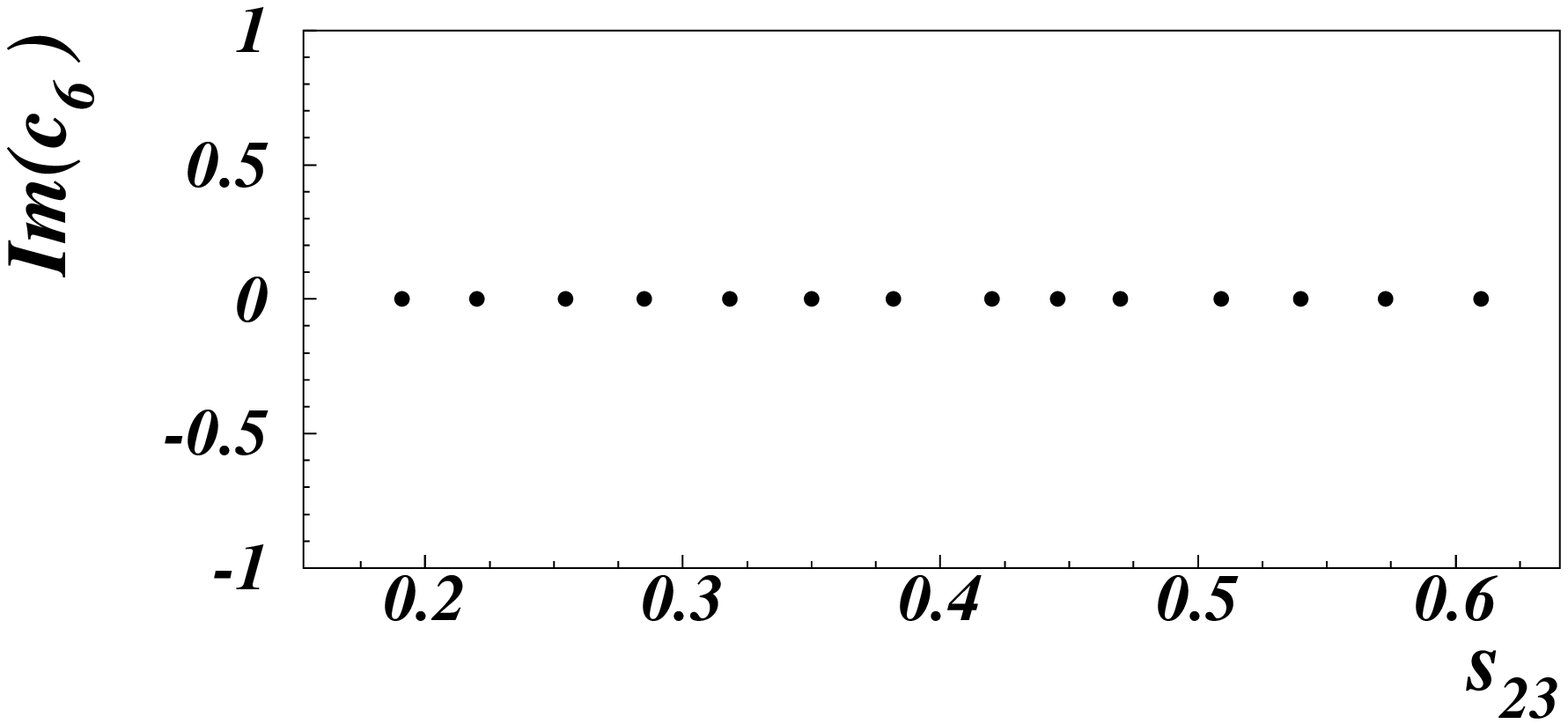}
\end{minipage}
\begin{minipage}{6.5cm}
\includegraphics[width=6cm]{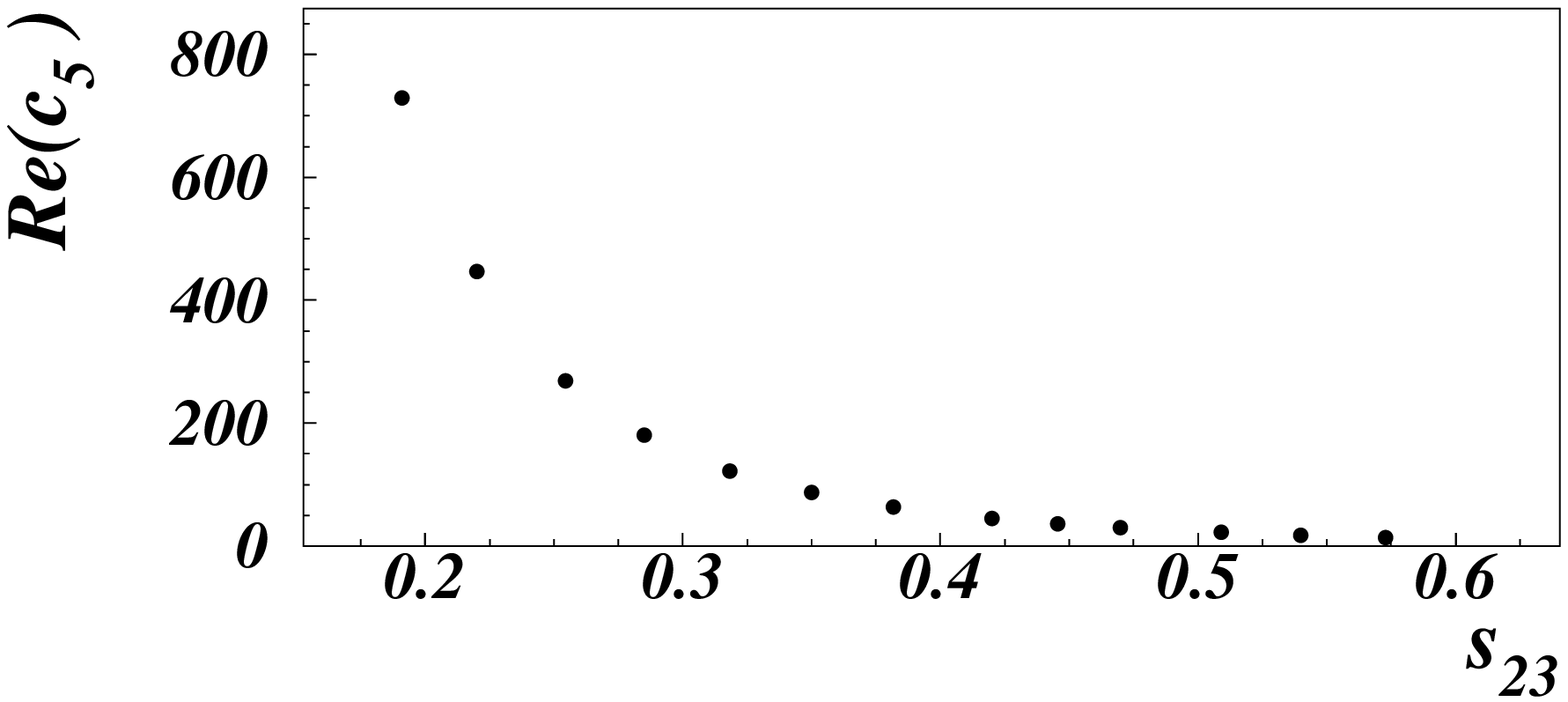}
\end{minipage}
\begin{minipage}{6.5cm}
\includegraphics[width=6cm]{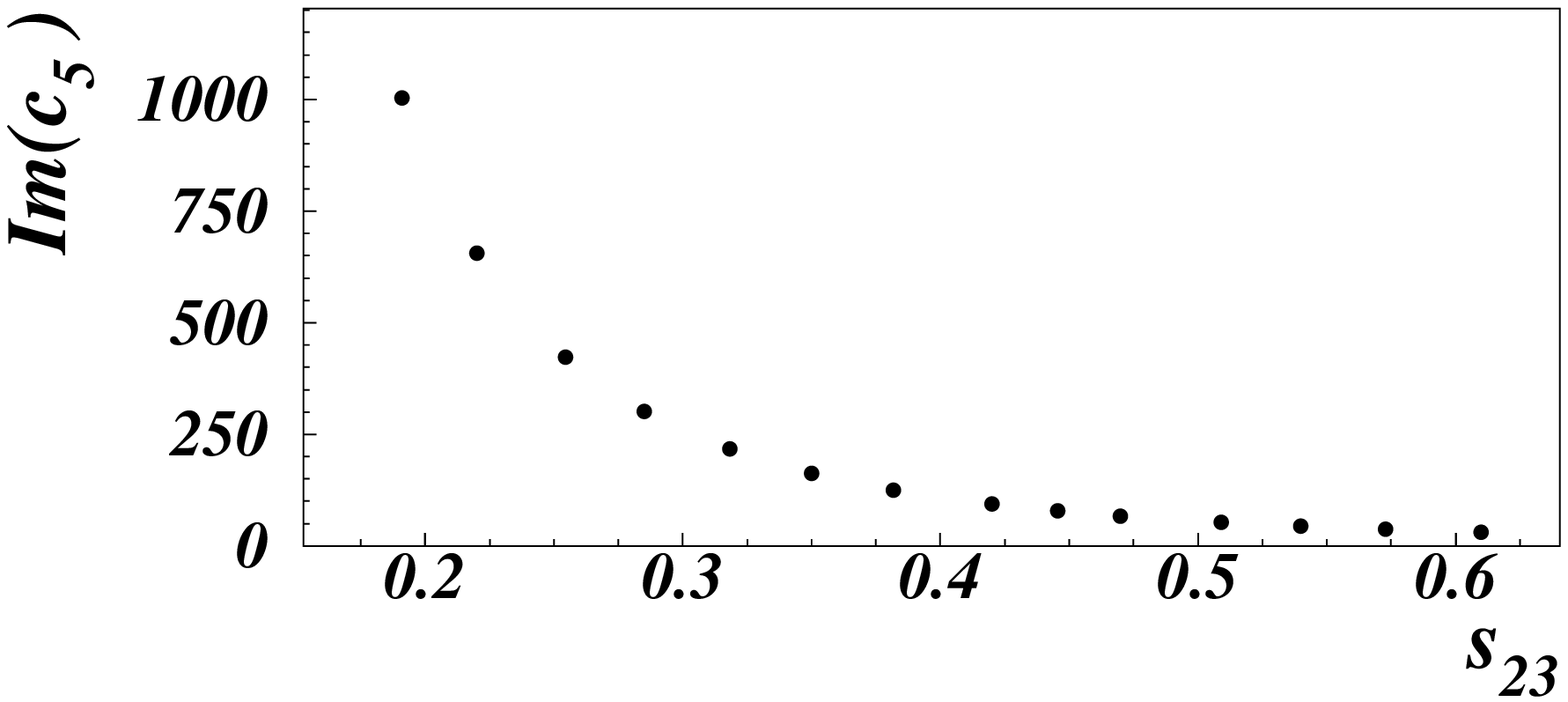}
\end{minipage}
\begin{minipage}{6.5cm}
\includegraphics[width=6cm]{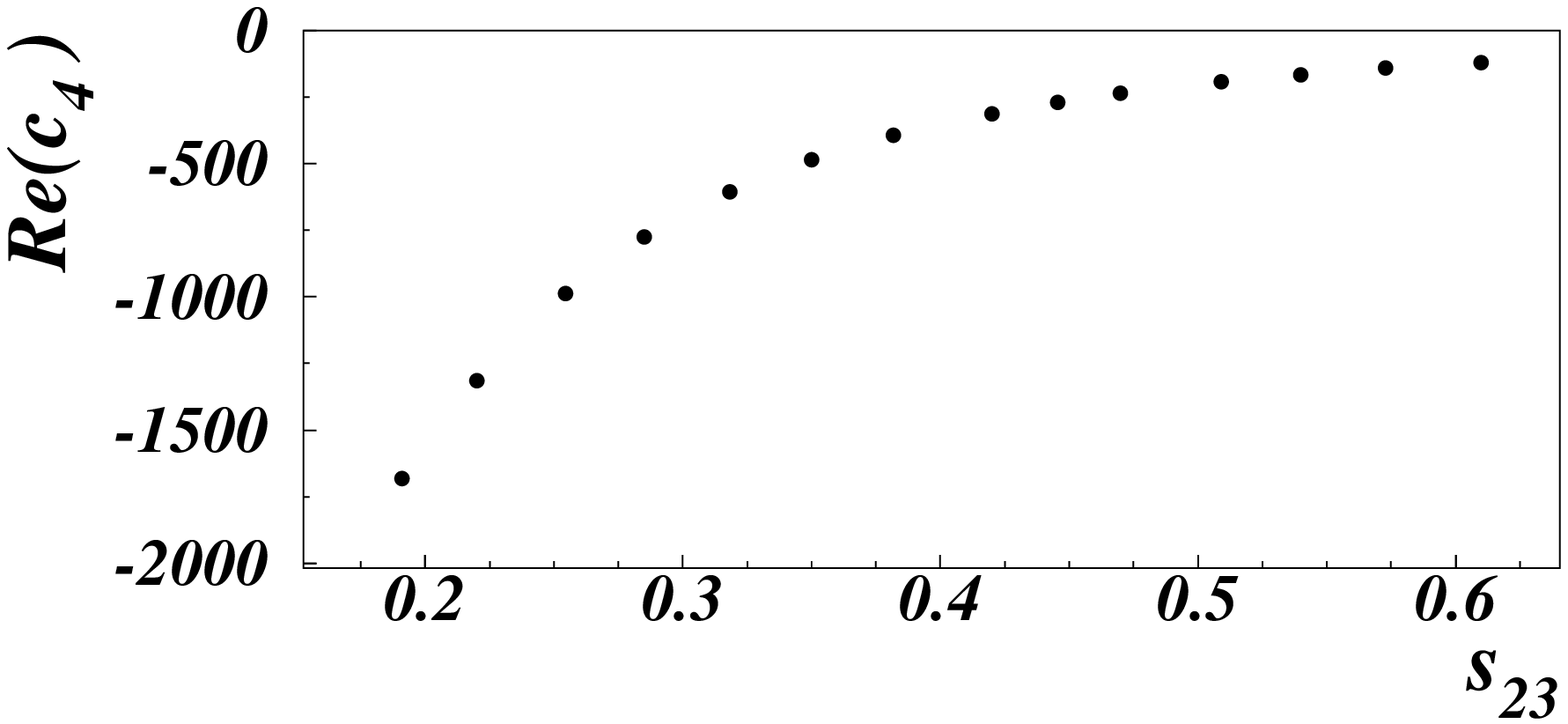}
\end{minipage}
\begin{minipage}{6.5cm}
\includegraphics[width=6cm]{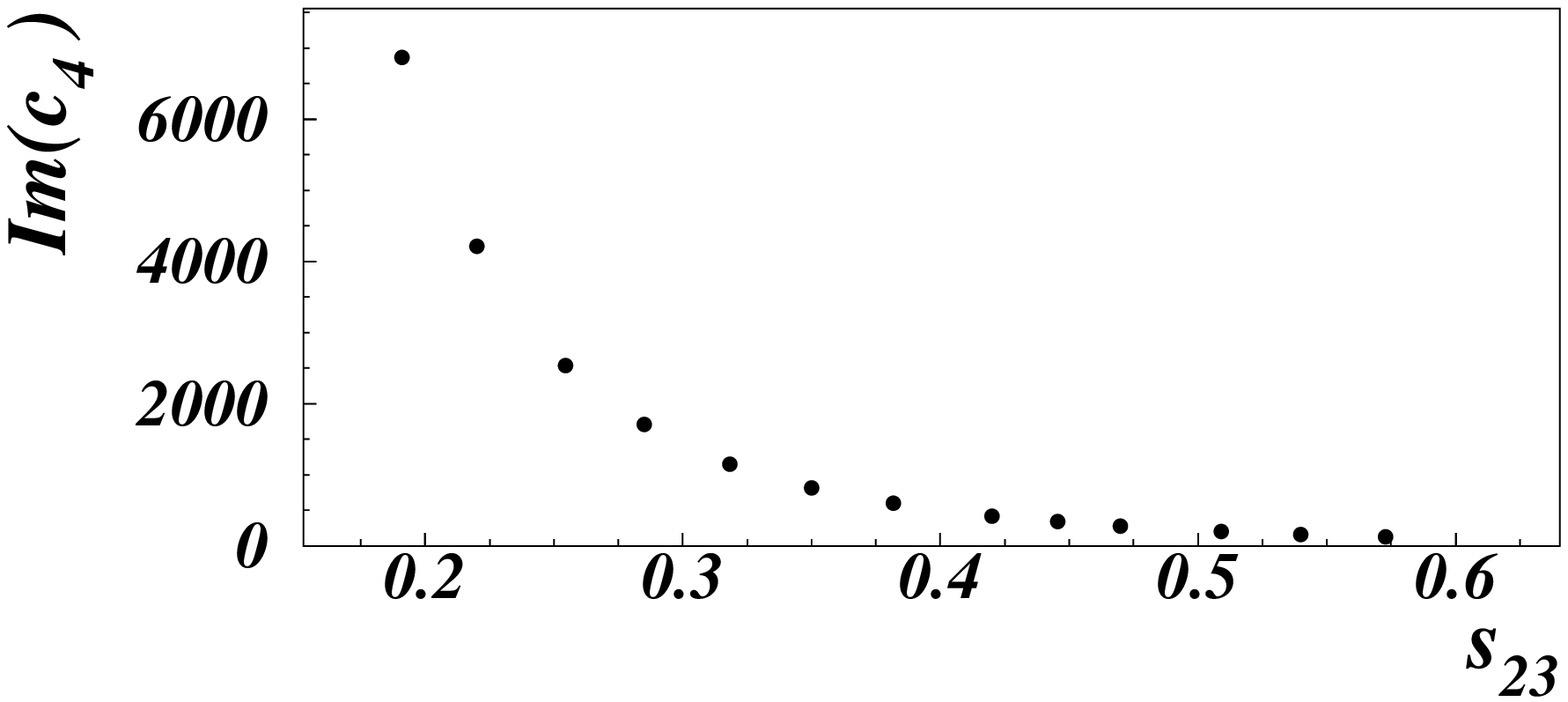}
\end{minipage}
\begin{minipage}{6.5cm}
\includegraphics[width=6cm]{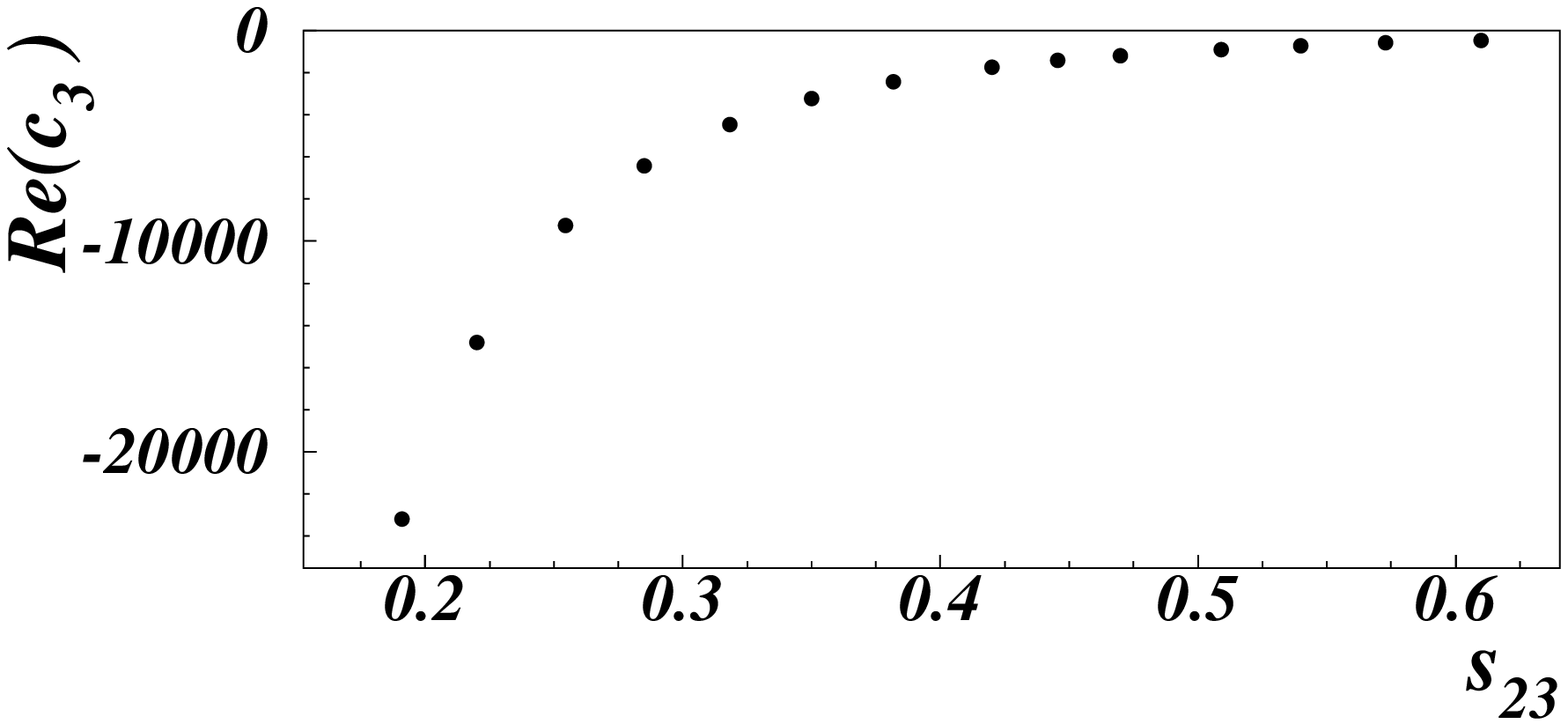}
\end{minipage}
\begin{minipage}{6.5cm}
\includegraphics[width=6cm]{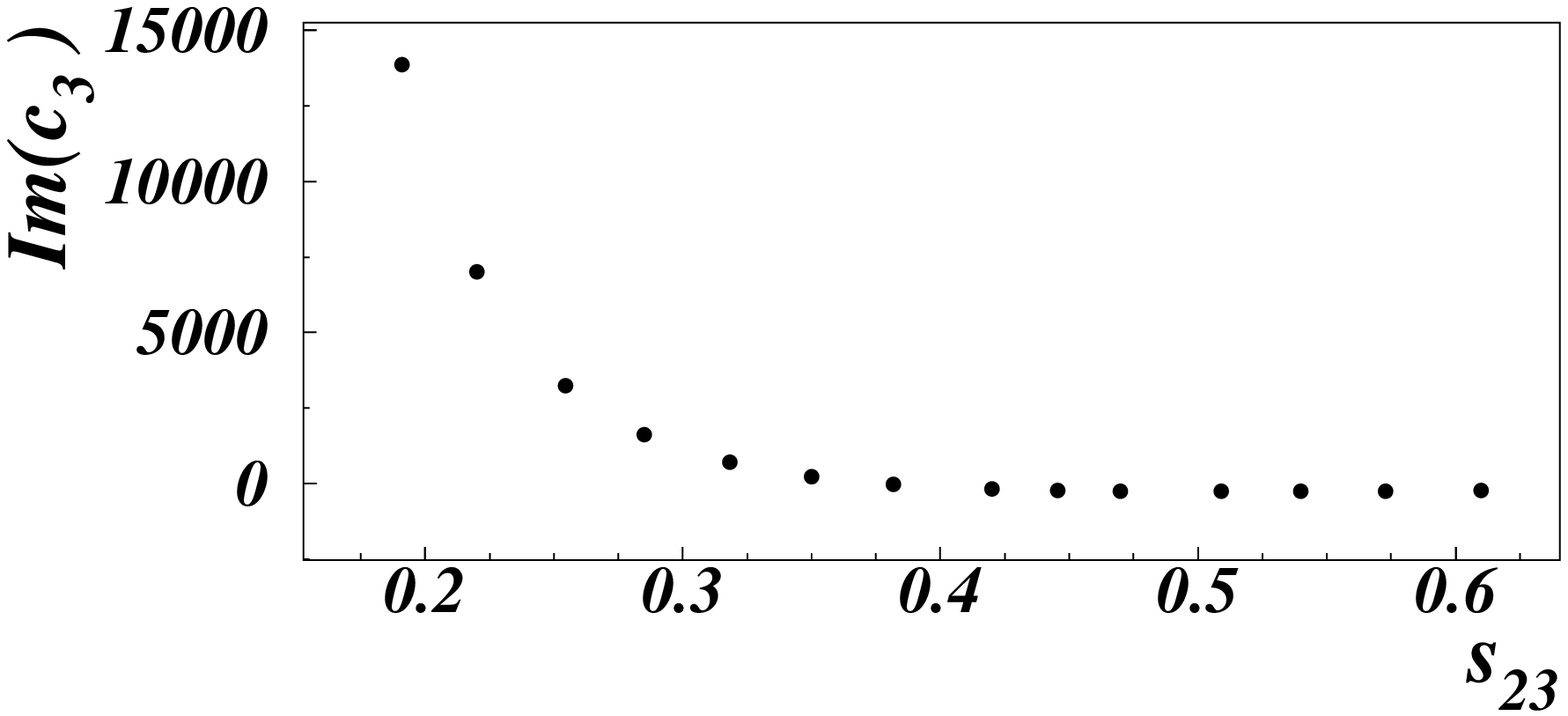}
\end{minipage}
\begin{minipage}{6.5cm}
\includegraphics[width=6cm]{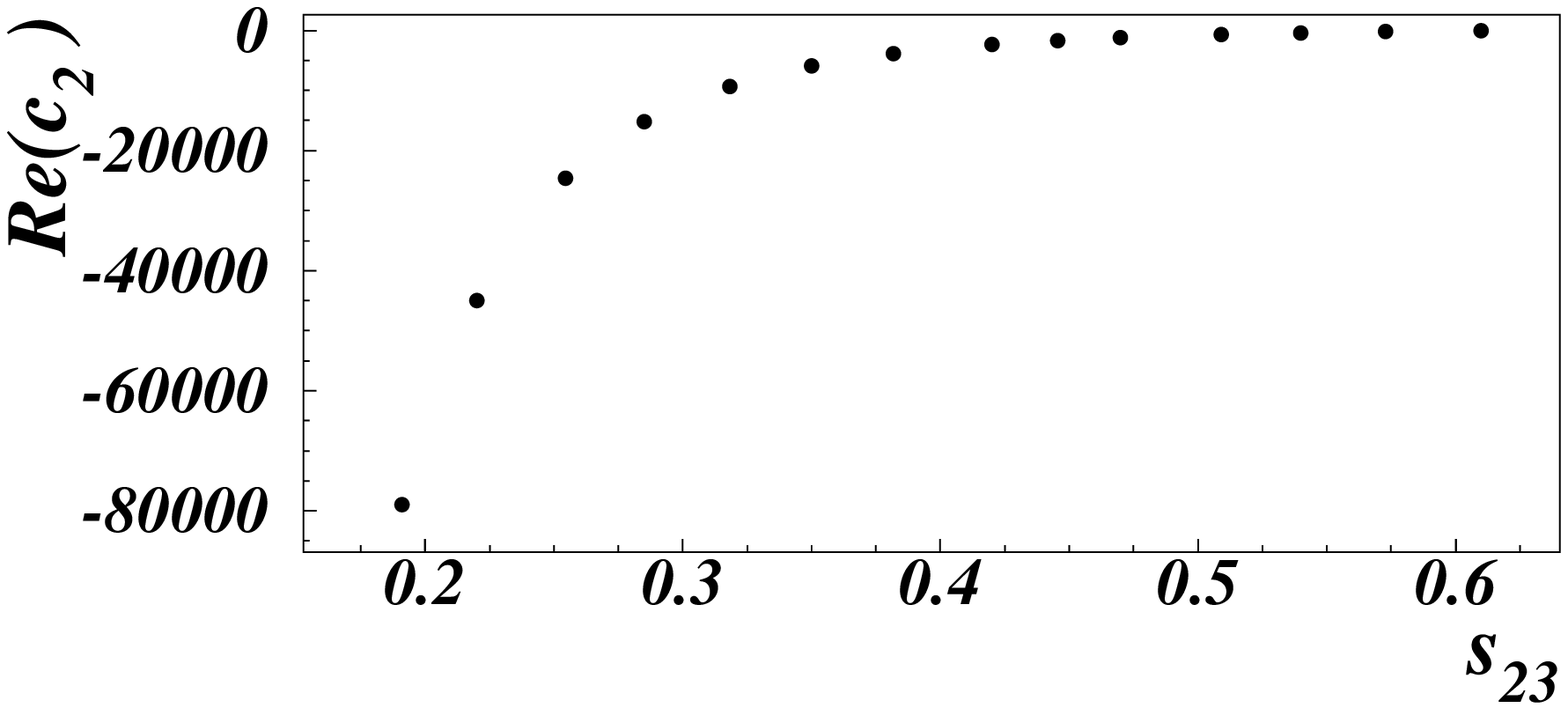}
\end{minipage}
\begin{minipage}{6.5cm}
\includegraphics[width=6cm]{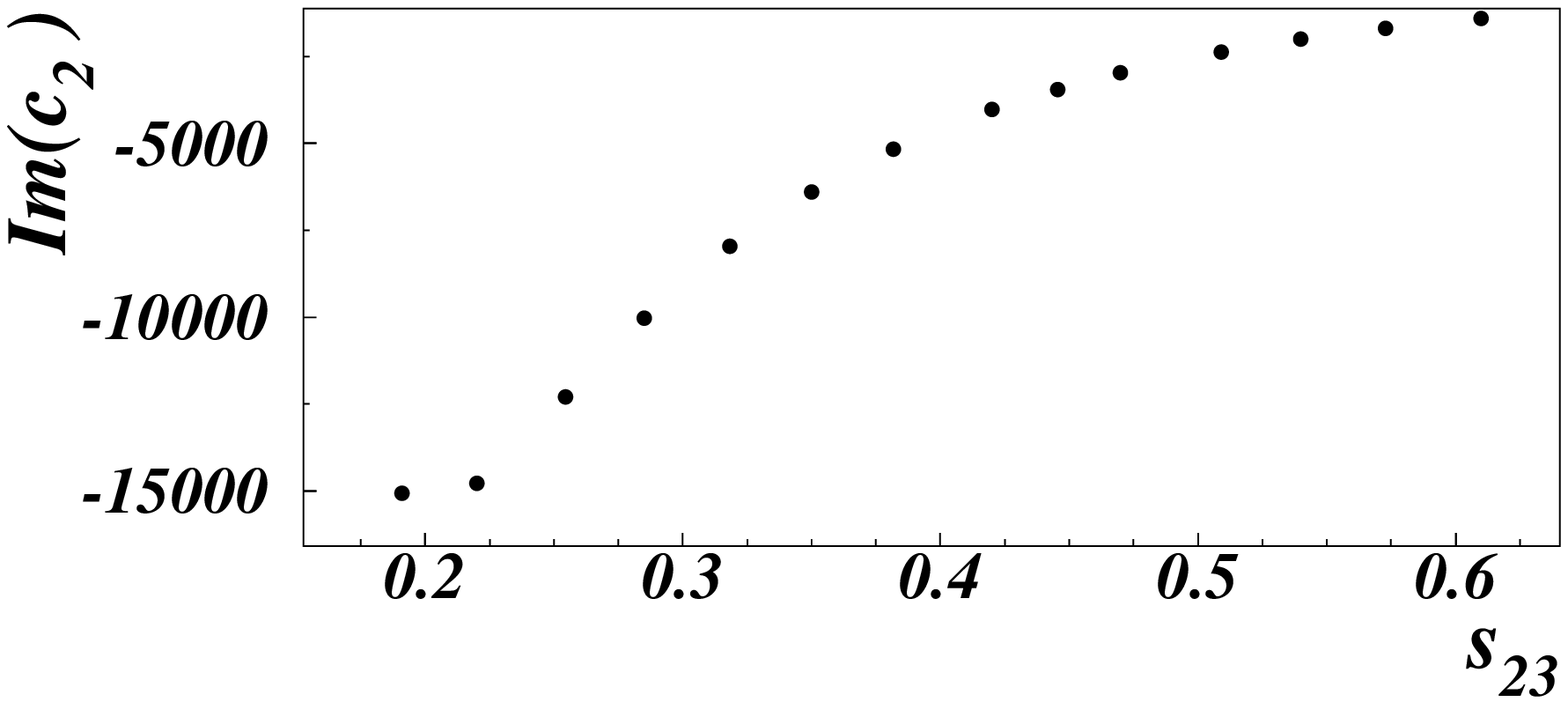}
\end{minipage}
\begin{minipage}{6.5cm}
\includegraphics[width=6cm]{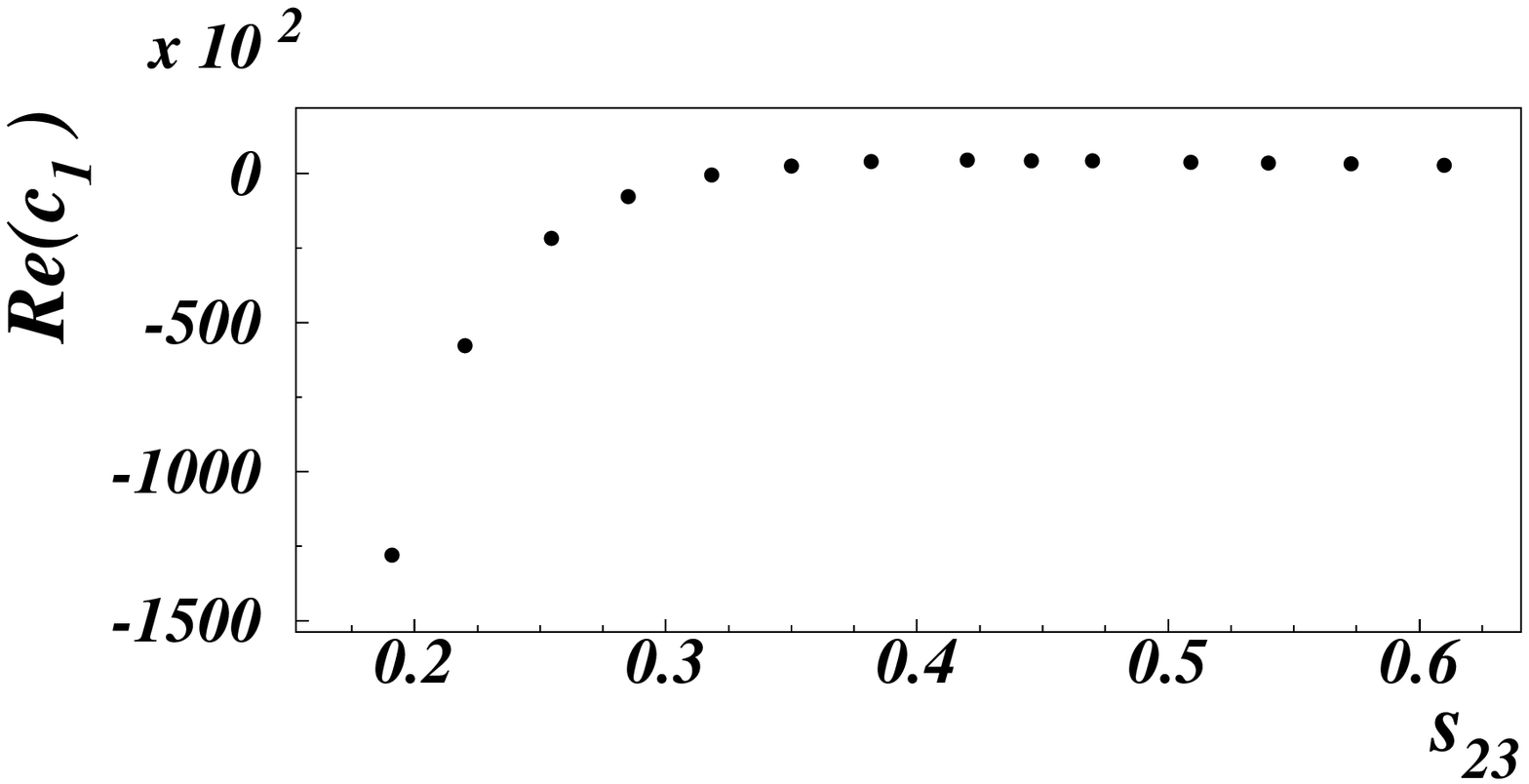}
\end{minipage}
\begin{minipage}{6.5cm}
\includegraphics[width=6cm]{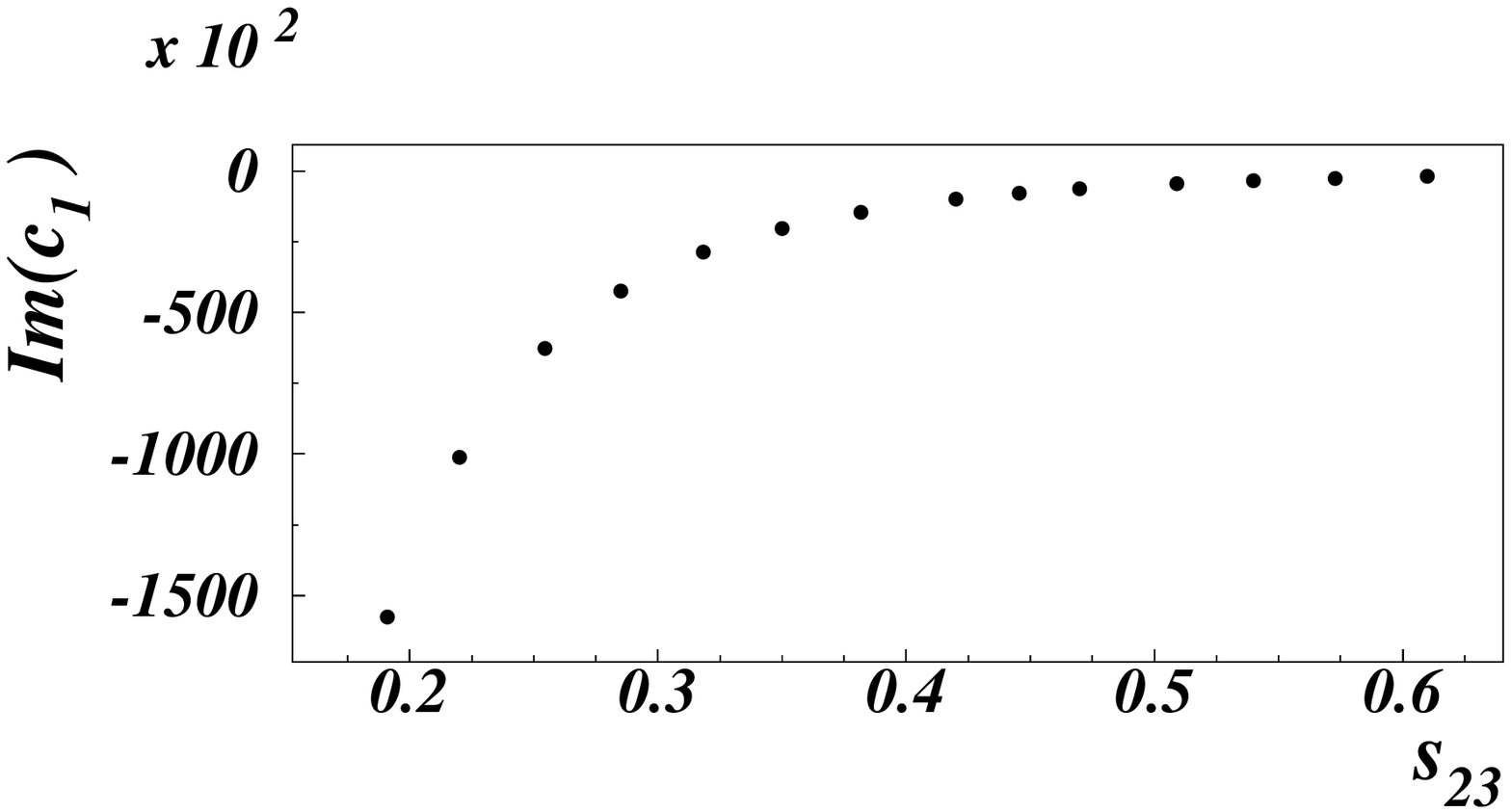}
\end{minipage}
\begin{minipage}{6.5cm}
\includegraphics[width=6cm]{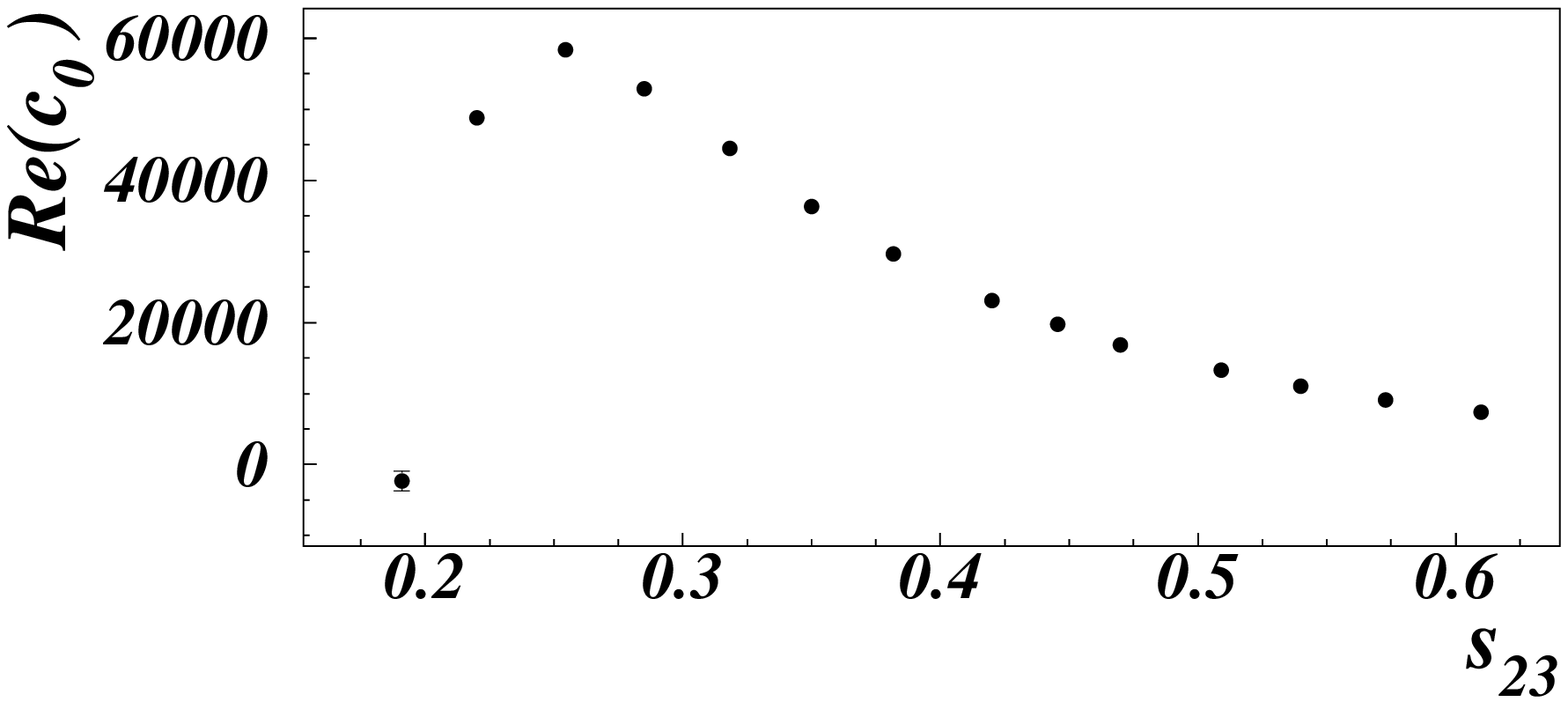}
\end{minipage}
\begin{minipage}{6.5cm}
\includegraphics[width=6cm]{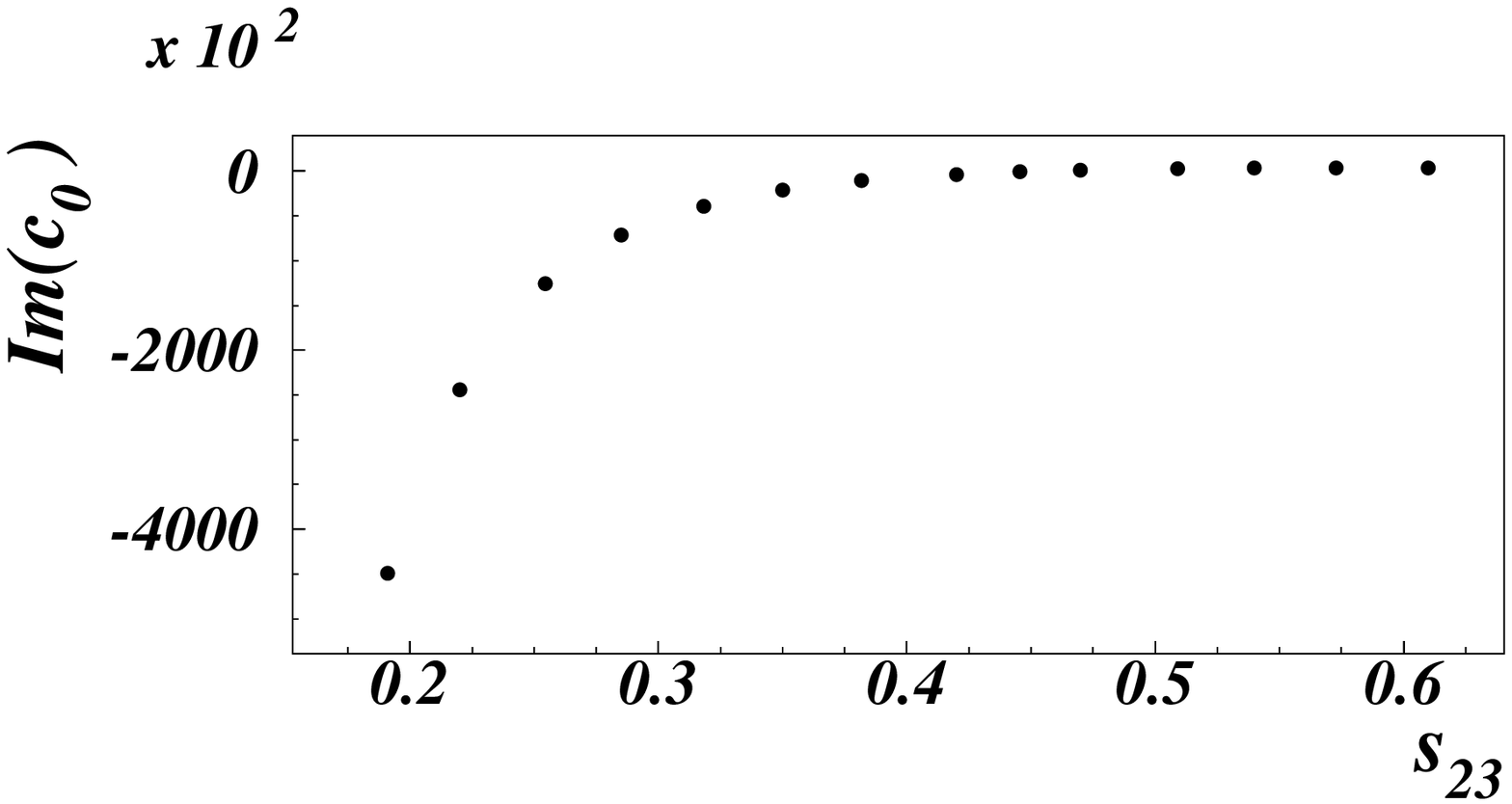}
\end{minipage}
\end{center}
\caption{Results for the triple box with one massive leg in the 
physical region of the decay of a massive particle, 
$p_4\rightarrow p_1+p_2+p_3$ 
(Figure \ref{fig:3Bdecay}). We show the results for fixed values of 
$p_4^2=1$  and $s_{13}=3/10$ as a function of the remaining invariant
$s_{23}$.}
\label{fig:1off3BM}
\end{figure}

We present results for this novel integral in the physical region
of the decay of a heavy particle,
$p_4\rightarrow p_1+p_2+p_3$ (Figure \ref{fig:3Bdecay}) as we did
in the two loops case. In Figure \ref{fig:1off3BM} we show the
the corresponding coefficients of the expansion in 
$\epsilon$ as a function of the invariant $s_{23}$ 
for fixed values of the other two variables: $p_4^2=1$ and $s_{13}=3/10$.
With the notable exception of the left-most point in the real part 
of the constant term, the error bars are all contained in the 
plotted points. They are typically of the order of $1\%$ or better 
for the finite parts after 50 minutes, in average, per point on a
2.8GHZ CPU.

% Diagram for the triple-box with one massive particle

% Results for the triple box with one massive particle

\begin{figure}
\begin{center}
\begin{minipage}{6.5cm}
\includegraphics[width=6cm]{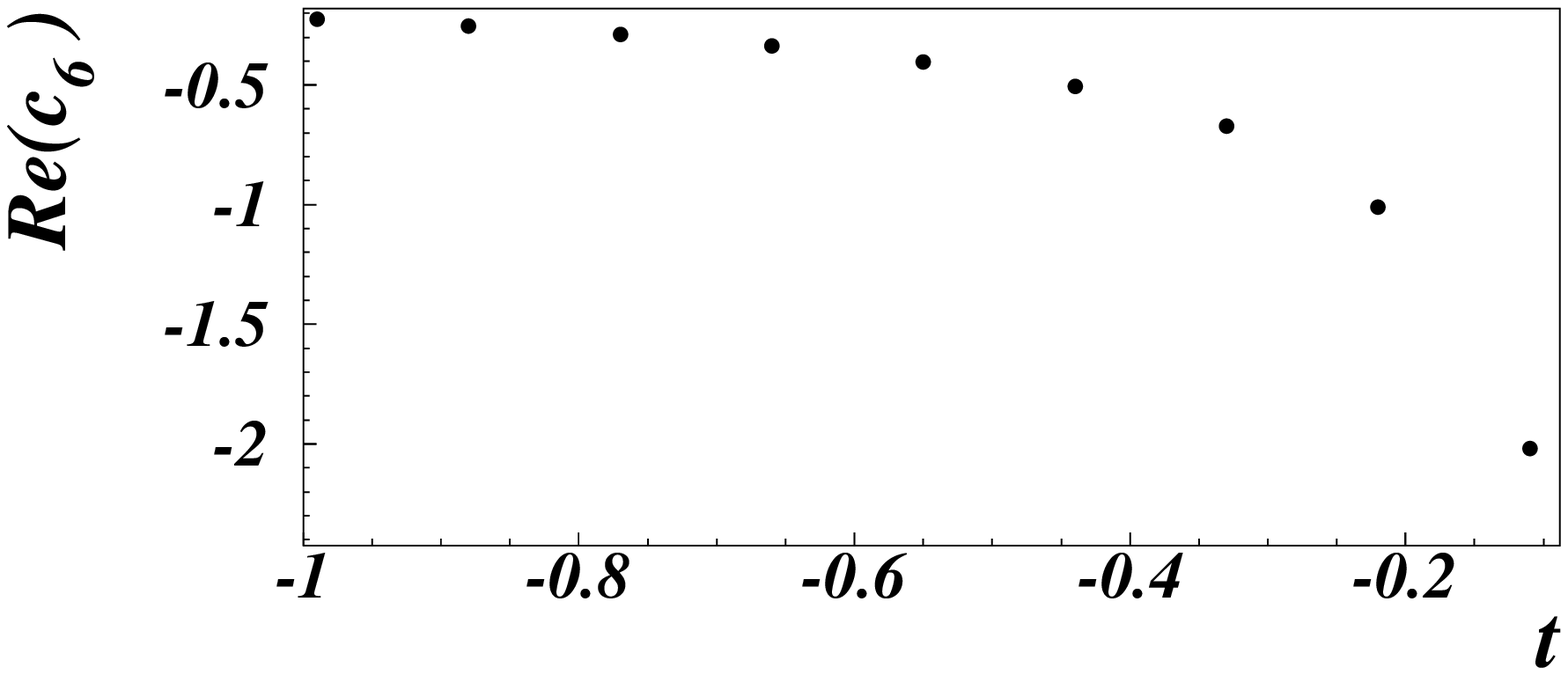}
\end{minipage}
\begin{minipage}{6.5cm}
\includegraphics[width=6cm]{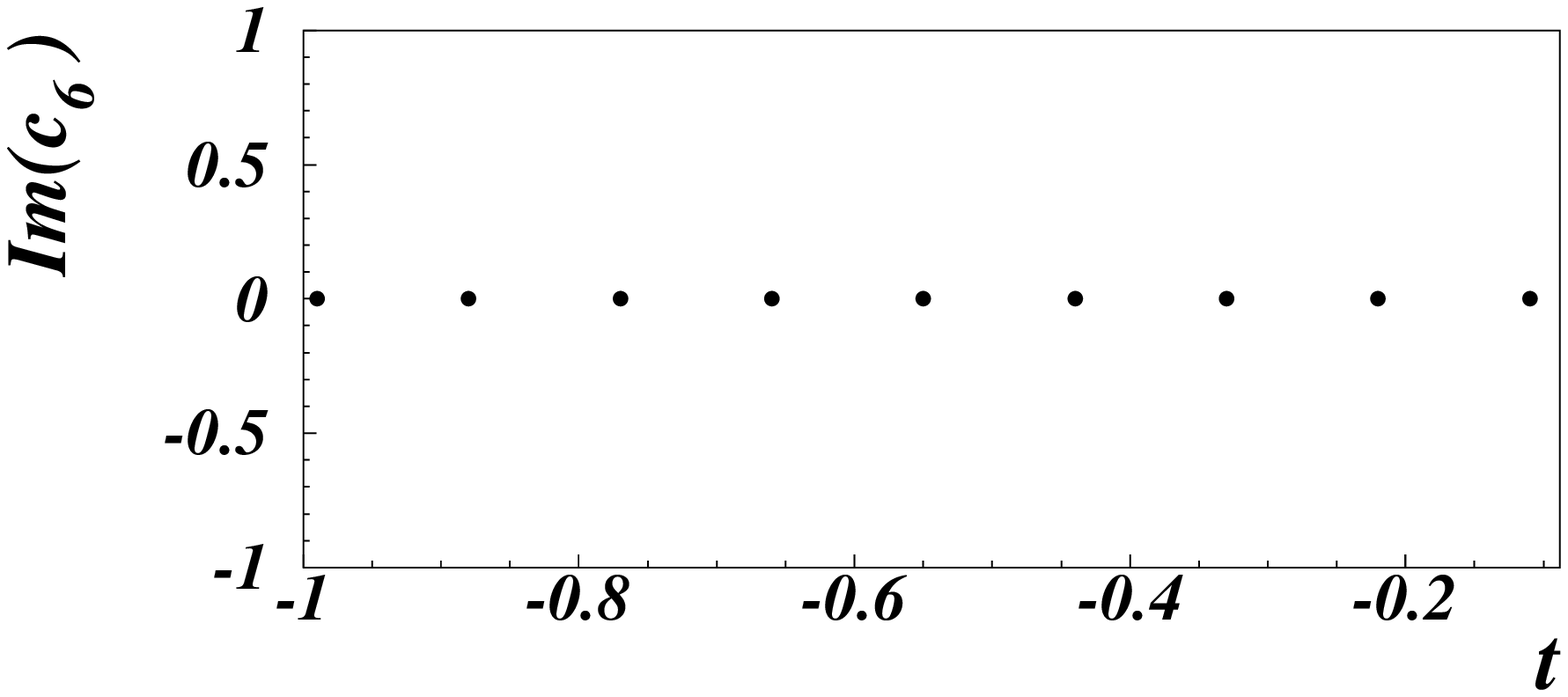}
\end{minipage}
\begin{minipage}{6.5cm}
\includegraphics[width=6cm]{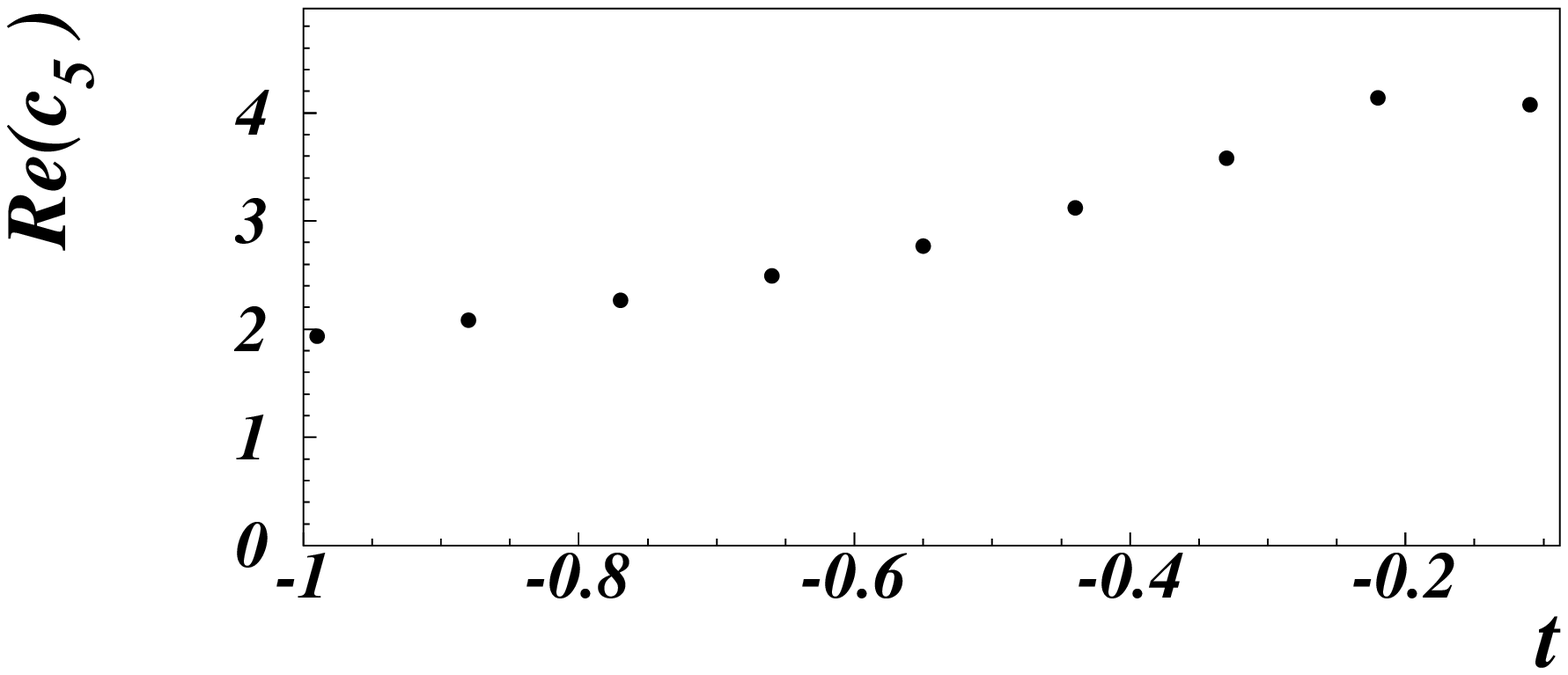}
\end{minipage}
\begin{minipage}{6.5cm}
\includegraphics[width=6cm]{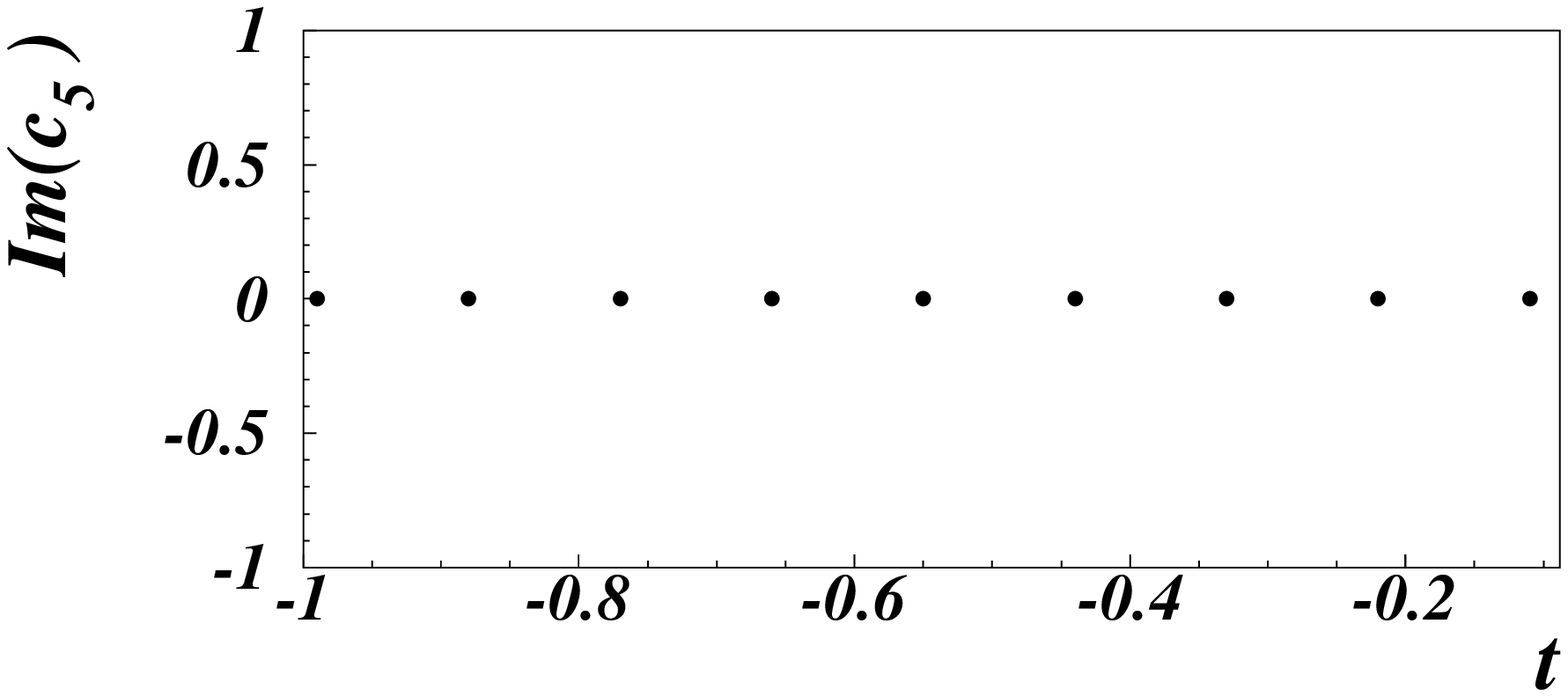}
\end{minipage}
\begin{minipage}{6.5cm}
\includegraphics[width=6cm]{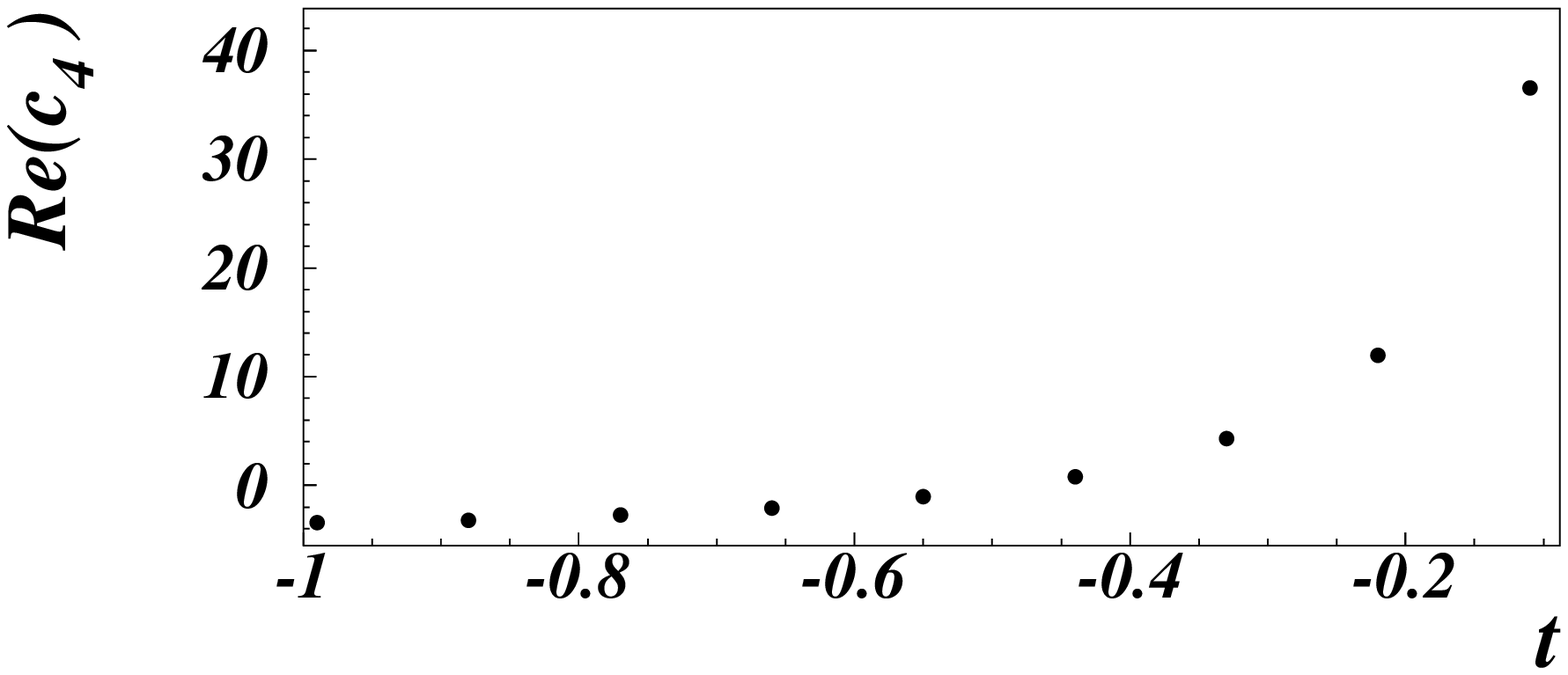}
\end{minipage}
\begin{minipage}{6.5cm}
\includegraphics[width=6cm]{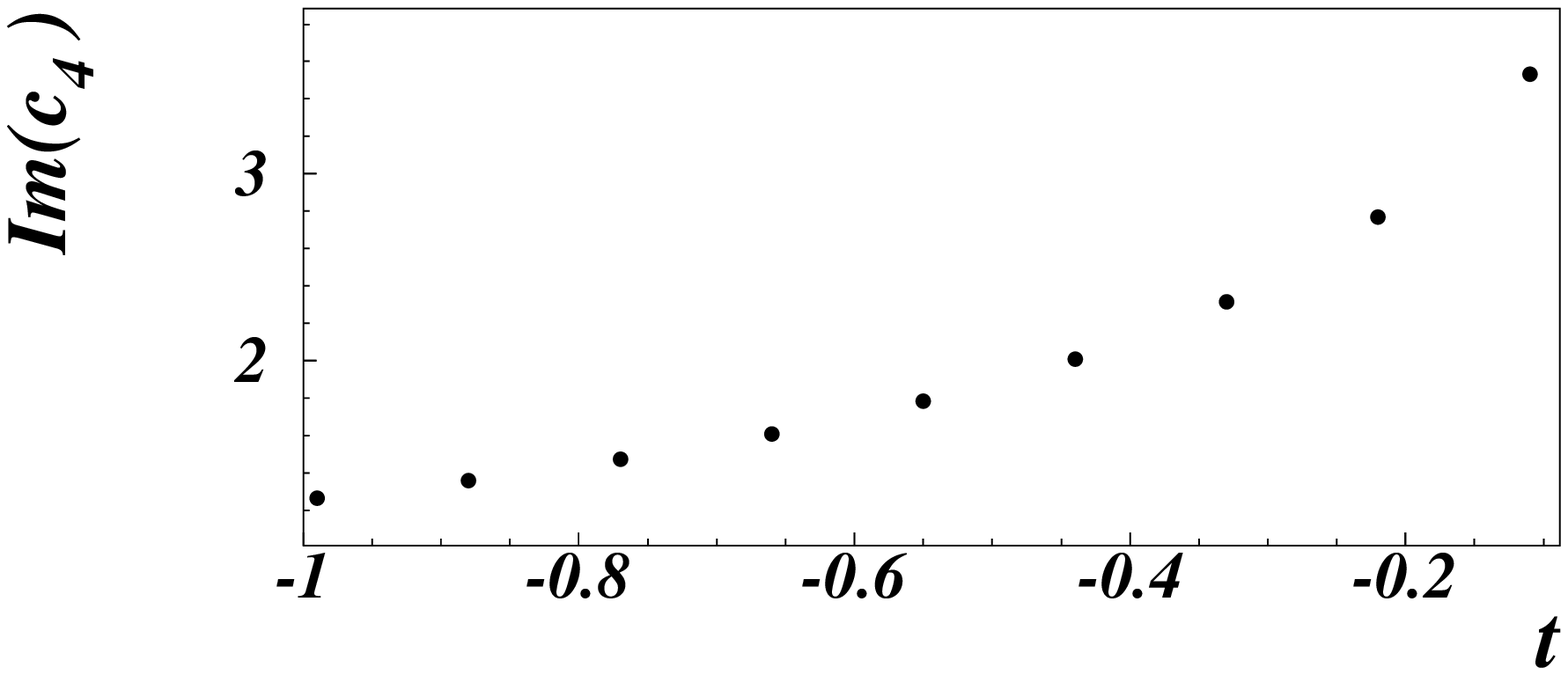}
\end{minipage}
\begin{minipage}{6.5cm}
\includegraphics[width=6cm]{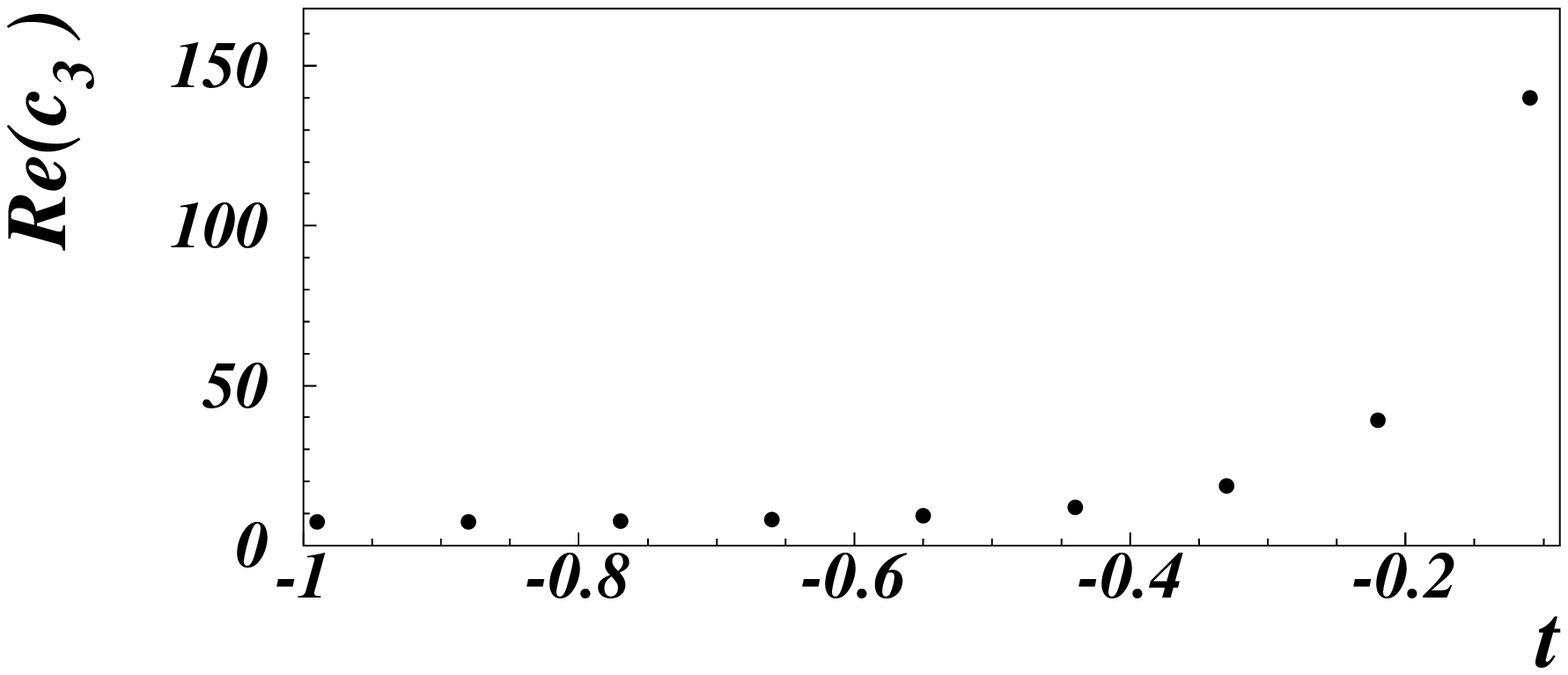}
\end{minipage}
\begin{minipage}{6.5cm}
\includegraphics[width=6cm]{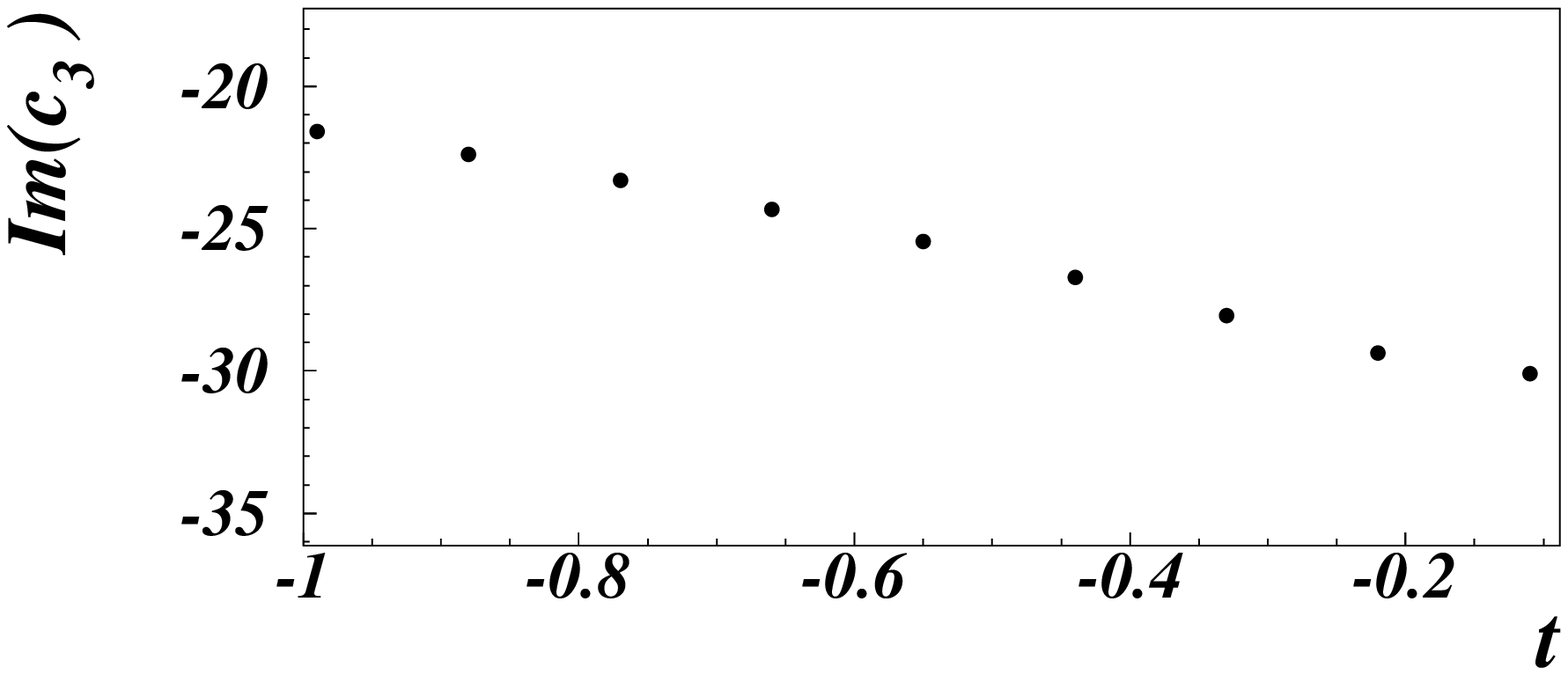}
\end{minipage}
\begin{minipage}{6.5cm}
\includegraphics[width=6cm]{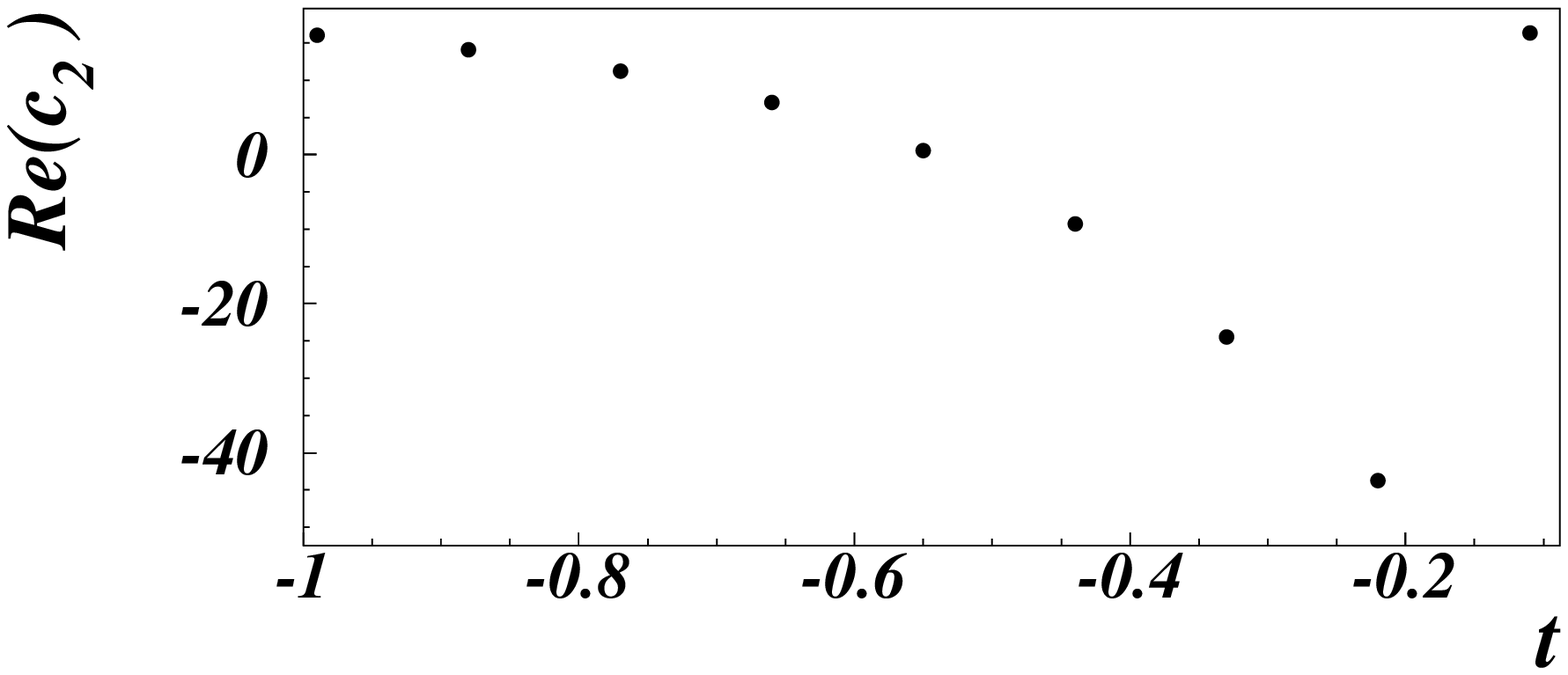}
\end{minipage}
\begin{minipage}{6.5cm}
\includegraphics[width=6cm]{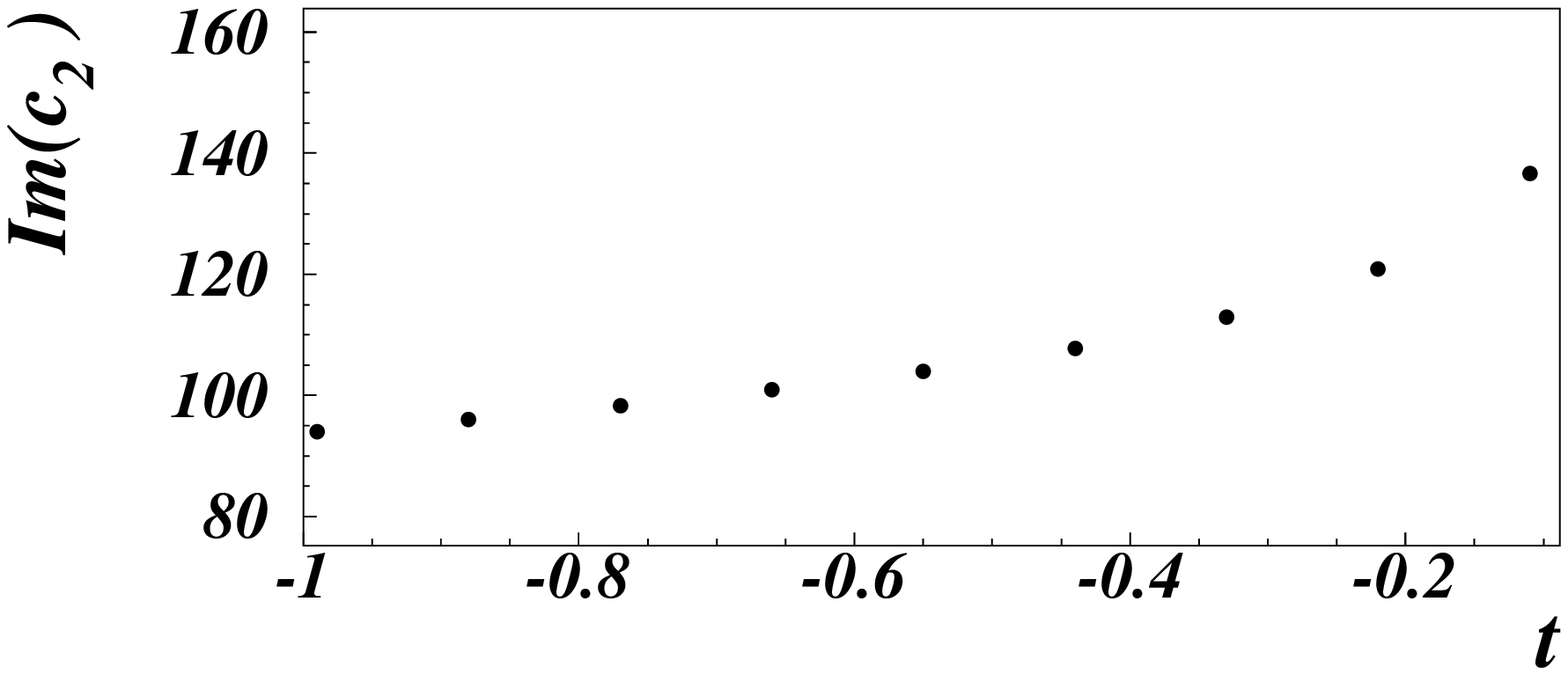}
\end{minipage}
\begin{minipage}{6.5cm}
\includegraphics[width=6cm]{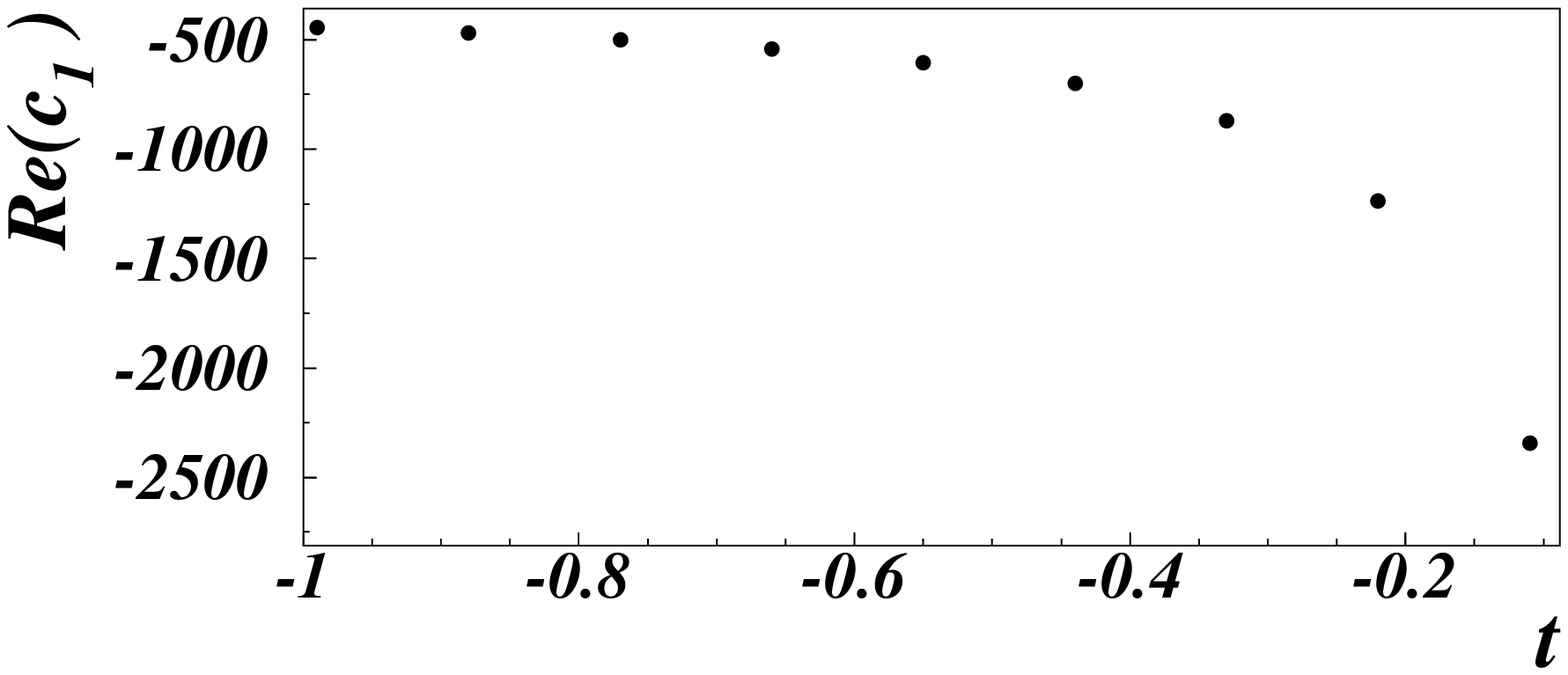}
\end{minipage}
\begin{minipage}{6.5cm}
\includegraphics[width=6cm]{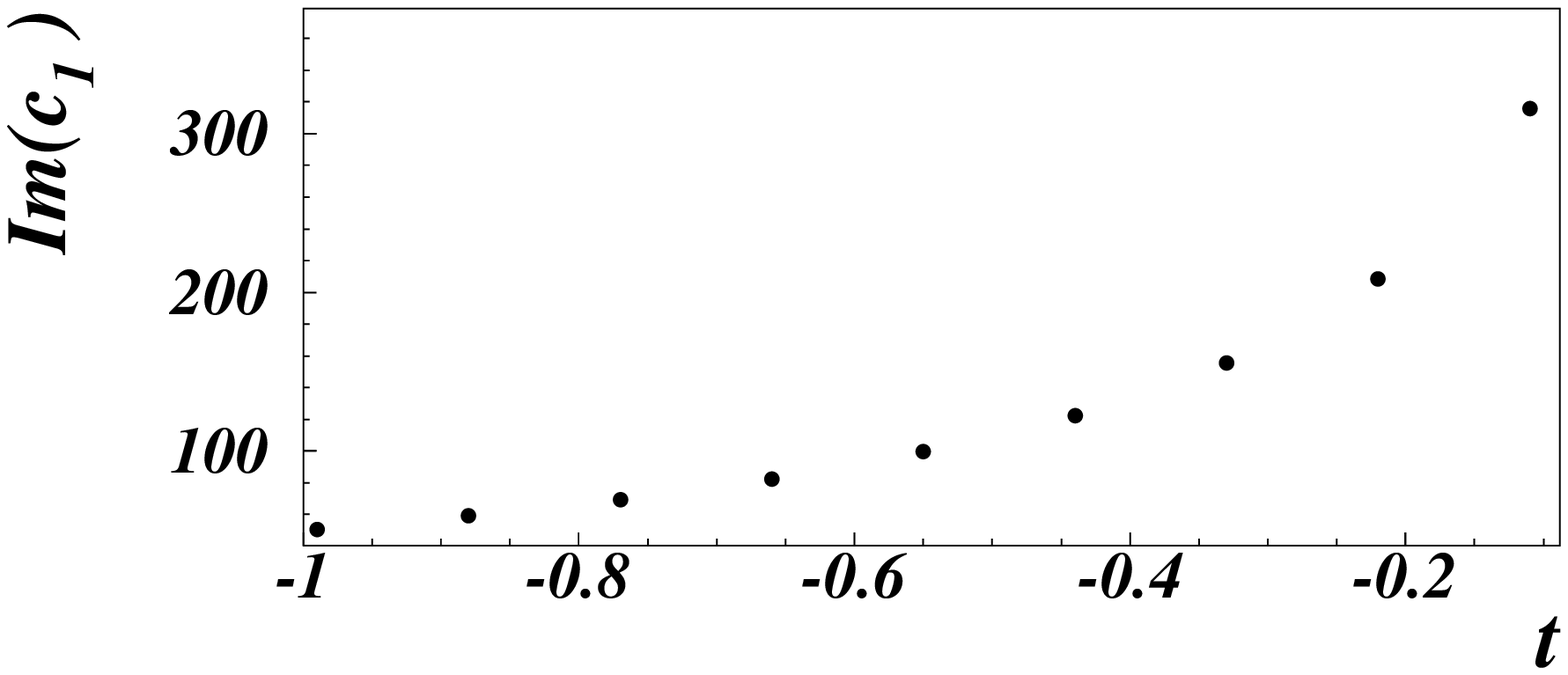}
\end{minipage}
\begin{minipage}{6.5cm}
\includegraphics[width=6cm]{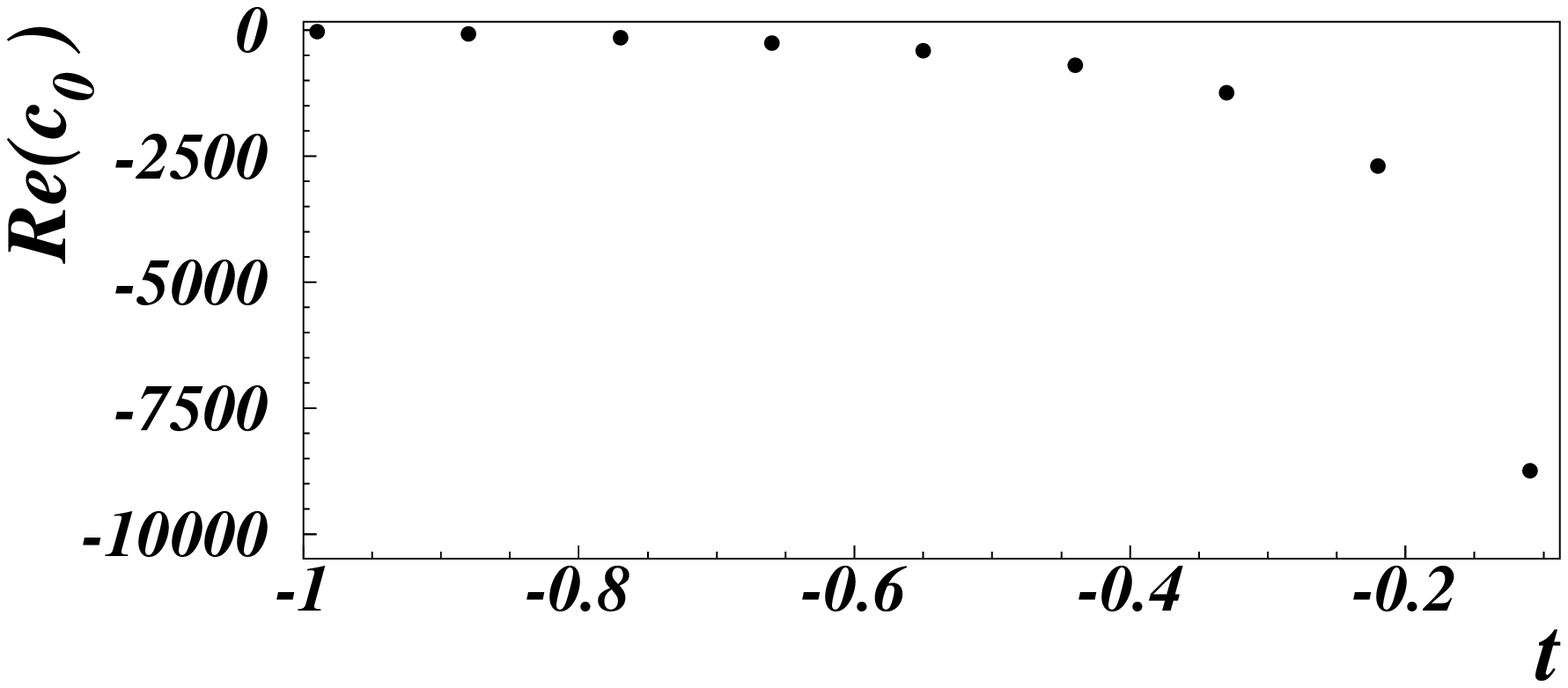}
\end{minipage}
\begin{minipage}{6.5cm}
\includegraphics[width=6cm]{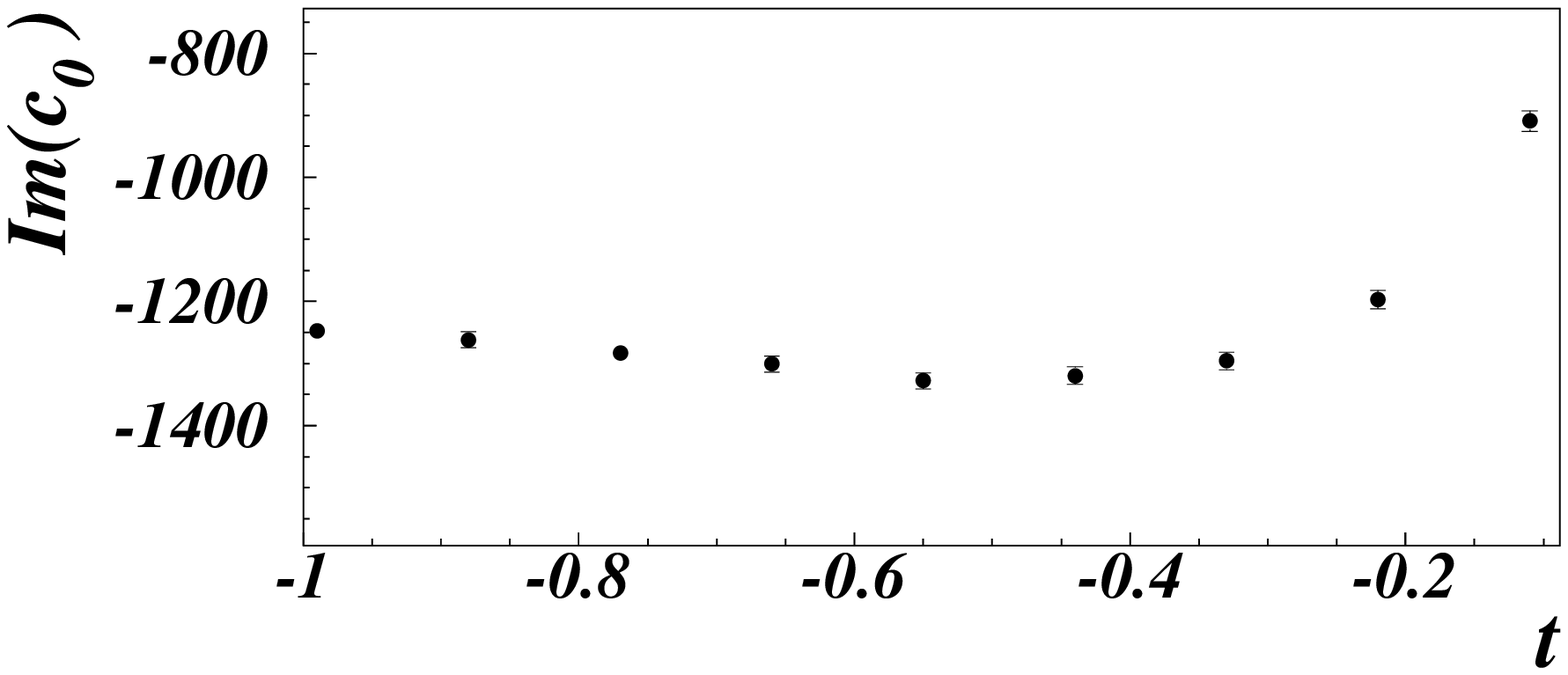}
\end{minipage}
\end{center}
\caption{Results for the  triple box with 
one leg off-shell in the physical
region for a $2\rightarrow 2$ process with the massive leg on the final state. 
We plot the real 
and imaginary parts as a function of the invariant $t=(p_2-p_3)^2$ 
for fixed value of $s=(p_1+p_2)^2=1$ and
$p_4^2=1/10$.}
\label{fig:1off3BM_scatt}
\end{figure}

\section{Conclusions and Outlook}\label{sec:conclusions}

In this paper, we have introduced a new method for  the numerical evaluation 
of arbitrary loop integrals.  We first derive Mellin-Barnes representations 
by either decomposing directly the Feynman representation, 
or by using one-loop representations as building blocks  for 
representations of multi-loop integrals. 
For a given representation, we find values of the 
space-time dimension and the powers of the propagators which yield a  
well defined representation. Then we use automated programs to perform 
the analytic continuation to values 
of the parameters for which the integral may develop divergences. Finally, 
we expand in $\epsilon$ and compute the integral 
coefficients of the expansion numerically. 

The method is based on the earlier work of 
Smirnov~\cite{smirnov} and Tausk~\cite{tausk}. The novel features in our 
publication are: (i) the automation of the procedure for the $\epsilon$
expansion,
(ii) the numerical evaluation of the coefficients of the 
expansion, avoiding cumbersome summations of multiple infinite series
and the  analytic continuation in the arguments of polylogarithms,  
(iii) and the efficient generalization of the method to tensor integrals. 
We believe that these new features render our method suitable for the 
practical evaluation of multi-loop amplitudes in gauge theories. 

We have performed a number of explicit calculations to verify our algorithms 
and demonstrate the power of our method. We first recalculated 
complicated loop integrals which are only recently known in the literature. 
Namely, we evaluated the planar and cross on-shell 
double boxes, the planar double-box with one leg off-shell, and the triple 
planar box with on-shell legs. In all cases, we have found an excellent 
agreement between our numerical results and the known analytic results. 
The analytic results require complicated analytic continuations of 
polylogarithms to non-Euclidean kinematic regions. With our method, we were 
able to compute the integrals in all kinematic regions effortlessly.  

In this paper, we have presented results for loop-integrals which were 
previously unknown. At two-loops, we have computed the double-box integral 
with two adjacent legs off-shell for independent external mass-parameters. 
This is one of the most complicated two-loop box integrals that enters 
the evaluation of two-loop amplitudes for heavy boson pair production 
at colliders. At three-loops, we have computed the planar 
triple-box with one off-shell leg; this integral emerges in 
$e^+e^- \to 3 \mbox{ jets}$, or the production 
of a single heavy boson in association with a jet at hadron colliders at
NNNLO in QCD. To the best of our knowledge, it is the first time that 
a two-loop box with four kinematic scales or a three-loop box with three
scales are ever computed in all physical regions. 

Many processes with six external legs are particularly important at the 
LHC. At present, there has been no NLO calculation of a cross-section 
for a  hadron collider process with six external states, due to the lack 
of efficient methods for evaluating loop amplitudes. 
In this paper, we applied our 
method to the evaluation of tensors through rank six 
for the hexagon topology. The purpose of this application was two-fold; 
first, we wanted to demonstrate that we were able to extend the method 
to tensor integrals economically, and second, to setup programs that 
are efficient for the evaluation of multi-scale one-loop amplitudes
(e.g. six parton QCD amplitudes). We have found that the evaluation of the
tensor integrals is not significantly more complicated than the scalar 
integral. The programs that we developed can be 
immediately used for the evaluation of the six parton one-loop QCD 
amplitudes.

Our technical goals for the current study were to build the necessary programs,
implement an efficient book-keeping platform, and to examine the scaling 
of computing intensity in applications with diverse features. We 
anticipate that our method can be improved and generalized even further, 
when we consider issues that were only briefly examined in this first study. 
For example, in preliminary investigations, we have found
that it is possible to exploit further the Cauchy theorem.  
If we specify a kinematic region, it turns out that 
some of the numerical integrals can be approximated very accurately 
by the sum of a finite number of residues. 
We expect that we can improve substantially our efficiency 
if we replace some  of the integrations which involve kinematic 
scales with  appropriate finite sums of such residues. 
In addition, we are investigating hybrid approaches, combining 
our method  with ideas in Ref.~\cite{diff_nnlo}.

The applications of our method are numerous. We believe that it will be 
particularly useful for precision calculations of observables in collider 
physics. We plan to apply our techniques to key NLO and NNLO calculations 
for the LHC and to answer more formal questions on the perturbative behavior
of gauge theories in the near future.

\section*{Acknowledgements}
We are indebted to Vladimir Smirnov,  Bas Tausk, 
Simon Schwarz and Jeppe Andersen 
for extensive discussions on the topic. 
We would like to thank Aude Gehrmann-de Ridder, Thomas Gehrmann,  
Gudrun Heinrich, Zoltan 
Kunszt, Zoltan Nagy, and Bas Tausk for their suggestions 
and for reading the manuscript.  
We are grateful to Thomas Gehrmann for providing us results and numerical 
routines from Refs.~\cite{thomas_int,harmpol1,harmpol2}, and Daniel 
Ma\^itre for
discussions about his program to evaluate harmonic 
polylogarithms~\cite{maitre}. 
This work was supported in part by the Swiss National Science Foundation 
(SNF) under contract number 200020-109162 and by the Forschungskredit der 
Universit\"at Z\"urich.

\end{document}